\documentclass[manuscript]{aastex}

\shorttitle{The P-L relation of RSGs in the SMC} \shortauthors{Yang \& Jiang}

\begin{document}

\title{The Period-Luminosity Relation of Red Supergiant Stars in the Small Magellanic Cloud}

\author{Ming Yang and B.~W. Jiang}
\affil{Department of Astronomy, Beijing Normal University, Beijing 100875, China; {myang@mail.bnu.edu.cn,
bjiang@bnu.edu.cn}}

\begin{abstract}

The characteristics of light variation of RSGs in SMC are analyzed based on the nearly 8-10 year long data collected by the ASAS and MACHO projects. The identified 126 RSGs are classified into five categories accordingly: 20 with poor photometry, 55 with no reliable period, 6 with semi-regular variation, 15 with Long Secondary Period (LSP) and distinguishable short period and 30 with only LSP. For the semi-regular variables and the LSP variables with distinguishable short period, the K$_{\rm S}$ band period-luminosity (P-L) relation is analyzed and compared with that of the Galaxy, LMC and M33. It is found that the RSGs in these galaxies obey similar P-L relation except the Galaxy. In addition, the P-L relations in the infrared bands, namely the 2MASS JHK$_{\rm S}$, \emph{Spitzer}/IRAC and \emph{Spitzer}/MIPS 24 $\mu {\rm m}$ bands, are derived with high reliability. The best P-L relation occurs in the \emph{Spitzer}/IRAC [3.6] and [4.5] bands. Based on the comparison with the theoretical calculation of the P-L relation, the mode of pulsation of RSGs in SMC is suggested to be the first overtone radial mode.
\end{abstract}


\section{Introduction}

Red Supergiant stars (RSGs) are massive evolved, He-burning, extreme Population I stars with mass range roughly between 10-30 M$_{\sun}$. They have relative low effective temperatures of about 3450 $\sim$ 4100 K (spectral type of late-K $\sim$ M), enormous radii up to 1500 R$_{\sun}$ and high luminosities of about $2000 \sim 30000$ ${\rm L_{\sun}}$ \citep{Levesque05, Massey08, Levesque10}. The very large radii make them one class of the largest stars in the universe in terms of physical size, although not the most luminous or most massive stars. Due to the large radii, their gravity is not strong enough to hold the matter in the outer layer, which gives rise to large mass loss rate (MLR) in the order of magnitude of about $10^{-6}$ to $10^{-4}$ M$_{\sun}{\rm yr}^{-1}$. This strong wind creates a thick dusty envelope that makes it very hard to determine their physical parameters \citep{Josselin00, Massey05, Groenewegen09, Yoon10, Sargent11}. Further more, RSGs have long been known for their hundreds-of-days semi-regular light variation, as a SRc class in General Catalogue of Variable Stars (GCVS; \citealp{Kholopov85}), which may be caused by the radial pulsation in the fundamental, first or even possibly second overtone mode \citep{Stothers69, Wood83, Lovy84, Schaller90, Li94, Heger97, Bono00, Guo02}. Besides, they also show an inexplicable long-term variation with a duration up to 4000 days or longer, known as the Long Secondary Period (LSP) phenomenon \citep{Stothers72, Kiss06, Messina07, Percy09}. Up to now, the mechanism of LSP is still unclear. The models of binary, pulsation, convection cell and surface hot spot are studied but none of them fully agrees with all the observations and theoretical expectations \citep{Buscher90, Wilson92, Tuthill97, Groenewegen04, Wood04, Kiss06, Messina07, Haubois09, Nie10, Stothers10}. In addition to the semi-regular short-term and LSP variations, RSGs yet presents an irregular photometric variation that is a feature of the Lc class in GCVS. It can be explained by the large convective cells in the surface of the star that are proved by some recent simulations \citep{Schwarzschild75, Antia84, Kiss06, Chiavassa10, Arnett11}.

A period-luminosity (P-L) relation is found in the light variation of RSGs, which is important for the potential use of RSG as a very luminous candle. The P-L relation of variables was first discovered by \citet{Leavitt08} for the Cepheid variables and then applied by \citet{Hubble25, Hubble26} to demonstrate unambiguously that NGC 6822 and M33 were extragalactic systems. Since then, Cepheids have become a major tool to determine the extragalactic distance. RSGs, as one class of the most luminous stars, can extend the scale of distance even further than Cepheids. \citet{Glass79} discussed the probability of using luminous RSGs as an extragalactic distance indicator originally. Then, \citet{Feast80} used infrared photometry to derive the bolometric magnitudes and demonstrated the existence of P-L relation for RSGs in the Large Magellanic Cloud (LMC). \citet{Catchpole81} confirmed the relationship in the Small Magellanic Cloud (SMC) in the same way. Later, \citet{Wood83} made use of infrared JHK photometry and low-dispersion red spectrum to investigate the Long Period Variables (LPVs) in the LMC and SMC. They found that the LMC RSGs had a similar P-L sequence to the Asymptotic Giant Branch stars (AGBs) but were approximately one magnitude brighter in the K band. Afterwards, \citet{Kinman87}, \citet{Mould90} and \citet{Dambis93} utilized the infrared P-L relation of RSGs to estimate the distance to the extragalaxy M33. \citet{Pierce00} re-calibrated the P-L relations of RSGs in the Perseus OB1 (Per OB1) association, LMC and M33 in various bands and suggested a universal relation in these heterogeneous environments to measure the distance to M101 resulting in about 6.4~Mpc. \citet{Kiss06} obtained a P-L relation similar to that of AGBs in LMC based on the analysis of 48 Galactic RSGs with almost 60-year data. After all, we (Yang \& Jiang 2011, Hereafter Paper I) made use of visual time-series photometric data collected by the ASAS and MACHO projects that span nearly 8-10 years, and established the P-L relations of RSGs in LMC in the infrared bands including the 2MASS JHK$_{\rm S}$ bands and \emph{Spitzer}/IRAC and MIPS 24$\mu$m bands. Our study proved that the semi-regular RSGs variables and the LSP variables with distinguishable short period show a tight P-L relation in the infrared bands but relatively sparse in the V band which may be caused by the inhomogeneous interstellar extinction. The results are compared with other P-L relationships for RSGs and the P-L sequences of red giants in LMC. The discrepancy between various results may be due to sample selection, the error in the periods or the indicator of the luminosity. For further improvement, we extend our work from LMC to SMC and try to find a consistent result within our Galaxy and MCs based on similar data, similar criteria for selecting RSGs and analyzing methods.

\section{Sample Selection and Data Analysis}

A sample of RSGs pure and as large as possible in SMC is the basis to derive a reliable P-L relation. As a first step, we assembled the candidates of RSGs from previous works as our targets to consist a large sample. One should be cautious for the selection because RSGs are easily confused with AGBs since both are red and luminous while RSGs are a bit more luminous and less red. \citet{Wood83} suggested that M$_{\rm bol}=-7.1$ can be a criterion to discriminate AGBs and RSGs. But sometimes the most luminous AGBs known as super-AGBs can be so bright as M$_{\rm bol}$ up to $-8$ mag \citep{Siess06, Siess07, Eldridge07, Poelarends08}. On the other side, some low-luminosity RSGs surrounded by a thick dusty envelope can be darkened to the average luminosity level of AGB stars. In addition, previous works selected RSGs in different ways. Early in 1981, infrared photometry of 22 RSGs variables in the SMC was performed by \citet{Catchpole81}. Excluding two anomalous objects, their fitting to 20 targets derived a good P-L relation as \citet{Feast80} did for LMC. \citet{Wood83} produced a new catalog of 42 LPVs in SMC from a detailed analysis of their near-infrared photometry and low-dispersion red spectra. But according to their criterion for AGBs and RSGs, there were only 5 targets identified as RSGs, which is discrepant with the result of \citet{Catchpole81}. A simple and easy solution to this discrepancy is to shift the borderline between AGBs and RSGs from M$_{\rm bol} = -7.1$ to M$_{\rm bol} = -6.1$, which would enlarge the sample size to 22 objects and coincide with \citet{Catchpole81}. A big extension of the RSGs sample in SMC was made by \citet{Massey03} who selected 105 sources through multi-object spectroscopy of a sample of red stars previously identified by \cite{Massey02} in separating bona fide RSGs from foreground dwarfs \citep{Massey98}. They simultaneously made use of high-accuracy (${\rm <1\,km\,s^{-1}}$) radial velocities for all the candidates to confirm them as RSGs. Also based on this work, \citet{Levesque06} and \citet{Massey07} identified additional 9 sources. The combination of the three samples consists the most comprehensive and reliable sample of RSGs in SMC. Recently, \citet{vanloon08} identified 8 RSGs in SMC acoording to their K-band luminosity and the VLT/ISAAC 3-4 ${\rm \mu m}$ spectra that showed the first overtone SiO band at 4 $\mu {\rm m}$ in emission (see Fig.6 in their paper). They \citep{vanloon10} also presented \emph{Spitzer} far-infrared spectra between 52 - 93 $\mu {\rm m}$ by MIPS for 2 RSGs to investigate their infrared and dust properties.

Our preliminary sample is a combination of all the samples from \citet{Catchpole81}, \citet{Wood83}, \citet{Massey03}, \citet{Levesque06}, \citet{Massey07}, \citet{vanloon08} and \citet{vanloon10}, which brings about 166 objects. Considering the redundancy between some samples, the same procedure as in Paper I is taken, i.e. to adopt the sources for its first appearance in paper published and subtract previous sources from following papers. Consequently, the final sample consists of 140 stars with 20 targets from \citet{Catchpole81}, 6 from \citet{Wood83}, 101 from \citet{Massey03}, 8 from \citet{Levesque06}, 1 from \citet{Massey07}, 3 from \citet{vanloon08} and 1 from \citet{vanloon10}. However, the sample will be further purified by the brightness and color indexes later.

The photometric data in visual bands are taken from the databases of the All Sky Automated Survey (ASAS) \citep{Pojmanski02} and the MAssive Compact Halo Objects (MACHO) projects \citep{Alcock97}. The one-epoch near- and mid-infrared photometric data are retrieved from the Two Micron All Sky Survey (2MASS) Point Source Catalog (PSC) \citep{Skrutskie06} , and the \emph{Spitzer}/SAGE Legacy Program \citep{Meixner06} database.

\subsection{Color-Magnitude Diagrams and Two-Color Diagrams}
\subsubsection{Near-Infrared Color-Magnitude Diagram $J-K_{S}$ vs. K$_{S}$}

As we mentioned earlier, AGBs and RSGs are easily confused if judged only by their brightness. For more accurate identification of RSGs, effective temperature has to be taken into account. With both brightness and effective temperature, the color-magnitude diagrams (CMDs) and two-color diagrams (TCD) are powerful tools to identify RSGs. However, RSGs sometimes have thick dusty envelop due to large MLR whose extinction affects the position of RSGs in these diagrams. In order to avoid this extinction effect as much as possible, the near- and mid-infrared bands of 2MASS, \emph{Spitzer}/IRAC and MIPS are chosen because the extinction in infrared is much smaller than in visual. The infrared magnitudes of these candidates are taken from the SAGE-SMC IRAC Combined Epoch Catalog v1.5 that already combined the Epoch 1 and Epoch 2 (the SAGE project is composed of two-epoch observations separated by three months) IRAC images and cross-associated with the 2MASS PSC. For the MIPS 24 $\mu {\rm m}$ band, there is not yet any combined catalog released, the SAGE-SMC MIPS 24 $\mu {\rm m}$ Epoch 1 Catalog is then adopted and the missing targets are extracted from the Epoch 2 and Epoch 0 (the IRAC and MIPS observations from the S$^3$MC pathfinder survey of the inner 3 square degree of the SMC, \citealt{Bolatto07}). These two catalogs are cross-associated and with high-quality photometry as the selected subsets of the IRAC and MIPS Archive Catalog. More details about the SAGE catalogs can be found in \citet{Meixner06} and the SAGE Data Description Document\footnote{http://irsa.ipac.caltech.edu/data/SPITZER/SAGE/doc/}.

The procedure of identifying the optical RSGs in the IR catalogs is the same as done in Paper I. A search radius of 1$\arcsec$ is used for the SAGE catalogs that agrees with the nominal pointing accuracy of \emph{Spitzer}. The closest and brightest counterpart is identified from all the objects within the search circle. The sources that have null results in the 2MASS or IRAC or MIPS bands are retained instead of being removed. Fig. \ref{smc} shows the spatial distribution of all the 140 sample stars superposed on the \emph{Spitzer}/SAGE 8 $\mu$m mosaic image. It can be seen that the sources are mostly located in the `bar', probably because that the SMC `bar' contains more star-forming regions and young stars than the `wing'. Table \ref{140tab} lists the coordinates and infrared brightness of all the sample targets, where the `---' symbol means the data is lacking in the corresponding band. In total, 38 measurements are lacking for 13 objects, i.e. No.34 and 132 without infrared data, No.61 with only the [24] band data, No.15 and 73 with no JHK$_{\rm S}$ band data, No.40 without the [4.5] and [8.0] band data, No.2, 27 and 55 without the [4.5] band data, No.7, 19, 86 and 108 without the [24] band data. These targets will be absent in further analysis of CMD and TCD when related. The RSGs candidates are compared with the 3654 massive stars (M $\geqslant$ 8 M$_{\sun}$) in the SMC from \citet{Bonanos10} whose infrared properties have been carefully investigated. They are collected from literatures and have been identified by the same criteria in the SAGE infrared data as ours.

The brightness is a key criterion in identifying RSGs. In Paper I, the bolometric magnitude is converted from the K$_{\rm S}$ band magnitude by a simple shift of 3 magnitudes as $\rm m_{bol}=m_{K}+3$ suggested by \citet{Josselin00} (the difference of the brightness between the K and K$_{\rm S}$ bands is very small and ignored in this work). This conversion was used for Galactic RSGs but reliable only under specific condition, i.e. the circumstellar and interstellar extinctions are already corrected. Considering that the metallicity of SMC is much lower than the Galaxy and no correction of the interstellar extinction that is comparable to the observational uncertainty (the extinction in the J and K$_{\rm S}$ band is about 0.08 and 0.03 mag respectively if $\rm E(B-V)=0.1$ \citep{Schwering91} is adopted), a different conversion is needed from the K$_{\rm S}$ band magnitude to the bolometric magnitude, especially when there are some heavily dust-enshrounded RSGs. Thus, a new linear relation between infrared bands and bolometric magnitude is derived from 113 targets in total, 72 from \citet{Massey03}, 37 from \citet{Levesque06} and 4 from \citet{Levesque07} respectively. The bolometric magnitudes of these sources are based on the spectral fitting since there may be some large (and in some cases, uncertain) reddening if bolometric magnitudes are derived from $(\rm V-K)_{0}$ \citep{Levesque06}. Fig.~\ref{band-mbol} shows the result of fitting at a form of $\rm m_{\lambda}=a\times m_{bol}+b$ which indicates a moderate discrepancy from $\rm m_{bol}=m_{K}+3$ shown as a dashed line in the K$_{\rm S}$ band. Fitting parameters are listed in Table~\ref{band-mbol-table} which will be used in determining the limits of brightness in the CMDs.

Hereafter, the ID number in Table~\ref{140tab} is used to code the targets in CMDs and TCDs for convenience. Fig.~\ref{figkjk} is the near-infrared CMD, the K$_{\rm S}$ versus $\rm J- K_{S}$ of all targets. In this diagram, most targets are located in a region of $7.5 < \rm K_{S} < 10$ and $0.5 < \rm J- K_{S} < 1.6$, which have similar range in $\rm J- K_{S}$, but fainter in the K$_{\rm S}$ band compared to LMC. The fainter K$_{\rm S}$ brightness can be partly explained by the larger distance of SMC, and also possibly the fewer number of very massive RSGs in SMC.

The boundaries of luminosity and color index are set up in the same way as in Paper I with a modification in the boundaries of the K$_{\rm S}$ magnitude. The new boundaries of the K$_{\rm S}$ magnitudes are 7.6 and 11.0 shown as the dashed horizontal lines in Fig.\ref{figkjk}, which is determined based on the mass-luminosity relation for massive stars (9-30 M$_{\sun}$), $\rm L/L_{\sun}=(M/M_{\sun})^\gamma$ with $\gamma$ very close to 4.0 \citep{Stothers71}. From the luminosity, the bolometric magnitude is calculated by $ \rm M_{bol}=4.74-2.5\log(L/L_{\sun})$ and the K$_{\rm S}$ magnitude is derived from our newly derived relationship between $\rm m_{K_{S}}$ and $\rm m_{bol}$ as listed in Table~\ref{band-mbol-table} with a distance modulus of 18.91 for SMC \citep{Hilditch05}. The corresponding mass of the most targets ranges from 13 M$_{\sun}$ to 30 M$_{\sun}$, a bit higher than used (10-25 M$_{\sun}$) in Paper I, which is consistent with the expected mass range of RSGs although lacking the low-mass (9-13 M$_{\sun}$) RSGs. For the color index $\rm J-K_{S}$, the limits of 0.5 and 1.6 mag correspond to the lower observational boundary defined by \citet{Josselin00} and the boundary of carbon-rich stars by \citet{Hughes90} respectively. With the restrictions on both the K$_{\rm S}$ band magnitude and the color index $\rm J-K_{S}$, 130 targets (92.9\%) are qualified.

Other suggestions on the criteria for RSGs are checked as well. The red limit of 1.2 mag in $\rm J-K_{S}$ is determined from the evolutionary synthesis models of RSGs in near-infrared by \citet{Origlia99} (see Figure 3 and 11 in their paper; although their model fitted the metallicity of LMC, there should be no remarkable difference to shift the limit of 1.2 mag for SMC). The lower limit of $\rm K_{S}=9.5$ mag is suggested to convert from $\rm M_{bol}=-7.1$ mag to distinguish the AGBs and RSGs by \citet{Wood83}, nevertheless there are alternative values of M$_{\rm bol}$ to discriminate RSGs from AGBs, such as M$_{\rm bol}=-7.5$ mag by \citet{vanloon99}, M$_{\rm bol}=-8.0$ mag by \citet{Groenewegen09}. These two criteria (K$_{\rm S}=9.5$ mag and $\rm J-K_{S}$=1.2 mag) are shown in the diagram as dotted lines. The modified limit of K$_{\rm S} < 9.5$ mag picks out 111 targets (79.3\% of the sources), and the limit $\rm J-K_{S} <$ 1.2 mag picks out 126 targets (90\% of the sample), both of which are satisfied by most of the objects. There are 103 targets within both the K$_{\rm S}$ magnitude and color index limits, i.e. 73.6\% and most of them are in the \citet{Massey03} catalog that confirmed the reliability of their sample. However, any dust-enshrouned RSGs will be so red that mimics the colors of AGBs. There is no effective way to distinguish very dusty RSGs from dusty AGBs with IR photometry solely, but they should have different features from the less-dusty counterparts \citep{Boyer11}, and they may also have different characteristics in the optical bands.

In Fig.\ref{figkjk}, five outliers are marked by their ID numbers shown as Nos.3, 22, 27, 52 and 86. Among them, No.3, No.52 and No.86 are too faint to satisfy the lower limit of the K$_{\rm S}$ band luminosity. No.22 exceeds the red borderline at $\rm J-K_{S}=1.6$ mag that may be caused by the heavy dust envelopes, and will be further investigated in MIR CMD since targets will have high luminosity in MIR if a large amount of dust exists. Although No.27 is a bit above the luminosity upper limit of a 30 M$_{\sun}$ star, it is not excluded considering the uncertainty in the determination of the luminosity at the high end of the mass and the proximity to the borderline.

Similarly, \citet{Boyer11} divided the NIR CMD into several sub-regions for various types of evolved stars as shown in Fig.~\ref{figkjk2}, where the $\rm J-K_{S}$ color cuts shown as solid lines follow \citet{Cioni06a, Cioni06b}, with C-rich AGBs (C-AGBs) defined by $\rm K_{S} < K0$ and $\rm K_{S}> K2$, {O-rich AGBs (O-AGBs) defined by $\rm K_{S}< K0$ and $\rm K1 < K_{S}< K2$, extreme-AGBs (x-AGBs) mostly by $\rm K_{S} < K_{S}$-band tip of the red giant branch (TRGB) (shown as dotted line) and $\rm J-K_{S}>2$ (see Figure 5 in \citealt{Boyer11}). Concerning RSGs, \citet{Boyer11} classified RSGs by restricting the the branch width to $\Delta({\rm J-K_{S}})=0.2$ mag and leaving a 0.05 mag gap from the O-AGBs shown as dashed lines in our diagram. The various regions defined by \citet{Boyer11} in the CMD are contoured in red, blue and purple solid lines for RSGs, O-AGBs and C-AGBs respectively. For a better view of distinguishing the RSGs from AGBs, all the AGBs, i.e. O-AGBs, C-AGBs and x-AGBs from Table 4 of \citet{Boyer11}, are overplotted in this diagram as well. It is clear that the AGB stars are less luminous and redder than the RSGs. With these two restrictions, most of our targets apparently locate in the RSGs area (the branch width probably needs to be extended blueward) and only some dusty ones or possible AGBs left out of the region. Almost all AGBs in Figure 5 of \citet{Boyer11} are fainter than the limit of $\rm K_{S}=9.5$ mag (shown as dotted line) except one x-AGB. This confirms that our determination of the separative limit by converting from the bolometric magnitude is acceptable. The discrepancy between our and Boyer's results is that the magnitude range of their RSGs sample is twice of ours and has a downward tendency blueward to the K$_{\rm S}$-TRGB. Meanwhile, our sample has a higher luminosity tip which indicate that more luminous sources are included. But according to the limit of $\rm M_{bol}=-7.1$ mag and simple theoretical calculation of RSGs mass-luminosity relation (shown as dashed line in the diagram), Boyer's sample may contain lots of AGBs. In a word, the sample of Boyer has too low luminosity and too narrow and blue color index. In order to fit our sample, their region of RSGs should be moved redward 0.1 mag and upward 1.5 mag shown as a red dotted line in Fig.~\ref{figkjk2}. After such adjustment, the region looks reasonable for RSGs. The outliers in previous CMD are mostly in the AGBs area with No.3, NO.52 and No.86 possible being O-AGBs and No.22 belong an x-ABGs.

\subsubsection{Mid-infrared Color-Magnitude Diagrams and Two-Color Diagrams}

Most RSG candidates are already confirmed in the NIR CMDs. Considering RSGs are in the evolved phase and have large MLRs, MIR CMDs will shed light on the circumstellar dust of RSGs. Several papers have pointed out that RSGs have circumstellar dust features from 8 to 12$\mu$m \citep{Hagen78, Skinner88, Josselin00, Ohnake08, Verhoelst09}, with the 9.7$\mu$m silicate feature prominent.

The color-magnitude diagram of [3.6]/$\rm J-[3.6]$ for all targets is shown in Fig.~\ref{fig36j36}. The 3.6 $\mu$m band is the most sensitive IRAC band comparable to the 2MASS K$_{\rm S}$ band. The [3.6] band is relatively free of dust and molecular emission/absorption \citep{Verhoelst09}. The diagram of [3.6] versus $\rm J-[3.6]$ reveals mainly the property of the stellar photosphere, where the region of AGBs set by \citet{Bonanos10} is shown as black solid line. The three outliers in previous NIR CMDs, Nos.3, 52 and 86, are within the region of AGB stars, so they may be AGB stars instead of RSGs. The region of RSGs by \citet{Boyer11} is shown by red solid line. Also shown are the limits of the [3.6]-band magnitude derived from the mass of RSGs and the bolometric magnitude as dashed horizontal lines. No.27 is still above the 30 M$_{\sun}$ line. The red edge of color index $\rm J-[3.6]$ is 2.6 mag, i.e. the reddest edge of AGBs by \citet{Bonanos10} since the dust-producing rates of RSGs are similar to AGB stars although RSGs are less numerous \citep{Boyer11}. Again, No.22 is beyond this limit, indicating a really cold dust component. The blue limit is $\rm J-[3.6]=0.6$ mag, i.e. the bluest color of RSGs region set by \citet{Boyer11}. With both the brightness and color limits, the outliers are the same as previous selection in the J/$\rm J-K_{s}$ diagram, and it is also the same that most of our targets are settled in the RSGs region.

Fig.~\ref{fig8k8} and Fig.~\ref{fig8j8} are CMDs of [8.0] versus $\rm K_{S}-[8.0]$ and $\rm J-[8.0]$. Following previous procedures, the [8.0] magnitude limits are set between $[8.0]=6.8$ mag (for a mass of 30 $\rm M_{\sun}$) and $[8.0]=9.9$ mag (13 $\rm M_{\sun}$) in both diagrams, while these limits cannot be strict since the the [8.0] radiation comes not only from the photosphere but also the warm dust around the stars. The effect of dust is seen in the [8.0]/$\rm K_{S}-[8.0]$ diagram, where $\rm K_{S}-[8.0]$ becomes large with the increased [8.0] brightness, which can be fitted linearly by $\rm [8.0]=(-2.40\pm0.09)(K_{S}-[8.0])+(9.59\pm0.08)$ with $\sigma=0.56$ shown as a dotted line. Consistently, the red limit is set at $\rm K_{S}-[8.0]=1.2$ mag, the cross-point of the fitted line with the upper limit of luminosity at [8.0]. The blue limit is $\rm K_{S}-[8.0]=0.1$ mag since no targets bluer than this. It's interesting to notice that the 3645 massive stars in SMC cluster in our two color limits (0.1 and 1.2 mag in $\rm K_{S}-[8.0]$) respectively, while RSGs distribute quite evenly between the limits. As expected, there are more targets lying out of these restrictions, and most of the outliers are beyond the red limit, which can be explained by the dust emission. Besides the outliers Nos.3, 22, 27, 52, 86 already showing up in previous CMDs, there are new outliers Nos.2, 4, 23, 35, 55, 60, 77, 130, 140 all with $\rm K_{S}-[8.0] >1.2$ mag. In the [8.0]/$\rm J-[8.0]$ diagram, the stellar classifications of \citet{Boyer11} are shown as red solid lines. The brightness and color limits are set in the same way as in Fig.~\ref{fig8k8}, except the linear fit is changed to be $\rm [8.0]=(-2.1\pm0.07)(J-[8.0])+(11.54\pm0.10)$ with $\sigma=0.57$ that results in the limits of $\rm J-[8.0]$, 2.3 and 1.0 mag in the red and blue respectively. The dashed line at $[8.0]=8.4$ mag is the upper limit of AGBs in the [8.0] band (see Figure 4 in \citealt{Boyer11}). The outliers, Nos.2, 3, 4, 22, 23, 27, 35, 52, 55, 77, 86, 130, 140, are the same as before, No.60 is no longer outlying and the new outlier No.97 is marked.

The MIPS [24] band reflects the status of colder dust in comparison with the [8.0] band \citep{Blum06, Bonanos09, Bonanos10, Boyer11}. Fig.~\ref{fig24824} shows the [24]/$[8.0] - [24]$ CMD. The correlation of the brightness in [24] with the color index $[8.0]-[24]$ is very tight since both reflect the radiation of the dust. The linear fit for this relationship is $[24]=(-1.96\pm0.05)([8.0]-[24])+(9.03\pm0.06)$ with $\sigma=0.45$. With the brightness limit in the [24] band between 5.0 and 10.3 mag for the mass range of RSGs, the color index $[8.0]-[24]$ is correspondingly set to -0.4 and 2.1 mag. Considering the uncertainty in converting the mass to the brightness and the inclusion of the dust emission in the [24] band, the upper limit of the [24] band cannot be taken too strictly. The sources Nos. 27 and 131 are thus kept in the sample even though they lie above the upper limit line. In fact, the color index can also be influenced by the dust in addition to the photosphere, five targets, Nos. 22, 23, 35, 77, 130, with $[8.0]-[24]$ redward of the limit, are neither excluded. On the contrary, their $[8.0]-[24]$ indexes explain their deviation from the majority in the NIR CMDs. For the same reason, Nos.3 and 52 previously classified as outliers are inside the limit box in this diagram.

Similar to the CMDs, TCDs involving IR colors are good indicators to classify stellar populations. Combined CMD with TCD, the characteristics of RSGs in IR bands should be clear and distinguishable from other populations.

Fig.~\ref{figjkk8} and Fig.~\ref{fig36808024} are TCDs used in \citet{Bonanos10} and Fig.~\ref{fig36454580} is used in Paper I. The black solid and dashed lines represent the Black Body (BB) at various temperatures and a model with 3500 K BB plus 250 K dust from \citet{Bonanos10} respectively (see Section 4 of \citealt{Bonanos09} for detailed description). Caution should be called that the spectra of RSGs dominated by emission lines make the blackbody not a perfect approximation to the atmosphere spectral energy distributions (SEDs). In Fig.~\ref{figjkk8}, most targets are located close to the point of the 3500 K BB. But there is still a discrepancy between the model and the targets that the targets are a bit bluer than the BB line, which may be due to the dust emission in the [8.0] band. The outliers are the same as in [8.0]/$\rm K_{S}-[8.0]$ and not marked in this diagram.

Fig.~\ref{fig36454580} shows the $[3.6]-[4.5]$ vs. $[4.5]-[8.0]$ TCD. Due to the depression at the [4.5] band by the CO lines \citep[see][]{Verhoelst09}, most targets deviate from the BB model significantly. The obvious outliers in this diagram are Nos.4, 22, 23, 35, 85, 130. In Fig.~\ref{fig36808024}, the $[3.6]-[8.0]$/$[8.0]-[24]$ TCD, many targets are below the 3500 K BB plus 250 K dust model line, which can be explained by the emission of silicate dust in the [8.0] band.

\subsubsection{Comparison of RSGs in SMC and LMC}

Paper I studied the RSGs in LMC. As the nearest two galaxies, both SMC and LMC present a possibility to spot the objects individually. With different metallicity, Z$_{\rm SMC}\sim$0.004 and Z$_{\rm LMC}\sim$0.008 \citep{Glatt10}, the comparison between them will reveal the effect of metallicity on RSGs. \citet{Origlia99} showed that higher metallicity brings about redder $\rm J-K_{S}$ by using instantaneous burst and continuous star formation models for the Galaxy and LMC (see Fig.3 of their paper). The histogram of $\rm J-K_{s}$ in Fig.~\ref{figlmcsmcjk} confirms this conclusion. The peak value of $\rm J-K_{S}$ for RSGs in SMC is about 0.1 mag bluer than in LMC. On the other hand, the bolometric magnitude hence the K$_{\rm S}$ band absolute magnitude is hardly dependent on the metallicity with a difference between LMC and SMC smaller than 0.1 mag \citep{Levesque06}. With a distance modulus of 18.50 for LMC \citep{Alves04} and 18.91 for SMC, the observed K$_{\rm S}$ magnitude is converted to the absolute magnitude M$_{\rm K_{\rm S}}$ in Fig.~\ref{figlmcsmcmk}, where No.22 with $\rm J-K_{S}=2.79$ mag is not included for an efficient view, and a shift of 0.1 mag towards red is added for the RSGs in SMC to compensate for the metallicity difference. It can be seen that most RSGs are mixed together in this diagram irrespective of their location in LMC or SMC. But comparing with 33 targets in LMC, there are only 9 targets in SMC with $\rm M_{K_{S}}<-10.84$ mag corresponding to a mass of 25 M$_{\sun}$, which implies fewer high-mass RSGs in SMC. In addition, there are only 25 SMC targets with $\rm J-{K_{S}}>1.2$ mag while 48 such targets in LMC after the 0.1 mag shift is corrected. This result confirms that the dust produced by RSGs are less in the SMC than in the LMC \citep{Boyer11}.

Given all these characteristics in color and brightness, the situation is complicate since some targets lack data in specific bands and some targets are within the limits in this criterion while beyond the limits in that criterion. In Fig.~\ref{fig36808024}, most targets that appear to be outliers in other diagrams actually are not far from the BB plus dust model, while Nos.4 and 130 are and thus excluded. Also excluded are Nos.3, 52 and 86 for their too low luminosities and no much dust to account for the low luminosity. In addition, Nos.34, 61 and 132 lack almost all bands data. At last, 132 targets are identified as RSGs and kept for further analysis.

From the CMDs and TCDs, it can be seen that almost all massive sources from \citet{Bonanos10} that locate in the same regions as RSGs are included in our sample. This indicates that our sample is nearly complete for the high-mass RSGs. But according to previous discussion, Boyer's sample may contain more low-mass RSGs, which needs further investigation since the lower mass limit of RSGs is somewhat uncertain.

\subsection{Period Determination}

The photometric data are collected from the ASAS and MACHO projects in the same way as in Paper I. Most of the photometry data comes from ASAS because the RSGs, except those heavily dust-enshrouded with extra visual extinction, easily get saturated in the MACHO observation due to their great intrinsic luminosity \citep{Massey05}. The saturation limit is about 13 mag in the Kron-Cousins V band and about 14 mag in the R band (Kem Cook, private communication) for the MACHO project. The photometric precision of MACHO is about 0.02 mag \citep{Alcock99} and of ASAS about 0.05 mag \citep{Pojmanski02}.

The way to handle the photometric data also follows Paper I. For ASAS, only the 10-year long data in the standard Johnson V band is analyzed while the dropped I band data covers only about 500 days. The search radius is fixed to be 30$\arcsec$ to retrieve the photometric data from the project web site and follow-up visual inspection of DSS image is performed to exclude confusions. Afterwards only the counterpart within 10$\arcsec$ of the position is accepted (Grzegorz Pojmanski, private communication). The criteria exclude 22 targets with poor measurement, 4 targets with more than two sources within the 10$\arcsec$ circle and 1 target with no counterpart. With the exclusion of 27 sources from just identified 132 candidates, 105 targets are left with a reliable association with the ASAS data. The original ASAS data is further processed for a better quality. First, a 3-sigma clipping is applied to remove the deviating measurements \citep{Hoaglin83}. Then the least-square (Savitzky-Golay) polynomial smoothing filtering is performed to the light curve \citep{Press92}. Finally a 10-day bin is done to smooth out the short-term light variation. The process of handling the ASAS photometric data is shown in Fig.~\ref{asas_lightcurve_process} for No.12 with original light curve, smoothed light curve and binned light curve respectively.

For the MACHO data, the non-standard two-color photometric system are transformed to the standard Kron-Cousins V and R system in the same way as in Paper I. With a radius of 3$\arcsec$ to search for the MACHO counterpart, 8 targets from the 132 candidates are found. One of them has the ASAS counterpart, three does not have and the others have poor ASAS photometry. Because of the low temperature and thick envelope, RSGs are usually much brighter in the R band than in the V band that makes R band easily get saturated and useless. In the process to convert the template magnitude into the standard system, the measurements with air mass larger than 2.0 are deleted. The measurements with photometric error bigger than 0.2 mag are also dropped. One more important thing concerning the MACHO data is the morphology of light curve. Since the definition of RSGs variables is {\it semi-regular with amplitudes of about 1 mag} \citep{Kiss06}, the targets with regular periodic variation or/and amplitude larger than 1 mag are excluded. The left column of Fig.~\ref{macho_outliers} exhibits the light curve of the three outliers in the MACHO database, i.e. Nos. 3, 52 and 86, with regular variation and large amplitude, which may be AGB stars with luminosity lower than RSGs shown in the CMDs. But there are 6 more sources in the 132 RSG candidates, Nos.15, 22, 23, 60, 66 and 87, with similar light curve of regular variation or/and large amplitude, three of which are shown in the right column of Fig.~\ref{macho_outliers}. Excluding these 6 sources, only Nos.46 and 59 are left as RSGs with reliable MACHO photometric data. Moreover, the sample of RSGs are reduced to 126 targets after excluding the 6 sources with regular or large amplitude light curves.

Combining the ASAS and MACHO data as a whole, there are 107 RSGs with available long-term visual photometry to derive the periods. The resource for the target is labeled by 'M' for MACHO and 'A' for ASAS in Table~\ref{140tab}.

RSGs are found to be variable for long time. But their variation is not regular that makes the period determination a tough task. On the other hand, a reliable determination of the period is the key to the P-L relation. In order to obtain the period as accurate as possible, four methods are tried to find the most consistent period.

The first method used is the Phase Dispersion Minimization (PDM) method \citep{Stellingwerf78}, while PDM2 takes place of the original version used in Paper I. PDM2\footnote{http://www.stellingwerf.com/} has many new features such as improved curve fits, suppressed subharmonics, and beta function statistics. This method is very useful in particular for data sets with gaps and non-sinusoidal variations. Since the light variation of RSGs is mostly non-sinusoidal and ASAS data have lots of gaps, it seems to be an appropriate technique. A sample of the PDM2 processing is shown in Fig.~\ref{pdm_process} for No.12 (for comparison, the same target is taken as the example in discussing other methods). Two significant periods are present in the diagram with 1485 d (LSP) and 351 d respectively. The significance levels of 0.05 and 0.01 are shown in dashed lines in the diagram.

The second method to derive the period of light variation is Period04 \citep{Lenz04}. Because the Long Secondary Period involved in the light variation of RSGs often exceeds the time span of the observation and leads to an uncertain period by PDM, Period04 \citep{Lenz04}, based on the Fourier transform and least-square fitting, can make cross-confirmation. The Period04 processing of No.12 is shown in Fig.~\ref{p04_process} (see our Paper I for a detailed description of PDM and Period04 procedures). The results are similar to PDM2, with two periods at 361 d and 1553 d.

The third method is newly taken into use this time, i.e the CLEAN method. This method deconvolves the spectral window function from the ``dirty'' Discrete Fourier Transform (DFT) spectrum by using a 1-D version of the iterative CLEAN algorithm \citep{Hoegbom74}. During each iteration, the window function is centered at the amplitude peak. Then the full amplitude of this peak is inserted into the complex array of ``CLEAN components''. To form a residual Fourier spectrum, the window function is scaled by the ``CLEAN components'' and subtracted from the input DFT. The process is repeated, with the residual spectrum from previous iteration used as the input spectrum. After several iterations, the ``CLEAN components'' is convolved with the Gaussian ``beam'', truncated at 5 sigma. Finally, the residual Fourier transform is added to this convolution to produce the CLEANed DFT spectrum. Through these steps, the information contained in the peak will be remained with the physical frequencies while their aliases or pseudo-aliases are decoupled. The example is shown in Fig.~\ref{clean_process}. The periods derived by the CLEAN method are 371 d and 1625 d.

It can be seen that the above three methods derive quite consistent results, all with two periods and agreeing with each other within about 10\%. But, there is still some uncertainty in the values, which may be caused by the variation of the period itself. The Weighted Wavelet Z-Transform (WWZ) \citep{Foster96} is used to evaluate the stability of the detected periods. The WWZ uses a standard Fourier fitting algorithm but with a sliding Gaussian weighting window whose width is determined by frequency and user-defined constant. The algorithm performs a sinusoidal wavelet to fit the data. The weight of each point is determined by the sliding window function, which means that the point close to the center of the window has heavy weights in the fitting. The window slides along the data set to represent the signal content of data as a function of time. The parameter \emph{c} which weighs the time resolution with frequency resolution is set to 0.005. The frequency range tested is from 0.0003 to 0.01 c/d with a step of 0.0001. An example is shown in Fig.~\ref{wwz_process}, where the long period actually has a variation from about 1400 d to about 1600 d that explains the consistency of the PDM2, Period04 and CLEAN methods.

\section{Period-Luminosity Relation}

Based on the analysis of the time-series photometric data, the 126 targets which have been identified by CMDs, TCDs and light curve morphology are divided into five categories. (1) The first category includes 20 RSGs. They are either too bright and saturated in photometry or in a crowded stellar field with a close companion unresolvable. These stars are not taken into account in further investigation. (2) The second category includes 55 RSGs. They are irregular variables with so complex light curve that no reliable period can be determined. They are neither considered for further study of the P-L relation. The other 51 RSGs are semi-regular or LSP RSGs which are involved in the determination of the P-L relation. (3) The third category includes 6 RSGs, being semi-regular variables with period statistical significance less than or equal to 0.05. They are all from \citet{Massey03}. (4) The fourth category includes 15 RSGs with LSP and distinguishable short period, 1 target from \citet{Catchpole81}, 11 from \citet{Massey03}, 2 from \citet{Levesque06} and 1 from\citet{Massey07}. (5) The fifth category includes 30 RSGs with LSP but no distinguishable short period. The relatively high-accuracy period can only be achieved for the third and fourth categories, i.e. the 21 RSGs with semi-regular variation or with distinguishable short period variation. It's them that are taken into account when deriving the P-L relation afterwards. Their periods are listed in Table \ref{srtab}, with also the period averaged over three methods and the dispersion most of which are shorter than 10 days. For an intuitive view, examples of irregular, semi-regular and LSP RSGs variation as well as the two light curves from the MACHO database are shown in Fig.~\ref{lightcurve}.

We also calculate the relation of period and amplitude for the RSGs in Table \ref{srtab}. Since there are two groups of period and amplitude derived by PDM and Period04 respectively, they are combined to make a robust linear fitting. The result is \\
\indent $\rm \bigtriangleup V = (0.76 \pm 0.14) \times \log P - (1.83 \pm 0.24)$\\
with $\sigma=0.10$, where $\bigtriangleup V$ is the amplitude in the $V$ band and $P$ is the period. Except few outlying points, most points roughly obey the relation. This is consistent with the general tendency of variables that the longer the period the greater the amplitude. The distribution of the amplitude and the period is shown in Fig.~\ref{figpa}.

With the period of variation determined, luminosity is the other key parameter in deriving the P-L relation and deserves as important attention as period. Historically, V band is the most common observed and used as the indicator of luminosity, but visual band suffers interstellar and/or circumstellar extinction severely.

Fortunately, the extinction influence is much smaller in near- or mid-infrared bands, e.g. the extinction in the GLIPMSE/IRAC bands is only about 25th of the visual extinction \citep{Gao09}. Therefore, the infrared bands, 2MASS JHK$_{\rm S}$, IRAC and MIPS 24 $\mu m$ band, are taken as the luminosity indicator while the V band is included only for comparison.

\subsection{K$_{\rm S}$-band Period-Luminosity Relation}

In Paper I, the P-L relation is determined for the LMC RSGs based on 47 semi-regular and LSP RSGs with distinguishable short period. Here a similar work is extended to the SMC, based on 6 semi-regular and 15 LSP RSGs with distinguishable short period listed in Table~\ref{srtab}. Fig.~\ref{smcplk} shows the robust linear fitting for the K$_{\rm S}$-band P-L relation of SMC RSGs, where filled and unfilled squares denote the semi-regular and LSP RSGs respectively. The fitting takes a form of $\rm m_{\lambda} = a \log P + b$ and the result is \\
\indent $\rm K_{S}=(-3.28\pm0.39) \times \log P+(17.25\pm0.64)$\\
With a dispersion of $\sigma=0.27$, the relation is very tight.

The P-L relation has been calculated for RSGs in the Galaxy and M33 by \citet{Kiss06} (13 targets) and \citet{Pierce00} (52 targets) respectively. For independence and consistency, we re-calculate the P-L relation in the K$_{\rm S}$ bands with some modification because the photometric data in infrared is only available in the K$_{\rm S}$ band for RSGs in the Galaxy and M33. The absolute K$_{\rm S}$-band magnitude of Galactic RSGs is taken from \citet{Kiss06} (see section 4.1 in their paper for detail description), but with addition of 3 targets, GP Cas, YZ Per and V500 Cas from \citet{Pierce00} using a distance modulus of 11.70 \citep{Southworth04} since they are all members of the Per OB1 association. For M33, the data of \citet{Pierce00} come originally from \citet{Kinman87} who investigated red long-period variables including both RSGs and AGBs. For the sake of simplicity, only the lower K$_{\rm S}$ band magnitude cut-off is set as K$_{\rm S}=14.86$ corresponding to the lower mass limit of RSGs (9 M$_{\sun}$) to the original dataset of \citet{Kinman87} for 54 targets. This excludes 14 targets and keeps 40. The distance moduli of LMC, SMC and M33 used to convert from the observed K$_{\rm S}$ to the absolute magnitude $\rm M_{K_{S}}$ are 18.50, 18.91 and 24.93 \citep{Bonanos06, U09} respectively.

After these modifications, a universal linear fitting of the K$_{\rm S}$ band P-L relation is performed for RSGs in the Galaxy, LMC, SMC and M33. The results are shown in Fig.~\ref{individualplk} and Table.~\ref{individualpl}. It can be seen that a tight P-L relation exists for RSGs in all these galaxies. In particular, the relation is quite consistent for the three extra-galaxies LMC, SMC and M33, while the Galaxy has a slight different slope and zero point which may be caused by the distance error. Combining all the data together, the linear fit results in \\
\indent $\rm M_{K_{S}}=(-3.68 \pm 0.18) \times \log P + (-0.62 \pm 0.30)$ \\
with $\sigma=0.31$. Also shown in Fig.~\ref{allplk} are the P-L relations of RSGs from \citet{Feast80}, \citet{Catchpole81}, \citet{Kinman87}, \citet{Pierce00} and \citet{Kiss06} by different line styles. The results of \cite{Feast80} and \cite{Catchpole81} obviously deviate from others, which may be caused by their impure sample contaminated by AGBs. The results of \citet{Kinman87}, \citet{Pierce00} and \citet{Kiss06} have quite close slope and intercept as ours.

Within the SMC, \citet{Kiss03}, \citet{Ita04}, \citet{Soszynski07aca} and \citet{Soszynski11aca} analyzed the P-L relations of LPVs found by the Optical Gravitational Lensing Experiment (OGLE) survey. The comparison shows that our result matches well with the extension of the AGB $a_{2}$ sequence of the OGLE Small Amplitude Red Giants (OSARGs) variables defined by \citet{Soszynski07aca}. For the OSARGs, the sequence $a_{2}$ is the second sequence (from longest to shortest periods) that is similar to the sequence B of \citet{Wood99} and identified as the first overtone radial pulsating red giants. But \citet{Kiss03} and \citet{Ita04} noticed an additional sequence (labeled $\rm C'$) between Wood's sequences B and C. If sequence C is identified as the fundamental mode of pulsation, sequence $\rm C'$ would be the first overtone and sequence B would be the second overtone, which makes it suspicious to identify the pulsation mode of RSG as the first overtone pulsation from the P-L relation. With the help of the theoretical model prediction, the pulsation mode of RSGs will be clearer as will be discussed in Section 3.3. It is also noticed that there are only four targets with the period shorter than 250 d, which may be caused by two reasons. One is the sample selection effect that the lower-mass RSGs with the shorter periods are missed. This can be seen also in CMDs of this paper that most targets have theoretical mass larger than 13 M$_{\sun}$. The other is the ambiguity of the relation between pulsation and convection for RSGs of different masses. Usually, the longer the period the greater the amplitude, vice versa. However, due to the existence of large convection cells in the surface of RSGs, they could produce unpredictable variation to even ``overwhelm'' the pulsation if the period is short enough, viz amplitude small enough. Thus no periodic variation can be observed. It can be seen in Fig.~\ref{figpa} of this paper and Fig.10 in Paper I that the amplitude decreases to an undetectable level compared to the observational error if period shorter than about 250 d ($\log \rm P < 2.4$).

\subsection{Other infrared P-L relations}

The photometric data are available in other infrared bands than the K$_{\rm S}$ band by space telescopes such as \emph{Spitzer}. The study of RSGs in LMC already found that the P-L relations exist in other infrared bands. The change of the brightness with period is thus investigated in the infrared bands, namely the 2MASS JHK$_{\rm S}$ bands, the \emph{Spitzer}/IRAC bands and the \emph{Spitzer}/MIPS 24 $\mu$m band, as well as the visual band V. Similar conclusions are drawn for RSGs in SMC as in LMC. The V band magnitude does not have clear correlation with the period because of the severe effect of extinction, in particular the RSGs in SMC are mostly located in the star-forming region as can be seen in Fig.~\ref{smc}. No effort is put into fitting the P-L relation in visual. On the contrary, there is little effect of extinction on the brightness in the infrared bands. Moreover, the amplitude of variation in infrared is much smaller than in visual. Thus, the one-epoch photometric result can be a good and stable indicator of the luminosity of the object. One just needs to be careful that the infrared brightness partly comes from the circumstellar dust instead of the stellar photosphere, which becomes more significant at the longer wavelengths, i.e. the [8.0] and [24] bands in our study. Nevertheless, a linear fitting is performed in all the related infrared bands in the form of $\rm M_{\lambda} = a \log P + b$ for LMC and SMC separately and for their combination. The slope $a$ and intercept $b$ with dispersions are listed in Table~\ref{pltab}, and the fitting diagrams are shown in Fig.~\ref{pl-semi-int1} and Fig.~\ref{pl-semi-int2}.

The diagrams tell that the P-L relationship does exist in the RSGs in SMC, the same as in LMC, in all infrared bands with the dispersion depending on the band. In the V band, the dispersion is so large that no relation is intended to determine. But the LMC RSGs are on average fainter than the SMC RSGs in the V band, which can be caused by the higher metallicity in LMC for more circumstellar dust extinction as can also be seen in the [8.0]- and [24]-related diagrams. In the J, [8.0] and [24] bands, the dispersions are the moderate, and the P-L relations are reasonably good. In the H, K$_{\rm S}$, [3.6], [4.5] and [5.8] bands, the dispersions are very small, and the P-L relations are very tight and reliable. The best relationship occurs in the [4.5] band, the same as for LMC in Paper I. This is reasonable since the radiation mainly comes from the stellar photosphere if the targets are cold and have no thick dust envelope and the extinction is very small in these bands. So the IRAC bands as well as the K$_{\rm S}$ band are recommended when adopting one-epoch photometry as the luminosity indicator.

Long secondary period is another period for those RSG with LSP. All the LSP targets have relative long period and small amplitude compared to the semi-regular RSGs, which means a different origin. Many models such as radial and non-radial pulsation, binary, stellar spot models, have been suggested but none of them receives general acceptance \citep{Wood09, Nie10}. Recently, \citet{Stothers10} has demonstrated that giant convection cells in the envelopes of massive red supergiants turn over in a time comparable in order of magnitude to the observed long secondary periods in these stars according to a theory proposed some years ago by \citet{Stothers71}. He has applied the theory successfully to the well observed RSGs of Betelgeuse, Antares and M-type giant, while more samples may need to prove this model. It will be interesting to know if LSP is related to luminosity. The amplitude and period of 45 SMC LSP are calculated in the same way as for the semi-regular RSGs by three methods. Because the observation spans only about 3000 days, 14 targets with period longer than this are discarded, while the results for the left 31 targets are listed in Table \ref{lsptab}. Fig.~\ref{lspsmc} shows their absolute magnitude in the K$_{\rm S}$ band versus the period, together with the LSP of RSGs in LMC and the short periods of RSGs in the four galaxies discussed earlier. Superposed in this diagram are the P-L relations of LPVs (solid lines) from \citet{Soszynski07aca} and various LPVs targets (points) from \citet{Soszynski11aca} in the SMC, where the symbols are the same as the original papers. With the addition of more and fainter objects in \citet{Soszynski11aca}, there are small shifts of the P-L relations by \citet{Soszynski07aca} from the locations of the objects. There is no clear P-L relation for these long secondary period. The major obstacle of LSP investigation is the need of very long-term observation and the irregularity of the variation. With the continuation of the ASAS project, a more reliable determination of the period and analysis of the mechanism are possible.

\subsection{Pulsation mode}

The exact mode of pulsation of RSGs is unclear although the radial modes are commonly accepted. In order to understand the pulsation mode of RSGs, the P-L relation derived from the observation is compared with the theoretical P-L relation. The observed bolometric magnitude needs to be converted to stellar luminosity to conveniently compare with the model. The reliable absolute bolometric magnitudes are collected from \citet{Levesque06}, \citet{Levesque07} and \citet{Levesque09} which applied the new RSGs effective temperature scales \citep{Levesque05} to LMC and SMC to derive M$_{\rm bol}$ that corrected the defect of the old effective temperature. The adopted M$_{\rm bol}$ is the value derived from spectral fitting other than (V$-$K)$_{0}$ which can be significantly affected by the extinction in the V band. This selection of M$_{\rm bol}$ picks out only 13 targets in Table.~\ref{srtab} in SMC and 13 targets in LMC.

The theoretical P-L relations of RSGs were predicted for the first three radial pulsation modes by \citet{Guo02} at three typical masses, 15, 20 and 30 M$_{\sun}$ based on a linear non-adiabatic model. In particular, they considered the relations under different metallicity that covers the cases of SMC, LMC and Galaxy. Similarly, \citet{Heger97} computed the P-L relations of RSGs from linear non-adiabatic models for four masses, 10, 12, 15 and 20 M$_{\sun}$, but without discrimination of the metallicity. The P-L relations by \citet{Guo02} are derived by simulating the evolutionary tracks of the stars in the H-R diagram with different chemical composition and a constant mixing-length parameter $\alpha=1.0$. Their results show that period increases almost monotonically in the RSG phase (see Table 3 and Figure 4-12 in their paper). Our results are compared with their P-L diagrams for $\rm Z=0.005$, $\rm Z=0.01$ and $\rm Z=0.02$, as the best approximations to the SMC ($\rm Z=0.004$), LMC ($\rm Z=0.008$) and Galaxy ($\rm Z=0.02$) respectively. In Fig.~\ref{guosmcpl} for the SMC case, our targets are mostly located in the range of mass between 20 and 30 M$_{\sun}$. From the observed fact that luminosity of RSG increases with period and the theoretically predicted P-L relations for three typical masses, our targets lie in the region of the theoretical first-overtone mode. The conclusion is similar in Fig.~\ref{guolmcpl} for $\rm Z=0.01$ for the LMC RSGs, with a slight shift to the region of the fundamental mode. Two objects may even belong to the fundamental-mode region. The shift in Fig.~\ref{guogalaxypl} for the Galaxy case is more significant. Half of the RSGs in the Galaxy from \citet{Kiss06} fits the P-L relations at the fundamental-mode and the other half at the first-overtone mode. When compared with the theoretical models of \citet{Heger97} in Fig.~\ref{hegerpl}, the conclusion is the same, i.e. most objects are consistent with the predicted P-L relation at the first-overtone mode except that two in LMC and half in the Galaxy are close to the fundamental-mode relation. Indeed, the identification of the first overtone mode is already implied in that the P-L relations of RSGs in both LMC and SMC agree very well with the extension of the AGB $a_{2}$ sequence corresponding to the first overtone radial mode of OSARGs \citep{Soszynski07aca}. However, as mentioned earlier and seen in Fig.~\ref{lspsmc}, the P-L relation of RSGs are not perfectly matched with the AGB $a_{2}$ sequence with the newly accumulating data while lies a kind of between the first- and second-overtone modes. This inconsistency may indicate that RSGs have their own P-L sequence, which deserves further investigation. It may be too exaggerated to argue that the pulsation mode is related to the metallicity only from these three cases of SMC, LMC and the Galaxy, while this deserves further study. Another approach to clarify the pulsation mode relies on the accurate determination of the absolute bolometric magnitude for more complete sample of RSGs, especially those in the Galaxy.

\section{Summary}

The P-L relation of RSGs has aroused much interest since 1980s due to the evidence that such luminous and pulsatile objects can lead to an extension of stellar standard candle. Using long-term optical survey and large-scale infrared survey, we have the opportunity to calibrate the P-L relation in several infrared bands for nearby Local Group galaxies and identify the pulsation mode with the help of the theoretical calculation.

The preliminary sample of RSGs in SMC is consisted of 140 stars assembled from previous works. The infrared counterparts are chosen in the 2MASS, \emph{Spitzer}/IRAC or MIPS catalogs by a one-arcsec position consistency and identified by the brightness and colors in the CMDs and TCDs. These procedures eliminate 5 targets for inconsistent color or luminosity and 3 targets for lack of brightness information, leaving 132 targets as the RSG candidates.

The time-series photometry data are taken from the databases by the ASAS and MACHO projects and further processed. The period of light variation is derived by the PDM, Period04, CLEAN and WWZ methods. According to the morphology of the light curve, 6 more sources are further identified as AGBs and removed from the RSGs sample. Based on the properties of light variation, the 126 RSGs are classified into five categories: 20 with poor photometry, 55 with no reliable period, 6 with semi-regular variation, 15 with LSP and distinguishable short period and 30 with only LSP.

For the semi-regular variables and the LSP variables with distinguishable short period, the K$_{\rm S}$ band P-L relation is analyzed and determined. This relation is compared with those in the Galaxy, LMC, and M33. It is found that the RSGs in these galaxies obey similar P-L relation except the Galaxy. In addition, the P-L relations in the infrared bands, namely the 2MASS JHK$_{\rm S}$ and \emph{Spitzer}/IRAC and \emph{Spitzer}/MIPS 24 $\mu m$ band, are derived with high reliability. The best P-L relation occurs in the \emph{Spitzer}/IRAC [3.6] and [4.5] bands. No clear P-L relation is found for the LSP RSGs.

From the P-L relation, the mode of pulsation of RSGs in SMC is discussed and found to be very possibly in the first overtone radial mode. When compared with the Galaxy and LMC, it seems there is a tendency that the pulsation mode depends on the metallicity.

\acknowledgments{The authors thank Prof. L. L. Kiss for kindly providing the data of the Galactic RSGs. This work is supported by the China's NSFC projects Nos.\,10973004 and 11173007. BWJ thanks the John Templeton Foundation for support.}

\clearpage

\begin{deluxetable}{ccccccccccccccc}
\tabletypesize{\scriptsize}
\rotate
\tablecaption{
 Infrared Catalog of the 140 SMC RSGs candidates
 \label{140tab}
}
\tablewidth{0pt}
\setlength{\tabcolsep}{0.05in}
\tablehead{
\colhead{No.} & \colhead{Name} & \colhead{R.A.(J2000)} & \colhead{Decl.(J2000)} &
\colhead{J} & \colhead{H} & \colhead{K$_{\rm S}$} & \colhead{[3.6]} & \colhead{[4.5]} & \colhead{[5.8]} & \colhead{[8.0]} &
\colhead{[24]} & \colhead{Spectral Type\tablenotemark{a}} & \colhead{Reference\tablenotemark{b}} & \colhead{Source\tablenotemark{c}}
}
\startdata
  1& HV11223                  & 00 32 01.61& -73 22 34.7&   11.166&  10.276&   9.971&   9.749&   9.878&   9.683&   9.557&   9.205&  -----  & CF81 &  A\\
  2& HV1349                   & 00 41 21.50& -72 50 15.0&   11.202&  10.446&   9.946&   9.284&   -----&   8.874&   8.255&   6.230&  -----  & CF81 &  A\\
  3& HV1366                   & 00 42 49.87& -72 55 11.4&   12.232&  11.431&  11.154&  10.733&  10.594&  10.539&  10.283&   9.844&  -----  & CF81 &  M\\
  4& HV1375                   & 00 42 52.30& -73 50 51.0&   11.300&  10.381&   9.832&   8.715&   8.365&   7.835&   7.204&   5.388&  -----  & CF81 &---\\
  5& [M2002]SMC005092         & 00 45 04.56& -73 05 27.4&    9.208&   8.377&   8.071&   7.765&   7.733&   7.491&   7.081&   5.185&  M2 I   & L06  &  A\\
  6& [M2002]SMC008324         & 00 47 16.84& -73 08 08.4&   10.456&   9.685&   9.507&   9.372&   9.462&   9.367&   9.267&   9.608&  K0: I  & M03  &  A\\
  7& [M2002]SMC008367         & 00 47 18.11& -73 10 39.3&   10.080&   9.261&   9.040&   8.838&   8.867&   8.708&   8.501&   -----&  K7 I   & M03  &  A\\
  8& [M2002]SMC008930         & 00 47 36.94& -73 04 44.3&    9.466&   8.601&   8.319&   8.039&   8.043&   7.814&   7.641&   7.235&  M0 I   & M03  &  A\\
  9& [M2002]SMC009766         & 00 48 01.22& -73 23 37.5&   10.208&   9.402&   9.213&   9.056&   9.219&   9.069&   8.961&   8.644&  K7 I   & M03  &  A\\
 10& [M2002]SMC010889         & 00 48 27.02& -73 12 12.3&    8.778&   7.990&   7.774&   7.571&   7.682&   7.463&   7.285&   5.628&  M0 I   & M03  &  A\\
 11& [M2002]SMC011101         & 00 48 31.92& -73 07 44.4&   10.316&   9.468&   9.237&   8.978&   9.091&   8.945&   8.879&   8.836&  K2.5 I & M03  &  A\\
 12& [M2002]SMC011709         & 00 48 46.32& -73 28 20.7&    9.510&   8.770&   8.530&   8.345&   8.471&   8.299&   8.205&   7.479&  K5-M0 I& M03  &  A\\
 13& [M2002]SMC011939         & 00 48 51.83& -73 22 39.3&    9.629&   8.864&   8.606&   8.296&   8.412&   8.116&   7.894&   5.936&  K2 I   & M03  &  A\\
 14& [M2002]SMC012322         & 00 49 00.32& -72 59 35.7&    9.280&   8.456&   8.276&   8.045&   8.126&   7.937&   7.799&   6.522&  K5-M0 I& M03  &  A\\
 15& CV78                     & 00 49 03.84& -73 05 19.9&    -----&   -----&   -----&   9.311&   9.193&   8.955&   8.760&   7.985&  -----  & WBF83&  M\\
 16& [M2002]SMC012572         & 00 49 05.25& -73 31 07.8&    9.204&   8.478&   8.258&   8.125&   8.236&   8.159&   8.118&   7.996&  -----  & M03  &  A\\
 17& [M2002]SMC012707         & 00 49 08.23& -73 14 15.5&   10.365&   9.560&   9.333&   9.140&   9.204&   9.077&   8.954&   8.845&  -----  & M03  &  A\\
 18& [M2002]SMC013472         & 00 49 24.53& -73 18 13.5&    9.111&   8.485&   8.301&   8.084&   8.013&   7.823&   7.484&   7.242&  K7: I  & M03  &  A\\
 19& [M2002]SMC013740         & 00 49 30.34& -73 26 49.9&   10.478&   9.623&   9.387&   9.241&   9.363&   9.219&   9.145&   -----&  K3 I   & M03  &  A\\
 20& [M2002]SMC013951         & 00 49 34.42& -73 14 09.9&   10.084&   9.262&   9.055&   8.870&   9.023&   8.859&   8.776&   8.485&  K3 I   & M03  &  A\\
 21& [M2002]SMC015510         & 00 50 06.42& -73 28 11.1&    9.590&   8.760&   8.535&   8.280&   8.401&   8.197&   7.965&   6.701&  K5 I   & M03  &  A\\
 22& IRAS F00483-7347         & 00 50 07.19& -73 31 25.2&   11.431&   9.777&   8.639&   6.620&   6.309&   5.507&   4.790&   2.591&  -----  & V08  &  M\\
 23& IRAS X0048-731           & 00 50 30.63& -72 51 29.9&   10.338&   9.394&   8.783&   8.643&   8.212&   8.030&   7.030&   4.740&  -----  & V08  &  M\\
 24& [M2002]SMC017656         & 00 50 47.22& -72 42 57.2&    9.839&   9.080&   8.921&   8.776&   8.861&   8.737&   8.647&   8.737&  K0-5 I & M03  &  A\\
 25& HV11295                  & 00 50 48.40& -72 52 29.0&   10.849&  10.004&   9.580&   9.267&   9.120&   8.948&   8.804&   7.910&  -----  & CF81 &  A\\
 26& [M2002]SMC018136         & 00 50 56.01& -72 15 05.7&    8.904&   8.107&   7.849&   7.631&   7.819&   7.609&   7.452&   5.850&  M0 I   & L06  &  A\\
 27& [M2002]SMC018592         & 00 51 03.90& -72 43 17.4&    8.385&   7.671&   7.425&   6.885&   -----&   6.706&   6.304&   4.605&  K0-2 I & M03  &  A\\
 28& [M2002]SMC019551         & 00 51 20.23& -72 49 22.1&   10.362&   9.593&   9.382&   9.239&   9.347&   9.233&   9.152&   9.162&  K2 I   & M03  &  A\\
 29& [M2002]SMC019743         & 00 51 23.28& -72 38 43.8&    9.678&   8.899&   8.730&   8.451&   8.466&   8.221&   7.947&   6.372&  K5 I   & M03  &  A\\
 30& [M2002]SMC020133         & 00 51 29.68& -73 10 44.3&    9.220&   8.410&   8.155&   7.953&   7.957&   7.693&   7.383&   5.509&  M0 I   & M03  &  A\\
 31& [M2002]SMC020612         & 00 51 37.57& -72 25 59.5&   10.451&   9.737&   9.569&   9.412&   9.511&   9.394&   9.302&   9.215&  K5 I   & M03  &  A\\
 32& [M2002]SMC021362         & 00 51 50.25& -72 05 57.2&    9.862&   9.017&   8.824&   8.674&   8.749&   8.573&   8.408&   8.210&  K5 I   & L06  &  A\\
 33& [M2002]SMC021381         & 00 51 50.46& -72 11 32.2&    9.967&   9.206&   8.999&   8.797&   8.844&   8.708&   8.609&   8.440&  K5 I   & L06  &  A\\
 34& HV11303                  & 00 52 09.00& -71 36 22.0&    -----&   -----&   -----&   -----&   -----&   -----&   -----&   -----&  -----  & CF81 &  A\\
 35& ISO-MCMS J005212.9-730852& 00 52 12.82& -73 08 52.8&   11.254&  10.468&   9.917&   8.800&   8.487&   8.268&   7.467&   4.734&  -----  & V10  &  A\\
 36& [M2002]SMC023401         & 00 52 25.36& -72 25 13.3&   10.358&   9.627&   9.425&   9.303&   9.370&   9.321&   9.192&   9.095&  K1 I   & L06  &  A\\
 37& [M2002]SMC023463         & 00 52 26.51& -72 45 15.6&    9.674&   8.956&   8.730&   8.328&   8.341&   8.101&   7.926&   7.643&  K0-5 I & M03  &  A\\
 38& [M2002]SMC023700         & 00 52 30.69& -72 26 46.8&   10.424&   9.656&   9.440&   9.325&   9.477&   9.341&   9.267&   9.106&  K0-2 I & M03  &  A\\
 39& [M2002]SMC023743         & 00 52 31.49& -72 11 37.3&   10.304&   9.570&   9.411&   9.230&   9.314&   9.197&   9.077&   9.100&  K2 I   & L06  &  A\\
 40& [M2002]SMC025550         & 00 53 02.85& -73 07 45.9&   10.500&   9.677&   9.453&   9.301&   -----&   9.303&   -----&   9.044&  K2 I   & M03  &  A\\
 41& [M2002]SMC025879         & 00 53 08.87& -72 29 38.6&    9.359&   8.624&   8.447&   8.139&   8.146&   8.021&   7.835&   6.995&  M0 I   & M03  &  A\\
 42& [M2002]SMC025888         & 00 53 09.04& -73 04 03.6&    9.262&   8.454&   8.217&   7.983&   8.043&   7.905&   7.713&   6.743&  K5-7 I & M03  &  A\\
 43& [M2002]SMC026402         & 00 53 17.81& -72 46 06.9&   10.276&   9.487&   9.271&   9.068&   9.178&   9.003&   8.856&   8.278&  K0-2 I & M03  &  A\\
 44& [M2002]SMC026778         & 00 53 24.56& -73 18 31.6&   10.188&   9.433&   9.222&   9.037&   9.150&   8.992&   8.892&   8.631&  K2 I   & M03  &  A\\
 45& [M2002]SMC027443         & 00 53 36.44& -73 01 34.8&    9.634&   8.873&   8.640&   8.388&   8.521&   8.347&   8.221&   7.360&  K7 I   & M03  &  A\\
 46& HV11329                  & 00 53 39.40& -72 52 39.0&   10.911&   9.987&   9.647&   9.493&   9.562&   9.416&   9.302&   9.076&  -----  & CF81 &  M\\
 47& [M2002]SMC027945         & 00 53 45.74& -72 53 38.5&   10.434&   9.820&   9.619&   9.496&   9.566&   9.443&   9.302&   9.214&  K3-5 I & M03  &  A\\
 48& [M2002]SMC030135         & 00 54 26.90& -72 52 59.4&   10.335&   9.690&   9.492&   9.236&   9.408&   9.161&   8.892&   8.048&  K2 I   & M03  &  A\\
 49& [M2002]SMC030616         & 00 54 35.90& -72 34 14.3&    9.308&   8.531&   8.341&   8.125&   8.199&   8.017&   7.768&   7.333&  K2 I   & M03  &  A\\
 50& [M2002]SMC032188         & 00 55 03.71& -73 00 36.6&    9.728&   9.052&   8.861&   8.765&   8.922&   8.782&   8.662&   8.358&  K5 I   & M03  &  A\\
 51& HV12122                  & 00 55 18.01& -72 05 31.9&   10.600&   9.714&   9.361&   8.911&   8.834&   8.558&   8.419&   8.376&  -----  & WBF83&  A\\
 52& HV1645                   & 00 55 19.76& -73 14 42.2&   12.663&  11.775&  11.421&  10.812&  10.674&  10.492&  10.290&  10.010&  -----  & CF81 &  M\\
 53& [M2002]SMC033610         & 00 55 26.82& -72 35 56.2&    9.803&   9.014&   8.762&   8.558&   8.604&   8.409&   8.195&   6.445&  K7 I   & M03  &  A\\
 54& [M2002]SMC034158         & 00 55 36.58& -72 36 23.6&    9.877&   9.123&   8.909&   8.727&   8.737&   8.543&   8.365&   7.185&  K2 I   & M03  &  A\\
 55& HV 838                   & 00 55 38.21& -73 11 41.1&   10.608&   9.860&   9.440&   8.369&   -----&   8.168&   8.148&   7.513&  -----  & CF81 &  A\\
 56& [M2002]SMC035231         & 00 55 55.10& -72 40 30.4&    9.995&   9.482&   9.343&   9.167&   9.142&   9.000&   8.935&   8.782&  K2 I   & M03  &  A\\
 57& [M2002]SMC035445         & 00 55 58.84& -73 20 41.4&    9.976&   9.177&   8.980&   8.834&   9.038&   8.855&   8.765&   8.673&  K1 I   & L06  &  A\\
 58& [M2002]SMC037994         & 00 56 43.55& -72 30 15.0&    9.779&   8.946&   8.691&   8.518&   8.599&   8.451&   8.354&   8.154&  K7 I   & M03  &  A\\
 59& HV11366                  & 00 56 54.77& -72 14 08.6&   11.066&  10.154&   9.870&   9.678&   9.795&   9.608&   9.518&   9.458&  -----  & CF81 &  M\\
 60& HV1719                   & 00 57 14.48& -73 01 21.3&   10.780&  10.117&   9.798&   9.091&   8.979&   8.734&   8.609&   7.913&  -----  & CF81 &A M\\
 61& [M2002]SMC041778         & 00 57 56.45& -72 17 33.3&    -----&   -----&   -----&   -----&   -----&   -----&   -----&   8.325&  K7 I   & M03  &  A\\
 62& [M2002]SMC042319         & 00 58 06.61& -72 20 59.8&   10.178&   9.364&   9.137&   8.987&   9.112&   8.955&   8.873&   8.637&  K2-5 I & M03  &  A\\
 63& [M2002]SMC042438         & 00 58 08.71& -72 19 26.7&   10.383&   9.596&   9.362&   9.192&   9.321&   9.206&   9.152&   9.078&  K2 I   & M03  &  A\\
 64& [M2002]SMC043219         & 00 58 23.30& -72 48 40.7&   10.173&   9.368&   9.114&   8.886&   9.063&   8.899&   8.838&   8.536&  K2 I   & M03  &  A\\
 65& [M2002]SMC043725         & 00 58 33.21& -72 19 15.6&   10.595&   9.774&   9.558&   9.391&   9.542&   9.416&   9.338&   9.239&  K5 I   & M03  &  A\\
 66& HV12149                  & 00 58 50.17& -72 18 35.6&    9.963&   9.022&   8.609&   8.548&   8.381&   8.192&   7.889&   6.878&  -----  & CF81 &  M\\
 67& [M2002]SMC045378         & 00 59 07.16& -72 13 08.6&   10.039&   9.205&   9.005&   8.807&   8.956&   8.814&   8.697&   8.694&  K3 I   & M03  &  A\\
 68& [M2002]SMC045850         & 00 59 16.90& -72 25 10.9&   10.150&   9.366&   9.199&   9.004&   9.131&   8.996&   8.929&   8.807&  K0-5 I & M03  &  A\\
 69& [M2002]SMC046497         & 00 59 31.33& -72 15 46.4&    9.432&   8.589&   8.308&   8.160&   8.145&   7.986&   7.788&   6.212&  K5-M0 I& M03  &  A\\
 70& [M2002]SMC046662         & 00 59 35.04& -72 04 06.2&    9.377&   8.627&   8.345&   8.160&   8.209&   8.001&   7.799&   6.142&  M0 I   & M03  &  A\\
 71& [M2002]SMC046910         & 00 59 40.58& -72 20 55.9&   10.227&   9.539&   9.330&   9.181&   9.258&   9.135&   9.021&   8.943&  K3-5 I & M03  &  A\\
 72& [M2002]SMC047757         & 01 00 00.63& -72 19 40.2&    9.294&   8.502&   8.240&   8.084&   8.104&   7.894&   7.711&   6.217&  M1 I   & M03  &  A\\
 73& [M2002]SMC048122         & 01 00 09.42& -72 08 44.5&    -----&   -----&   -----&   8.800&   8.931&   8.810&   8.692&   8.200&  K1 I   & M03  &  A\\
 74& [M2002]SMC049033         & 01 00 30.43& -71 58 24.7&    9.741&   9.016&   8.775&   8.589&   8.689&   8.563&   8.476&   8.322&  K5 I   & M03  &  A\\
 75& [M2002]SMC049428         & 01 00 40.32& -72 35 58.8&   10.240&   9.511&   9.284&   9.113&   9.225&   9.077&   9.021&   8.927&  K0-7 I & M03  &  A\\
 76& N371 R20                 & 01 00 41.56& -72 10 37.0&    8.955&   8.220&   7.963&   7.821&   7.827&   7.620&   7.446&   7.207&  K5-M0 I& WBF83&  A\\
 77& HV11417                  & 01 00 48.17& -72 51 02.1&    9.754&   8.806&   8.453&   7.891&   7.795&   7.568&   7.108&   4.686&  -----  & V08  &  A\\
 78& [M2002]SMC049990         & 01 00 54.13& -72 51 36.3&    9.517&   8.924&   8.677&   8.687&   8.682&   8.417&   8.248&   7.389&  K5 I   & M03  &  A\\
 79& [M2002]SMC050028         & 01 00 55.20& -71 37 52.9&    9.078&   8.289&   8.043&   7.458&   7.580&   7.176&   7.064&   5.406&  -----  & M07  &  A\\
 80& [M2002]SMC050237         & 01 01 00.31& -72 13 41.6&   10.413&   9.623&   9.459&   9.336&   9.440&   9.326&   9.250&   9.267&  K2-5 I & M03  &  A\\
 81& [M2002]SMC050348         & 01 01 03.26& -72 04 39.4&   10.290&   9.583&   9.363&   9.209&   9.298&   9.188&   9.099&   8.963&  K7 I   & M03  &  A\\
 82& [M2002]SMC050360         & 01 01 03.58& -72 02 58.5&   10.427&   9.570&   9.388&   9.259&   9.391&   9.262&   9.208&   9.135&  K5 I   & M03  &  A\\
 83& [M2002]SMC050840         & 01 01 15.99& -72 13 10.0&    9.386&   8.576&   8.368&   8.118&   8.187&   8.011&   7.803&   6.131&  M1 I   & M03  &  A\\
 84& [M2002]SMC051000         & 01 01 19.92& -72 05 13.1&   10.224&   9.416&   9.237&   9.104&   9.198&   9.065&   8.981&   8.887&  K5 I   & M03  &  A\\
 85& [M2002]SMC051265         & 01 01 26.89& -72 01 41.3&   10.226&   9.414&   9.235&   9.582&   9.201&   9.057&   8.981&   8.787&  K3-5 I & M03  &  A\\
 86& HV11427                  & 01 01 27.80& -72 48 29.0&   12.660&  11.821&  11.504&  10.914&  10.747&  10.412&  10.344&   -----&  -----  & CF81 &  M\\
 87& HV1865                   & 01 01 36.82& -72 44 41.1&   10.438&   9.550&   9.205&   8.877&   8.838&   8.524&   8.336&   8.147&  -----  & CF81 &  M\\
 88& [M2002]SMC051906         & 01 01 43.57& -72 38 25.1&   10.363&   9.596&   9.358&   9.203&   9.367&   9.229&   9.160&   9.076&  K5 I   & M03  &  A\\
 89& [M2002]SMC052334         & 01 01 54.16& -71 52 18.8&    9.689&   8.962&   8.695&   8.558&   8.669&   8.502&   8.365&   7.321&  K5-M0 I& M03  &  A\\
 90& [M2002]SMC052389         & 01 01 55.43& -72 00 29.5&    9.912&   9.104&   8.910&   8.727&   8.872&   8.717&   8.604&   8.143&  K7 I   & M03  &  A\\
 91& [M2002]SMC053557         & 01 02 23.71& -72 55 21.2&   10.008&   9.241&   9.024&   8.869&   9.063&   8.882&   8.815&   8.736&  K7 I   & M03  &  A\\
 92& [M2002]SMC053638         & 01 02 25.83& -72 38 56.9&   10.398&   9.563&   9.370&   9.180&   9.311&   9.157&   9.085&   9.153&  K2-7 I & M03  &  A\\
 93& [M2002]SMC054111         & 01 02 37.22& -72 16 25.1&    9.872&   9.149&   8.969&   8.811&   8.889&   8.773&   8.697&   8.573&  K0-5 I & M03  &  A\\
 94& [M2002]SMC054300         & 01 02 42.12& -72 37 29.1&   10.243&   9.437&   9.224&   9.031&   9.196&   9.073&   9.007&   8.951&  K0-5 I & M03  &  A\\
 95& [M2002]SMC054414         & 01 02 44.82& -72 01 51.9&   10.340&   9.574&   9.386&   9.217&   9.360&   9.238&   9.184&   8.740&  K3-5 I & M03  &  A\\
 96& [M2002]SMC054708         & 01 02 51.37& -72 24 15.5&   10.055&   9.296&   9.081&   8.957&   9.055&   8.882&   8.804&   8.574&  K1 I   & M03  &  A\\
 97& N371\#29                 & 01 03 02.38& -72 01 52.9&    9.901&   9.011&   8.619&   8.328&   8.136&   7.877&   7.530&   5.890&  M4.5 I & WBF83&  A\\
 98& [M2002]SMC055275         & 01 03 04.34& -72 34 12.8&    9.737&   8.890&   8.654&   8.433&   8.536&   8.311&   8.167&   7.974&  K7-M0 I& M03  &  A\\
 99& [M2002]SMC055355         & 01 03 06.43& -72 28 35.1&    9.836&   9.049&   8.842&   8.706&   8.654&   8.490&   8.289&   8.158&  K7-M0 I& M03  &  A\\
100& [M2002]SMC055470         & 01 03 08.88& -71 55 50.8&   10.427&   9.716&   9.548&   9.367&   9.436&   9.306&   9.208&   9.161&  K3-5 I & M03  &  A\\
101& [M2002]SMC055560         & 01 03 10.93& -72 18 32.9&   10.227&   9.444&   9.225&   9.114&   9.237&   9.069&   8.987&   8.852&  K3-5 I & M03  &  A\\
102& N371 C12                 & 01 03 12.98& -72 09 26.5&    9.642&   8.852&   8.594&   8.423&   8.423&   8.214&   7.960&   6.132&  K5-M0 I& WBF83&  A\\
103& [M2002]SMC055933         & 01 03 18.56& -72 06 46.2&   10.095&   9.357&   9.170&   8.932&   8.982&   8.804&   8.523&   7.622&  K3-5 I & M03  &  A\\
104& [M2002]SMC056389         & 01 03 27.61& -72 52 09.4&    8.802&   8.059&   7.779&   7.580&   7.647&   7.445&   7.240&   5.576&  K5-7 I & M03  &  A\\
105& [M2002]SMC056732         & 01 03 34.30& -72 06 05.8&    9.940&   9.122&   8.864&   8.683&   8.774&   8.630&   8.523&   8.437&  K5-M0 I& M03  &  A\\
106& HV11452                  & 01 03 36.92& -73 33 37.8&   10.887&  10.023&   9.689&   9.496&   9.480&   9.296&   9.056&   8.877&  -----  & CF81 &  A\\
107& [M2002]SMC057386         & 01 03 47.35& -72 01 16.0&   10.002&   9.216&   8.968&   8.854&   8.982&   8.862&   8.776&   8.734&  K1 I   & M03  &  A\\
108& [M2002]SMC057472         & 01 03 48.89& -72 02 12.7&   10.003&   9.207&   8.967&   8.836&   8.929&   8.814&   8.712&   -----&  K2 I   & M03  &  A\\
109& [M2002]SMC058149         & 01 04 02.77& -72 05 27.7&   10.247&   9.474&   9.257&   9.111&   9.243&   9.094&   9.021&   8.813&  K2-5 I & M03  &  A\\
110& [M2002]SMC058472         & 01 04 09.52& -72 50 15.3&   10.320&   9.578&   9.352&   9.261&   9.249&   9.038&   8.910&   8.156&  K0 I   & M03  &  A\\
111& HV1956                   & 01 04 15.46& -72 45 19.9&   10.474&  10.069&   9.910&   9.313&   9.321&   9.174&   9.028&   9.018&  K2 I   & CF81 &  A\\
112& [M2002]SMC058839         & 01 04 17.71& -71 57 32.5&   10.401&   9.625&   9.413&   9.224&   9.373&   9.229&   9.152&   9.091&  K2 I   & M03  &  A\\
113& HV1963                   & 01 04 26.63& -72 34 40.3&   10.638&   9.677&   9.409&   9.077&   9.210&   9.011&   8.916&   8.571&  -----  & CF81 &  A\\
114& [M2002]SMC059426         & 01 04 30.26& -72 04 36.1&   10.006&   9.174&   8.963&   8.698&   8.851&   8.688&   8.549&   8.053&  K5-7 I & M03  &  A\\
115& [M2002]SMC059803         & 01 04 38.16& -72 01 27.2&    9.093&   8.301&   8.097&   7.858&   7.860&   7.620&   7.383&   6.432&  K2-3 I & M03  &  A\\
116& [M2002]SMC060447         & 01 04 53.05& -72 47 48.5&   10.266&   9.403&   9.175&   9.018&   9.112&   8.916&   8.815&   7.860&  K2 I   & M03  &  A\\
117& HV12179                  & 01 05 01.76& -72 06 56.4&   10.951&  10.110&   9.851&   9.522&   9.409&   9.230&   9.008&   8.480&  -----  & WBF83&  A\\
118& [M2002]SMC061296         & 01 05 11.50& -72 02 27.5&   10.193&   9.437&   9.237&   9.061&   9.129&   8.959&   8.867&   8.882&  K7 I   & M03  &  A\\
119& [M2002]SMC062427         & 01 05 40.04& -71 58 46.4&   10.396&   9.641&   9.435&   9.288&   9.422&   9.306&   9.233&   9.203&  K5 I   & M03  &  A\\
120& [M2002]SMC063114         & 01 06 01.37& -72 52 43.2&    9.800&   9.037&   8.788&   8.579&   8.619&   8.485&   8.321&   7.997&  K5-7 I & M03  &  A\\
121& [M2002]SMC063131         & 01 06 01.72& -72 24 03.8&   10.511&   9.729&   9.563&   9.367&   9.454&   9.326&   9.241&   9.110&  K5 I   & M03  &  A\\
122& [M2002]SMC063188         & 01 06 03.21& -72 52 16.0&   10.202&   9.428&   9.187&   8.951&   9.050&   8.899&   8.793&   8.428&  K2 I   & M03  &  A\\
123& [M2002]SMC064448         & 01 06 40.21& -72 28 45.2&   10.407&   9.770&   9.583&   9.486&   9.522&   9.416&   9.329&   9.278&  K2-7 I & M03  &  A\\
124& [M2002]SMC064663         & 01 06 47.62& -72 16 11.9&    9.241&   8.520&   8.312&   7.995&   8.033&   7.786&   7.456&   5.946&  K7 I   & M03  &  A\\
125& [M2002]SMC066066         & 01 07 29.36& -72 30 45.7&   10.036&   9.266&   9.090&   8.948&   9.078&   8.920&   8.873&   8.846&  K5 I   & M03  &  A\\
126& [M2002]SMC066694         & 01 07 48.88& -72 23 42.4&    9.790&   8.977&   8.774&   8.691&   8.764&   8.609&   8.581&   8.508&  K5 I   & M03  &  A\\
127& [M2002]SMC067509         & 01 08 13.34& -72 00 02.9&   10.126&   9.344&   9.172&   9.047&   9.148&   9.038&   8.974&   8.918&  K1 I   & M03  &  A\\
128& [M2002]SMC067554         & 01 08 14.65& -72 46 40.8&   10.129&   9.396&   9.174&   8.989&   9.126&   8.959&   8.867&   8.498&  K5-7 I & M03  &  A\\
129& [M2002]SMC068648         & 01 08 52.08& -72 23 07.0&    9.683&   8.918&   8.716&   8.579&   8.730&   8.583&   8.506&   8.420&  K2 I   & M03  &  A\\
130& HV12956                  & 01 09 02.20& -71 24 10.0&   11.532&  10.853&  10.336&   9.414&   9.134&   8.568&   7.301&   3.143&  -----  & CF81 &  A\\
131& [M2002]SMC069886         & 01 09 38.08& -73 20 01.9&    8.849&   8.136&   7.793&   7.531&   7.373&   7.050&   6.758&   4.800&  K5-M0 I& L06  &  A\\
132& HV2112                   & 01 10 04.20& -72 36 54.0&    -----&   -----&   -----&   -----&   -----&   -----&   -----&   -----&  -----  & CF81 &  A\\
133& HV 859                   & 01 10 27.10& -72 35 48.0&   10.278&   9.522&   9.083&   8.579&   8.483&   8.258&   8.051&   7.990&  -----  & CF81 &  A\\
134& [M2002]SMC071507         & 01 10 50.25& -72 00 14.5&   10.554&   9.822&   9.595&   9.481&   9.607&   9.502&   9.444&   9.242&  K3-5 I & M03  &  A\\
135& [M2002]SMC071566         & 01 10 53.51& -72 25 40.0&   10.365&   9.594&   9.393&   9.194&   9.327&   9.201&   9.152&   8.979&  K7 I   & M03  &  A\\
136& [M2002]SMC081668         & 01 24 54.03& -73 26 49.2&   10.304&   9.549&   9.355&   9.133&   9.295&   9.153&   9.056&   8.883&  -----  & M03  &  A\\
137& [M2002]SMC081961         & 01 25 38.80& -73 21 55.6&    9.031&   8.217&   7.984&   7.942&   8.023&   7.859&   7.744&   7.059&  -----  & M03  &  A\\
138& [M2002]SMC082159         & 01 26 09.91& -73 23 15.4&   10.069&   9.351&   9.167&   9.023&   9.129&   9.007&   8.929&   8.911&  -----  & M03  &  A\\
139& [M2002]SMC083202         & 01 29 18.52& -73 01 59.3&    8.861&   8.178&   7.949&   7.816&   7.894&   7.722&   7.603&   7.465&  -----  & M03  &  A\\
140& [M2002]SMC083593         & 01 30 33.92& -73 18 41.9&    9.636&   8.955&   8.601&   8.020&   7.811&   7.509&   7.286&   5.706&  M2 I   & M03  &  A\\

\enddata
\tablenotetext{a}{Spectral types come from \citet{Massey03, Levesque06, Levesque07} and all have been updated.}
\tablenotetext{b}{Reference: 'CF81' for \citet{Catchpole81}, 'WBF83' for \citet{Wood83}, 'M03' for \citet{Massey03}, 'L06' for \citet{Levesque06}, 'V08' for \citet{vanloon08}, 'V10' for \citet{vanloon10}.}
\tablenotetext{c}{Photometry data resource: 'A' for ASAS, 'M' for MACHO and '---' for null.}
\end{deluxetable}

\clearpage


\begin{deluxetable}{cccc}
\tabletypesize{\scriptsize}
\tablecaption{
 Results of robust linear fitting for relationship between infrared and bolometric magnitude
 \label{band-mbol-table}
}
\tablewidth{0pt}
\tablehead{
 \colhead{Band} & \colhead{Slope $a$} & \colhead{Intercept $b$} & \colhead{$\sigma$}
}
\startdata
J          &   0.65$\pm$0.05 &   2.82$\pm$0.17 &   0.31\\
H          &   0.63$\pm$0.05 &   2.23$\pm$0.18 &   0.33\\
K$_{\rm S}$&   0.65$\pm$0.06 &   1.78$\pm$0.19 &   0.35\\
$[3.6]$    &   0.69$\pm$0.06 &   1.23$\pm$0.21 &   0.38\\
$[4.5]$    &   0.72$\pm$0.06 &   0.97$\pm$0.21 &   0.37\\
$[5.8]$    &   0.79$\pm$0.07 &   0.07$\pm$0.22 &   0.40\\
$[8.0]$    &   0.86$\pm$0.07 &  -0.86$\pm$0.24 &   0.44\\
$[24]$     &   1.45$\pm$0.14 &  -7.82$\pm$0.48 &   0.86\\
\enddata

\end{deluxetable}


\begin{deluxetable}{ccccccccccccccc}
\rotate
\tabletypesize{\scriptsize}
\tablecaption{
 Results of time-series analysis by PDM, Period04 and CLEAN methods for 6 semi-regular and 15 LSP RSGs
 \label{srtab}
}
\tablewidth{0pt}
\setlength{\tabcolsep}{0.05in}
\tablehead{ &
 \multicolumn{2}{c}{Method} &
 \multicolumn{3}{c}{PDM} &
 &
 \multicolumn{3}{c}{Period04}&
 {CLEAN}
 \\ \cline{4-6} \cline{8-10}

 \colhead{No.} &
 \colhead{$\langle$$m_{V}$$\rangle$} &
 \colhead{M$_{\rm bol}$}&
 \colhead{P (d)} &
 \colhead{A (mag)} &
 \colhead{$\Theta$} &
 &
 \colhead{P (d)} &
 \colhead{A (mag)} &
 \colhead{Phase} &
 \colhead{P (d)} &
 \colhead{$\langle$P$\rangle \pm \sigma$ (d)} &
 \colhead{Points}&
 \colhead{Type\tablenotemark{a}}&
 \colhead{Ref\tablenotemark{b}}
}
\startdata
5  & 13.11&   -8.48&   600&   0.33&  0.35&&  620&   0.22&  0.84&  623&   614$\pm$13&  208&   L&   L06\\
8  & 13.01&   -8.38&   593&   0.31&  0.62&&  603&   0.16&  0.13&  595&   597$\pm$5 &  208&   L&   L06\\
10 & 12.11&   -8.47&   572&   0.17&  0.41&&  566&   0.15&  0.01&  569&   569$\pm$3 &  214&   S&   L06\\
12 & 12.51&   -7.93&   366&   0.17&  0.46&&  358&   0.08&  0.14&  363&   362$\pm$4 &  214&   L&   L06\\
21 & 12.86&   -8.13&   388&   0.18&  0.55&&  381&   0.08&  0.75&  385&   384$\pm$4 &  212&   S&   L06\\
29 & 12.82&   ---  &   512&   0.30&  0.43&&  510&   0.31&  0.66&  520&   514$\pm$5 &  196&   S&   ---\\
30 & 12.42&   -8.39&   568&   0.23&  0.60&&  562&   0.09&  0.97&  569&   566$\pm$4 &  205&   L&   L06\\
37 & 12.61&   ---  &   620&   0.17&  0.38&&  618&   0.20&  0.32&  623&   620$\pm$3 &  205&   S&   ---\\
41 & 11.96&   -8.44&   531&   0.14&  0.44&&  513&   0.13&  0.78&  523&   522$\pm$9 &  197&   L&   L06\\
42 & 12.12&   ---  &   483&   0.13&  0.30&&  484&   0.07&  0.19&  484&   483$\pm$1 &  206&   L&   ---\\
50 & 12.57&   ---  &   309&   0.11&  0.61&&  313&   0.07&  0.08&  319&   313$\pm$5 &  200&   S&   ---\\
54 & 13.05&   -7.88&   419&   0.24&  0.40&&  427&   0.16&  0.73&  434&   426$\pm$8 &  186&   L&   L06\\
56 & 12.07&   ---  &   195&   0.11&  0.67&&  195&   0.04&  0.17&  195&   195$\pm$0 &  195&   S&   ---\\
72 & 12.51&   ---  &   517&   0.29&  0.46&&  522&   0.23&  0.11&  542&   527$\pm$13&  189&   L&   ---\\
79 & 12.36&   -8.80&   721&   0.44&  0.30&&  716&   0.42&  0.07&  723&   720$\pm$4 &  190&   L&   L09\\
83 & 12.61&   -8.12&   541&   0.16&  0.45&&  571&   0.11&  0.48&  566&   559$\pm$16&  187&   L&   L06\\
98 & 12.47&   ---  &   464&   0.21&  0.49&&  448&   0.08&  0.19&  448&   453$\pm$9 &  182&   L&   ---\\
111& 12.35&   ---  &   210&   0.34&  0.17&&  209&   0.29&  0.66&  210&   209$\pm$1 &  189&   L&   ---\\
115& 12.17&   -8.69&   541&   0.22&  0.33&&  534&   0.13&  0.22&  566&   547$\pm$17&  189&   L&   L06\\
131& 12.08&   -8.76&   728&   0.39&  0.26&&  728&   0.36&  0.08&  766&   740$\pm$22&  190&   L&   L06\\
140& 12.54&   -8.06&   506&   0.50&  0.46&&  499&   0.29&  0.12&  517&   507$\pm$9 &  175&   L&   L07\\
\enddata
\tablenotetext{a}{Type: 'S' for semi-regular RSGs, 'L' for LSP RSGs with distinguish short period.}
\tablenotetext{b}{'Ref' of M$_{\rm bol}$: 'L06' for \citet{Levesque06}, 'L07' for \citet{Levesque07}, 'L09' for \citet{Levesque09}.}
\end{deluxetable}
\clearpage


\begin{deluxetable}{cccc}
\tabletypesize{\scriptsize}
\tablecaption{
 K$_{\rm S}$-band P-L relations of RSGs in the Galaxy, LMC, SMC and M33
 \label{individualpl}
}
\tablewidth{0pt}
\tablehead{
 \colhead{Galaxies} & \colhead{Slope $a$} & \colhead{intercept $b$} & \colhead{$\sigma$}
}
\startdata
Galaxy      & -4.34$\pm$0.60&  0.91$\pm$0.98& 0.51\\
LMC         & -3.74$\pm$0.30& -0.38$\pm$0.51& 0.24\\
SMC         & -3.28$\pm$0.39& -1.66$\pm$0.64& 0.27\\
M33         & -3.59$\pm$0.41& -0.88$\pm$0.69& 0.31\\
Combination & -3.68$\pm$0.18& -0.62$\pm$0.30& 0.31\\
\enddata

\end{deluxetable}
\clearpage


\begin{deluxetable}{cccccccccccc}
\tabletypesize{\scriptsize}
\tablecaption{
 Near- and Mid-infrared P-L relations of RSGs in SMC and LMC
 \label{pltab}
}
\tablewidth{0pt}
\tablehead{
 \colhead{} &
 \multicolumn{3}{c}{SMC} &
 &
 \multicolumn{3}{c}{LMC}&
 &
 \multicolumn{3}{c}{SMC+LMC}
 \\ \cline{2-4} \cline{6-8} \cline{10-12}

 \colhead{Band} &
 \colhead{Slope $a$} & \colhead{intercept $b$} & \colhead{$\sigma$} &
 &
 \colhead{Slope $a$} & \colhead{intercept $b$} & \colhead{$\sigma$} &
 &
 \colhead{Slope $a$} & \colhead{intercept $b$} & \colhead{$\sigma$}
}
\startdata
J          & -2.57$\pm$0.35& -2.56$\pm$0.58& 0.25&& -3.61$\pm$0.38&  0.39$\pm$0.63& 0.30&& -3.15$\pm$0.25& -0.93$\pm$0.42& 0.28\\
H          & -3.08$\pm$0.39& -1.96$\pm$0.63& 0.27&& -3.58$\pm$0.33& -0.50$\pm$0.54& 0.26&& -3.30$\pm$0.23& -1.30$\pm$0.38& 0.26\\
K$_{\rm S}$& -3.28$\pm$0.39& -1.66$\pm$0.64& 0.27&& -3.75$\pm$0.30& -0.38$\pm$0.51& 0.24&& -3.55$\pm$0.22& -0.90$\pm$0.37& 0.25\\
$[3.6]$    & -3.11$\pm$0.37& -2.37$\pm$0.60& 0.25&& -3.97$\pm$0.28& -0.12$\pm$0.47& 0.22&& -3.78$\pm$0.22& -0.64$\pm$0.36& 0.24\\
$[4.5]$    & -3.29$\pm$0.35& -1.87$\pm$0.57& 0.24&& -4.34$\pm$0.25&  0.89$\pm$0.41& 0.20&& -4.10$\pm$0.20&  0.26$\pm$0.33& 0.22\\
$[5.8]$    & -3.58$\pm$0.37& -1.33$\pm$0.61& 0.26&& -4.53$\pm$0.28&  1.13$\pm$0.46& 0.22&& -4.41$\pm$0.22&  0.83$\pm$0.36& 0.25\\
$[8.0]$    & -4.03$\pm$0.39& -0.33$\pm$0.64& 0.27&& -5.33$\pm$0.39&  2.90$\pm$0.64& 0.31&& -5.36$\pm$0.30&  3.07$\pm$0.49& 0.30\\
$[24]$     & -7.73$\pm$1.04&  8.50$\pm$1.70& 0.72&& -7.82$\pm$0.69&  7.80$\pm$1.16& 0.56&& -9.36$\pm$0.66& 12.34$\pm$1.10& 0.74\\
\enddata

\end{deluxetable}


\begin{deluxetable}{cccccc}
\tabletypesize{\scriptsize}
\tablecaption{
 Period of LSP of 31 SMC RSGs obtained by PDM, Period04 and CLEAN
 \label{lsptab}
}
\tablewidth{0pt}
\tablehead{
 \colhead{No.} & \colhead{A (mag)} &\colhead{PDM (d)} & \colhead{Period04 (d)} & \colhead{CLEAN (d)} & \colhead{$\langle$LSP$\rangle \pm \sigma$ (d)}
}
\startdata
5     &0.19  &2460   &2436   &2380   &2425$\pm$41 \\
6     &0.06  &1828   &1816   &1870   &1838$\pm$28 \\
8     &0.19  &2406   &2445   &2510   &2453$\pm$53 \\
12    &0.07  &1588   &1553   &1625   &1588$\pm$36 \\
14    &0.14  &2697   &2656   &2598   &2650$\pm$50 \\
24    &0.02  &1720   &1549   &1613   &1627$\pm$86 \\
26    &0.17  &1627   &1606   &1636   &1623$\pm$15 \\
32    &0.04  &2024   &2017   &2182   &2074$\pm$93 \\
42    &0.13  &1848   &1811   &1793   &1817$\pm$28 \\
45    &0.11  &2582   &2474   &2633   &2563$\pm$81 \\
54    &0.11  &1283   &1295   &1302   &1293$\pm$10 \\
62    &0.08  &2589   &2545   &2566   &2566$\pm$22 \\
64    &0.07  &1615   &1601   &1624   &1613$\pm$12 \\
69    &0.10  &2549   &2538   &2507   &2531$\pm$22 \\
74    &0.04  &2010   &2045   &2002   &2019$\pm$23 \\
79    &0.36  &2089   &2007   &2145   &2080$\pm$69 \\
83    &0.11  &1849   &2004   &2151   &2001$\pm$151\\
84    &0.07  &2749   &2521   &2604   &2624$\pm$115\\
89    &0.09  &1620   &1610   &1640   &1623$\pm$15 \\
93    &0.04  &1420   &1309   &1434   &1387$\pm$68 \\
95    &0.09  &2533   &2546   &2582   &2553$\pm$25 \\
98    &0.08  &2212   &2294   &2230   &2245$\pm$43 \\
100   &0.06  &1385   &1377   &1447   &1403$\pm$38 \\
107   &0.04  &1702   &1740   &1807   &1749$\pm$53 \\
110   &0.13  &1482   &1574   &1613   &1556$\pm$67 \\
114   &0.04  &1827   &1891   &1845   &1854$\pm$33 \\
115   &0.10  &2507   &2540   &2582   &2543$\pm$38 \\
119   &0.07  &2414   &2537   &2575   &2508$\pm$84 \\
120   &0.05  &1154   &1209   &1184   &1182$\pm$28 \\
128   &0.04  &1849   &1746   &1860   &1818$\pm$63 \\
129   &0.03  &1620   &1600   &1628   &1616$\pm$14 \\

\enddata
\end{deluxetable}


\begin{figure}
\centering
\includegraphics[width=\textwidth, bb=85 225 480 545]{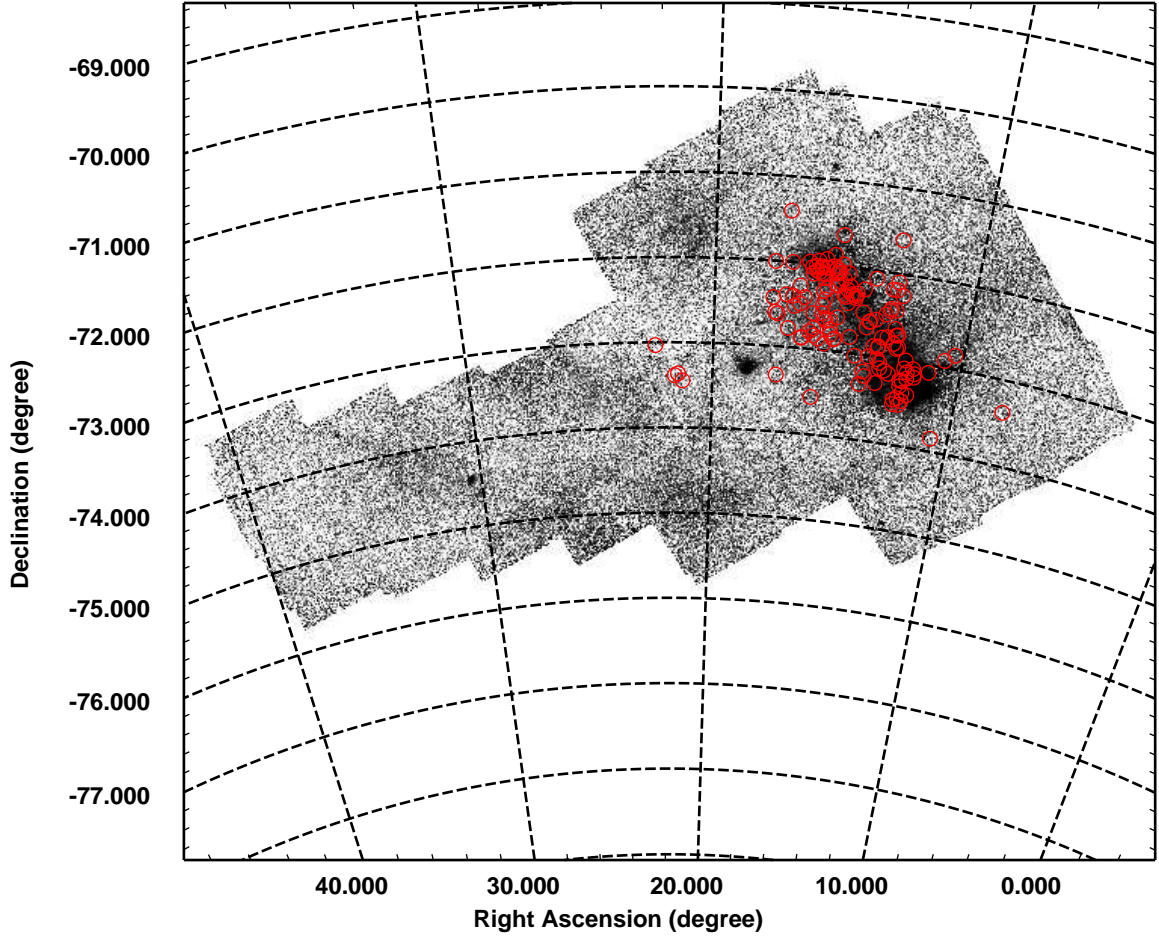}
\caption{
 The spatial distribution of all the 140 sample stars (red circles) superposed on the \emph{Spitzer}/SAGE 8 $\mu$m mosaic image, mostly in the bar region.
} \label{smc}
\end{figure}

\clearpage


\begin{figure}
\centering
\includegraphics[width=\textwidth, bb=55 380 550 720]{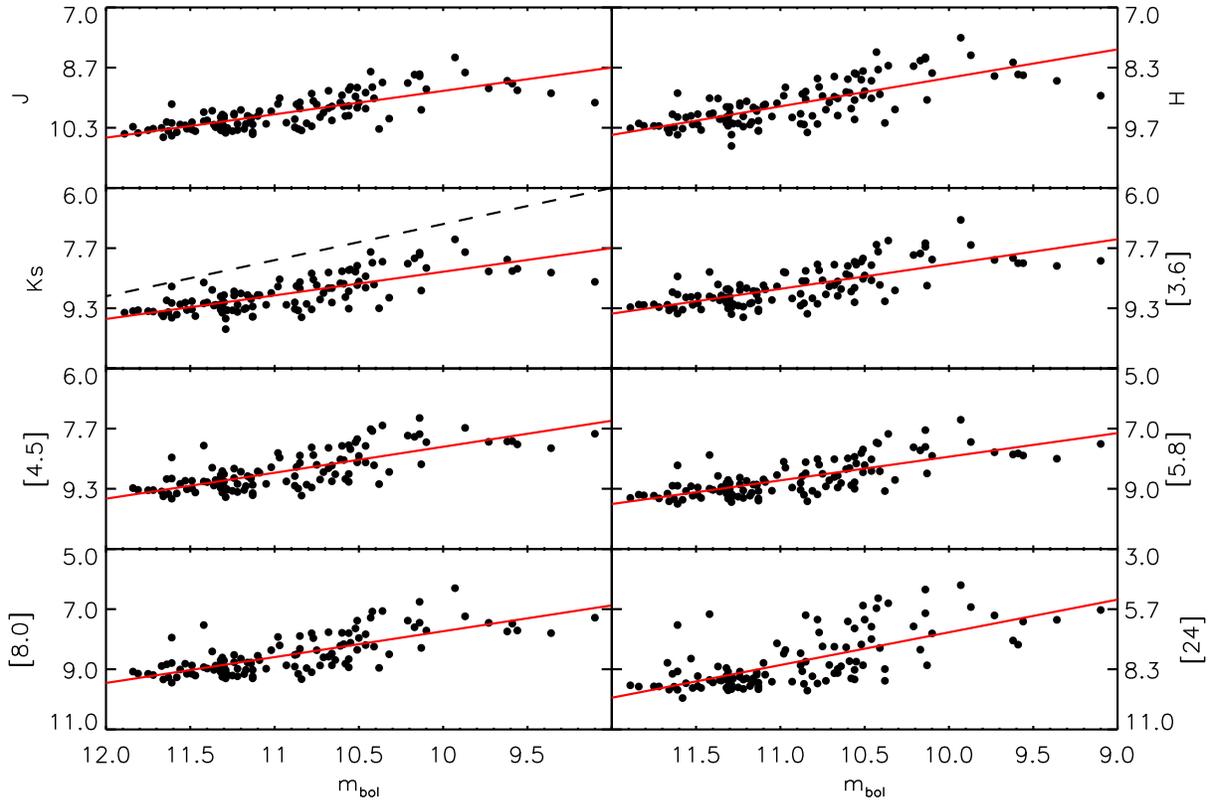}
\caption{
 The relationship between the infrared and bolometric magnitudes of the 113 reliable RSGs targets from \citet{Massey03}, \citet{Levesque06} and \citet{Massey07}. The red solid line shows the robust linear fitting. The result in the K$_{\rm S}$ band indicates a moderate discrepancy between ours and \citet{Josselin00}'s (black dash-line) possibly due to the influence of metallicity and extinction.
} \label{band-mbol}
\end{figure}

\clearpage


\begin{figure}
\centering
\includegraphics[width=0.95\textwidth, bb=50 300 550 740]{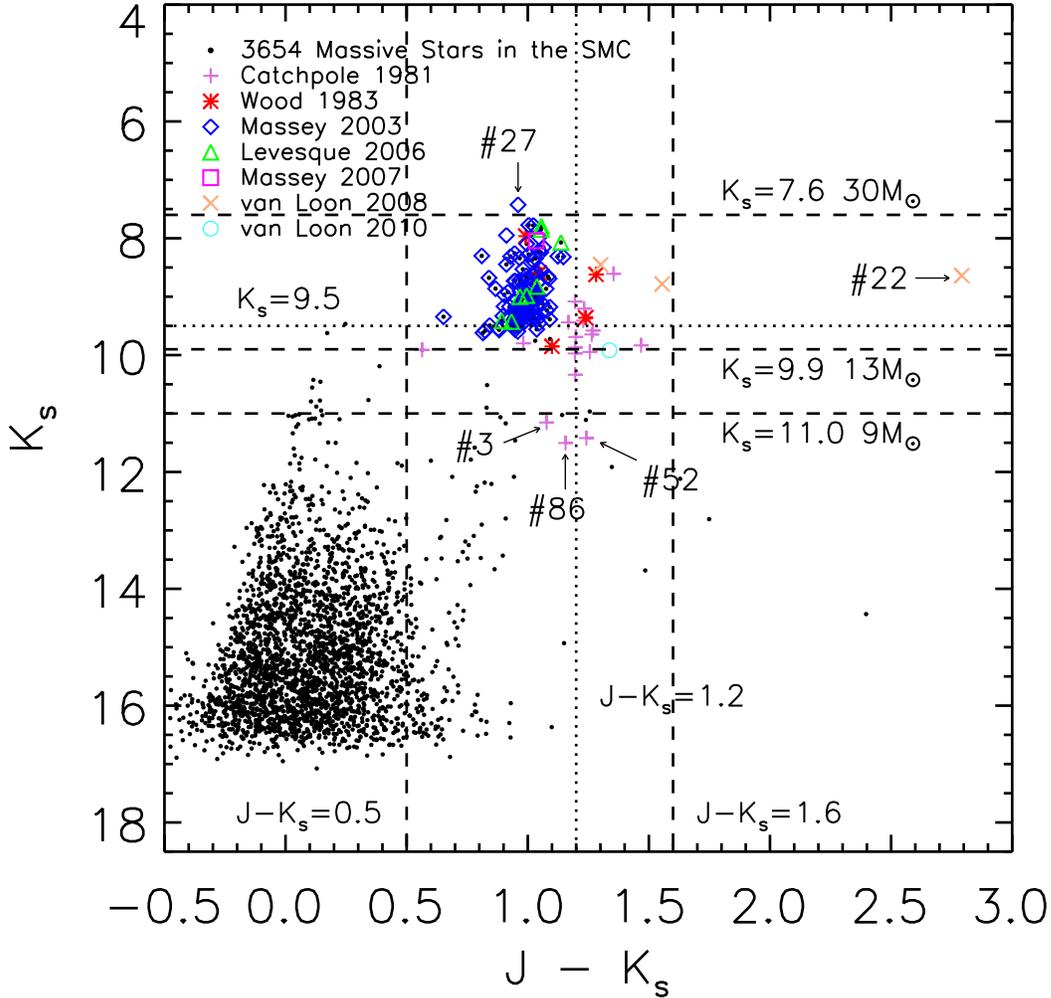}
\caption{
 The K$_{\rm S}$ vs. $\rm J - K_{S}$ CMD for all the targets. For comparison, the 3654 massive stars from \citet{Bonanos10} are added as background and denoted by black dots. Different symbols denote different resources. Most of the targets have occupied the region of 7.5 $<\rm K_{S} <$ 10 mag and 0.5 $<\rm J - K_{S} < $ 1.6 mag. The vertical dash-line at $\rm J - K_{S} = 1.6$ mag indicates the boundary of carbon-rich stars suggested by \citet{Wood83}. The other dash-line at $\rm J - K_{S} = 0.5$ mag shows the observational boundary of RSGs by \citet{Josselin00}. The horizontal dash-lines show the theoretical luminosity boundaries in K$_{\rm S}$ band with defined masses. The vertical dot-line indicates the limit of evolutionary synthesis models of RSGs in near-infrared by \citet{Origlia99}. The horizontal dot-line indicates the criterion of $\rm M_{ bol}=-7.1$ mag used to distinguish AGBs and RSGs by \citet{Wood83}. Obvious outliers are marked by their ID in Table~\ref{140tab}. The conventions of the symbols for resources are kept in following figures.
} \label{figkjk}
\end{figure}

\clearpage


\begin{figure}
\centering
\includegraphics[width=\textwidth, bb=50 280 550 750]{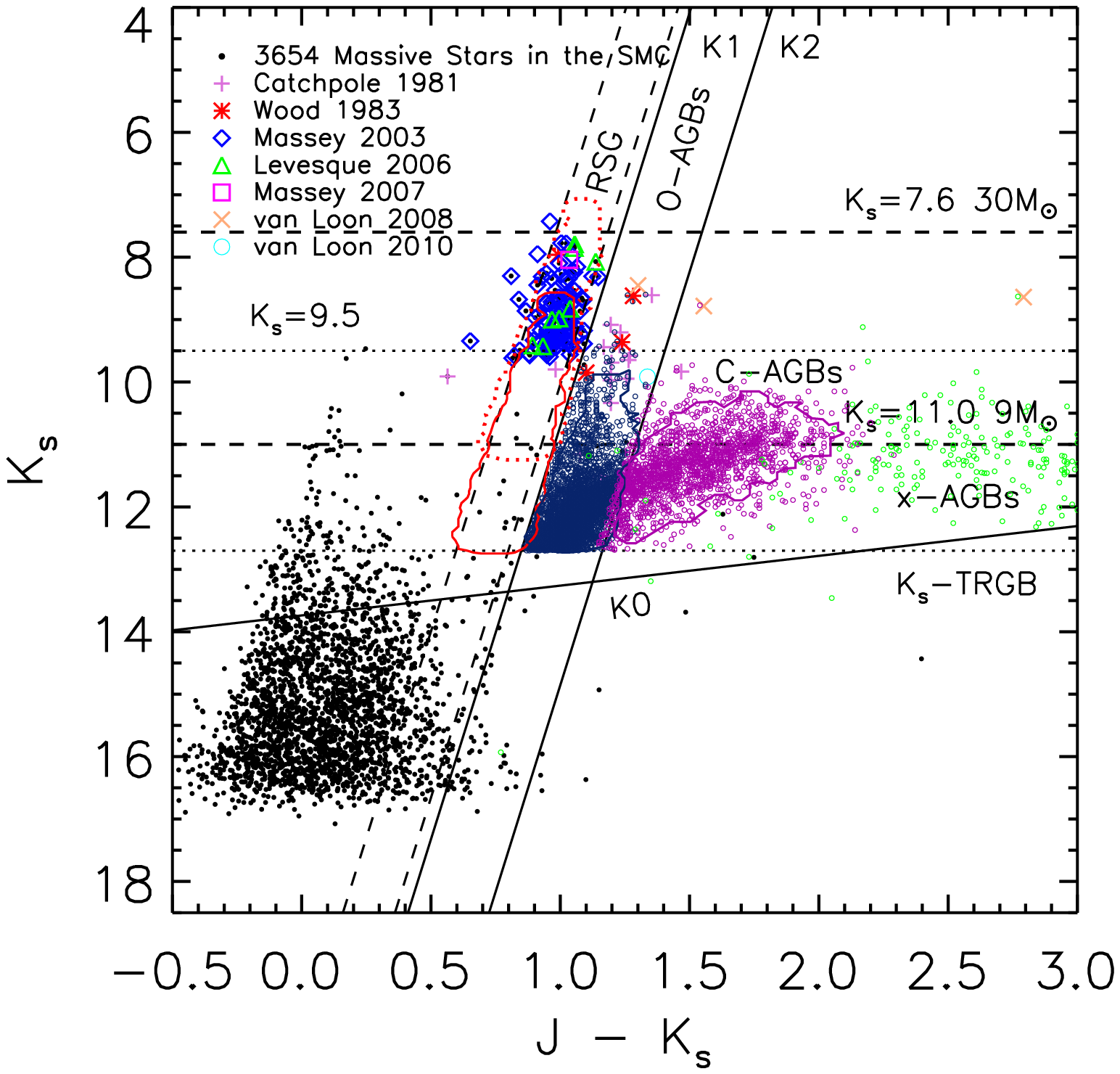}
\caption{
 The same as Fig.~\ref{figkjk} but for different classifications of evolved stars. The $\rm J-K_{S}$ cuts shown as solid lines are from \citet{Cioni06a, Cioni06b}, where C-AGBs are defined by K$_{\rm S} <$ K0 and K$_{\rm S}>$ K2, O-AGBs by K$_{\rm S}<$ K0 and K1 $<\rm K_{S}<$ K2. The K$_{\rm S}$-band tip of the red giant branch (TRGB) is denoted by dot-line. The profiles of contour maps are from \citet{Boyer11} with red, blue and purple solid lines for RSG, O-AGBs and C-AGBs respectively. For a better view of distinguishing RSGs from AGBs, all the O-AGBs (dark blue circle), C-AGBs (purple circle) and x-AGBs (green circle) from Table 4 of \citet{Boyer11} are overplotted. From the diagram, the RSGs sample of Boyer has too low luminosity and too narrow and blue color index, where the red dot-line shows the region of RSGs after moving redward 0.1 mag and upward 1.5 mag.
} \label{figkjk2}
\end{figure}

\clearpage


\begin{figure}
\centering
\includegraphics[width=\textwidth, bb=50 280 550 750]{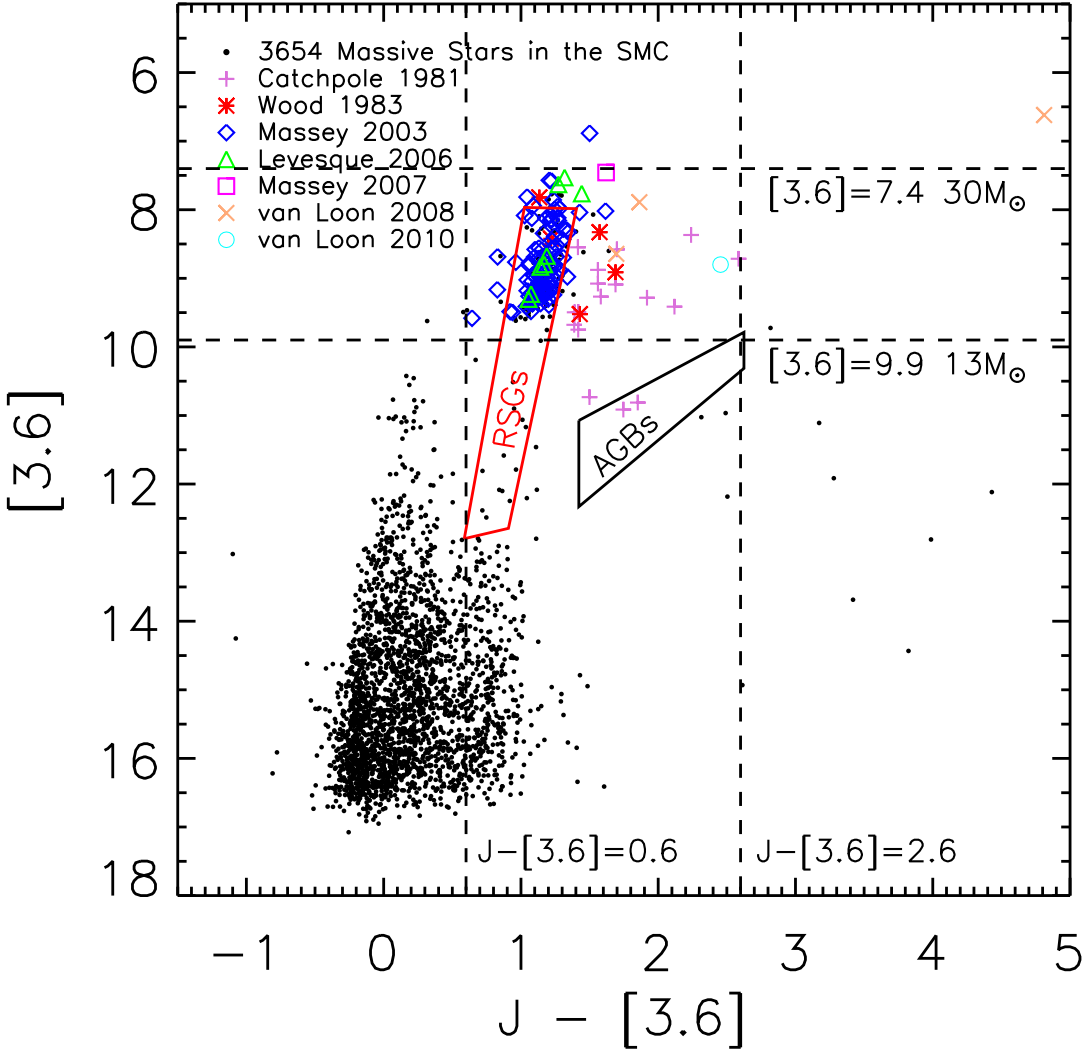}
\caption{
 The [3.6] vs. $\rm J-[3.6]$ diagram. The regions of AGBs by \citet{Bonanos10} and RSGs by \citet{Boyer11} are denoted by black and red solid lines respectively. The limits of [3.6]-band magnitude are shown as horizontal dash-lines according to the RSGs masses. The boundaries of color index are $\rm J-[3.6]=0.59$ mag and $\rm J-[3.6]=2.62$ mag, i.e. the bluest point of RSGs region by \citet{Boyer11} and the reddest edge of AGBs by \citet{Bonanos10}.
} \label{fig36j36}
\end{figure}

\clearpage


\begin{figure}
\centering
\includegraphics[width=\textwidth, bb=50 280 550 750]{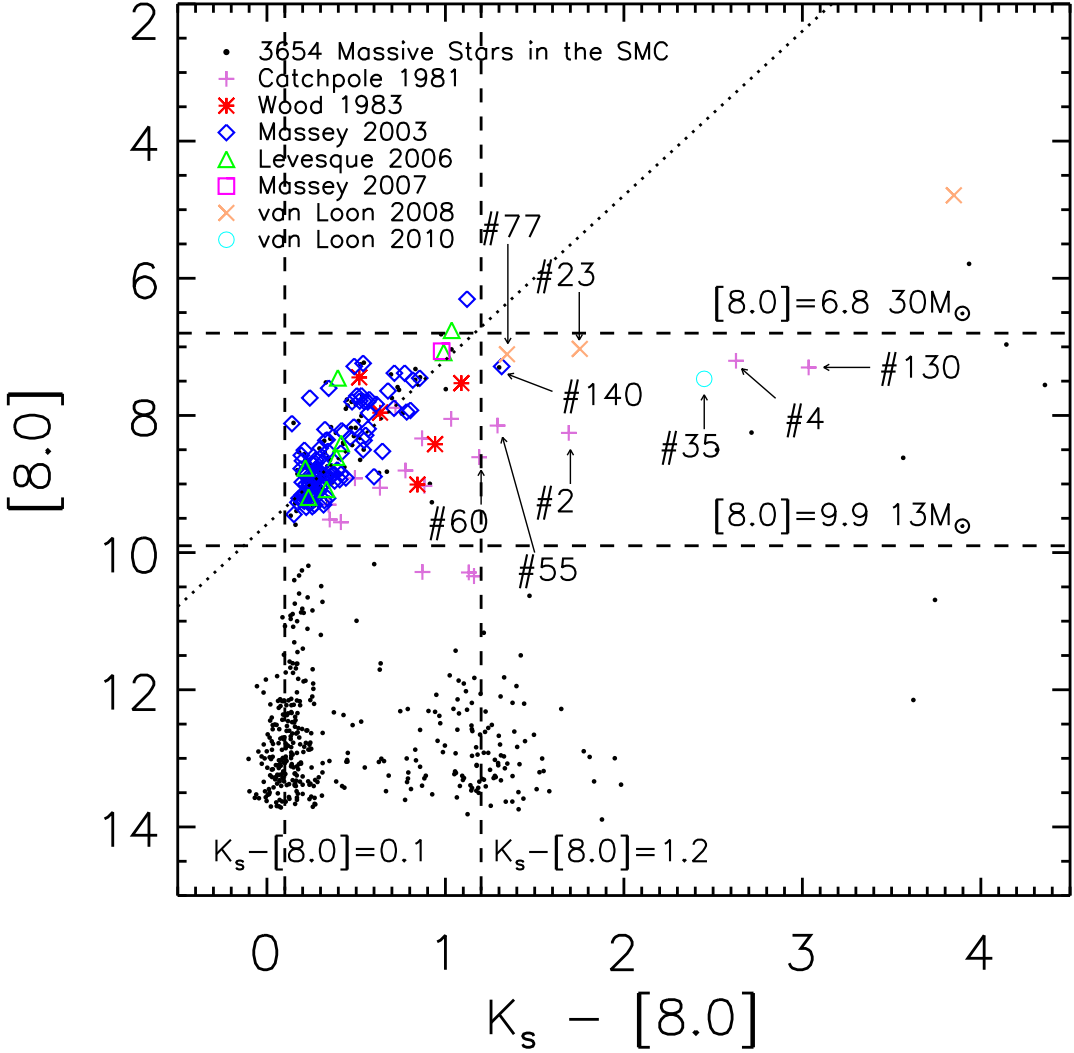}
\caption{
 The [8.0] vs. K$_{\rm S}-[8.0]$ diagram. The [8.0] magnitude limits are set up at $[8.0]=6.78$ mag (30 $\rm M_{\sun}$) and $[8.0]=9.90$ mag (13 $\rm M_{\sun}$). The K$_{\rm S}-[8.0]$ gets larger with increased [8.0], which can be fitted by a line as $\rm [8.0]=(-1.05\pm0.11)(K_{S}-[8.0])+(9.08\pm0.10)$ with $\sigma=0.71$ shown as a dot-line. According to this line, the red limit of K$_{\rm S}-[8.0]$ is decided as the cross-point of the line with the upper limit of luminosity. The blue limit of K$_{\rm S}-[8.0]$ is set at 0.1 mag since no targets bluer than it.
} \label{fig8k8}
\end{figure}

\clearpage


\begin{figure}
\centering
\includegraphics[width=\textwidth, bb=50 280 550 750]{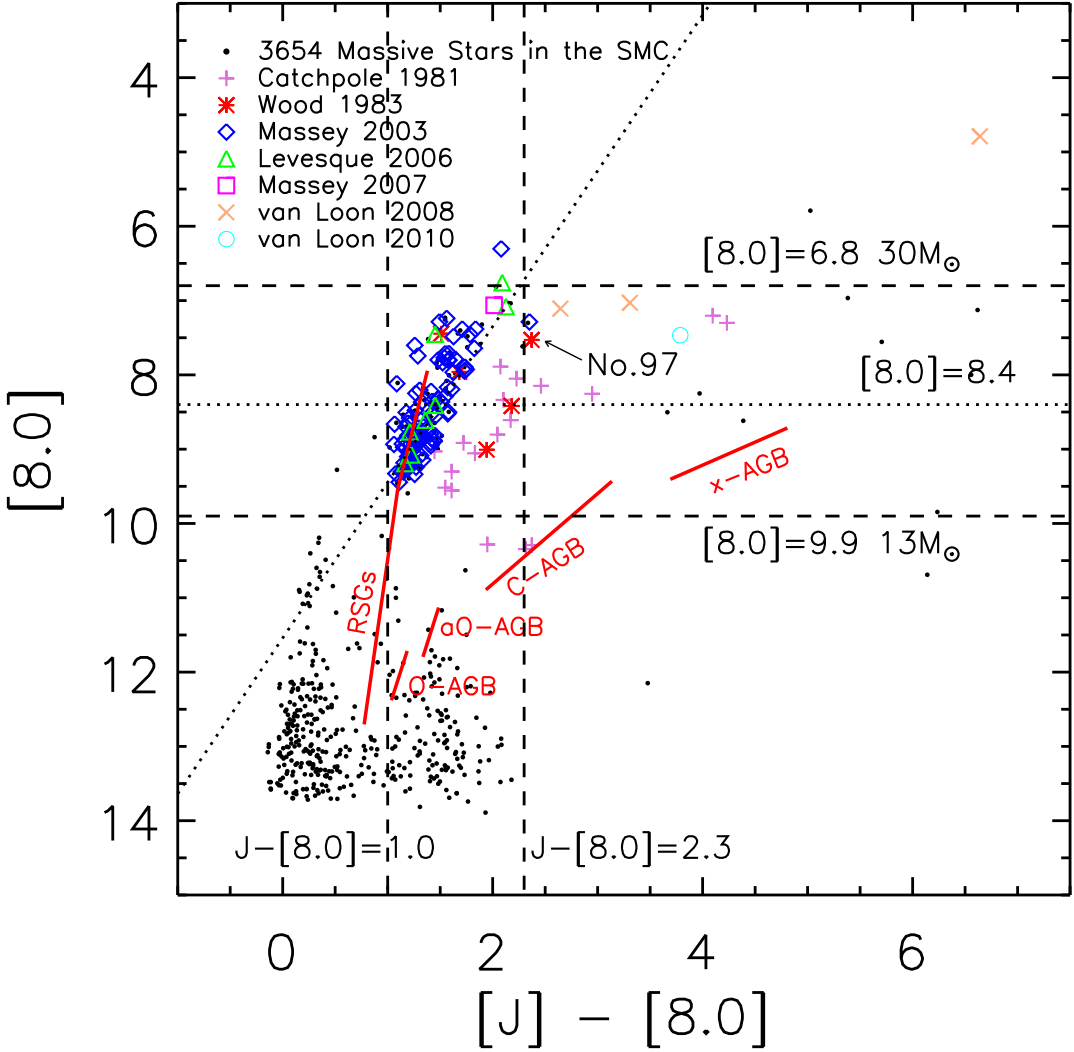}
\caption{
 The [8.0] vs. $\rm J-[8.0]$ diagram. As Fig.~\ref{fig8k8}, the linear fit $[8.0]=(-1.0\pm0.88)(\rm J-[8.0])+(10.08\pm0.13)$ with $\sigma=0.70$ is shown as a dot-line. The limits of $\rm J-[8.0]$ are set at 3.31 and 1.0 mag by the same way as in Fig.~\ref{fig8k8}. The dashed line at $[8.0] = 8.4$ mag is the upper limit of AGBs in the [8.0]-band (see Figure 4 in \citealt{Boyer11}).
} \label{fig8j8}
\end{figure}

\clearpage


\begin{figure}
\centering
\includegraphics[width=\textwidth, bb=50 280 550 750]{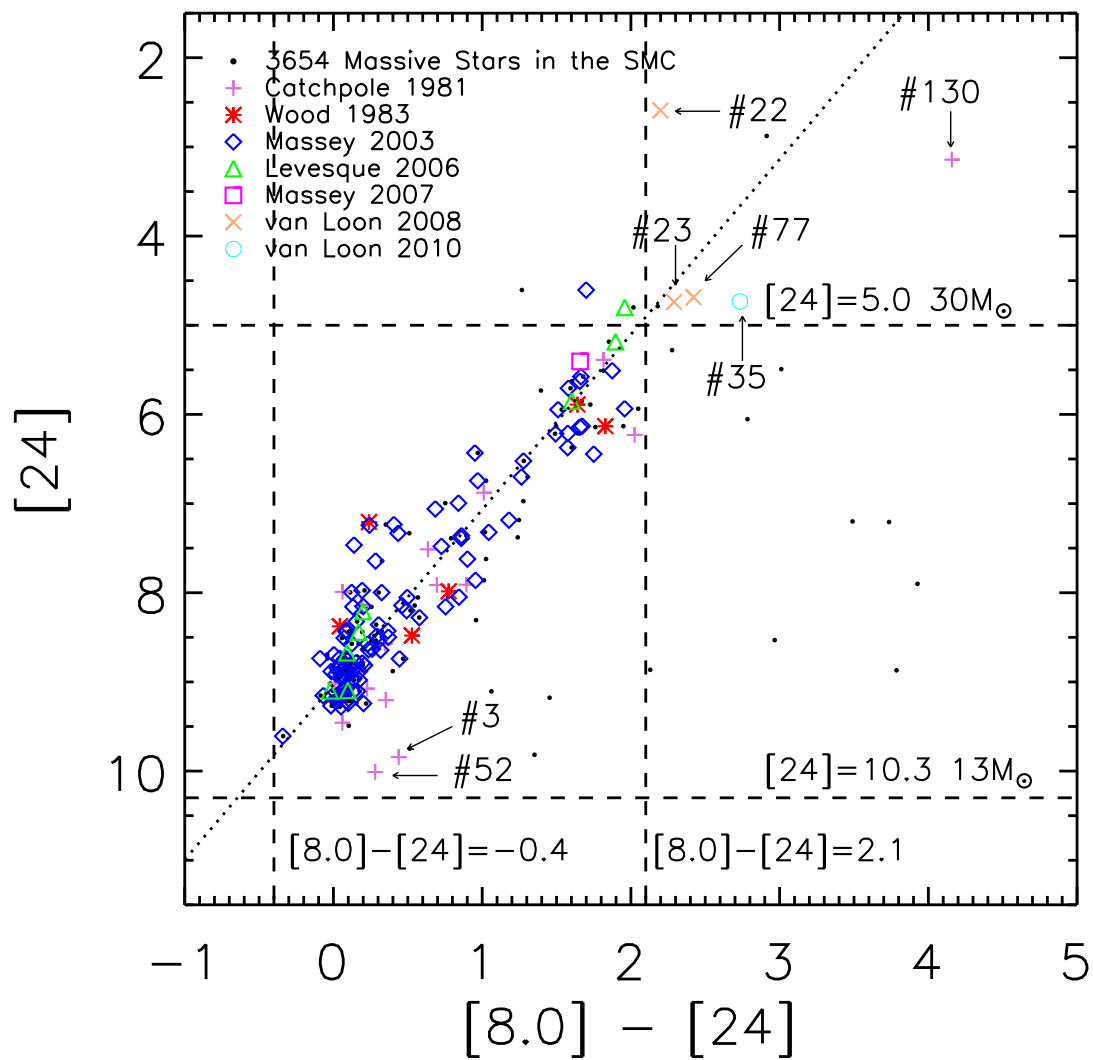}
\caption{
 The same as Fig.~\ref{fig8k8}, but for [24] vs. [8.0] - [24]. The correlation of the brightness in [24] with the color index $[8.0]-[24]$ is very tight since both reflect the radiation of the dust. The linear fit for this relationship is $[24]=(-1.96\pm0.05)([8.0]-[24])+(9.03\pm0.06)$ with $\sigma=0.45$.
} \label{fig24824}
\end{figure}

\clearpage


\begin{figure}
\centering
\includegraphics[width=\textwidth, bb=50 280 550 750]{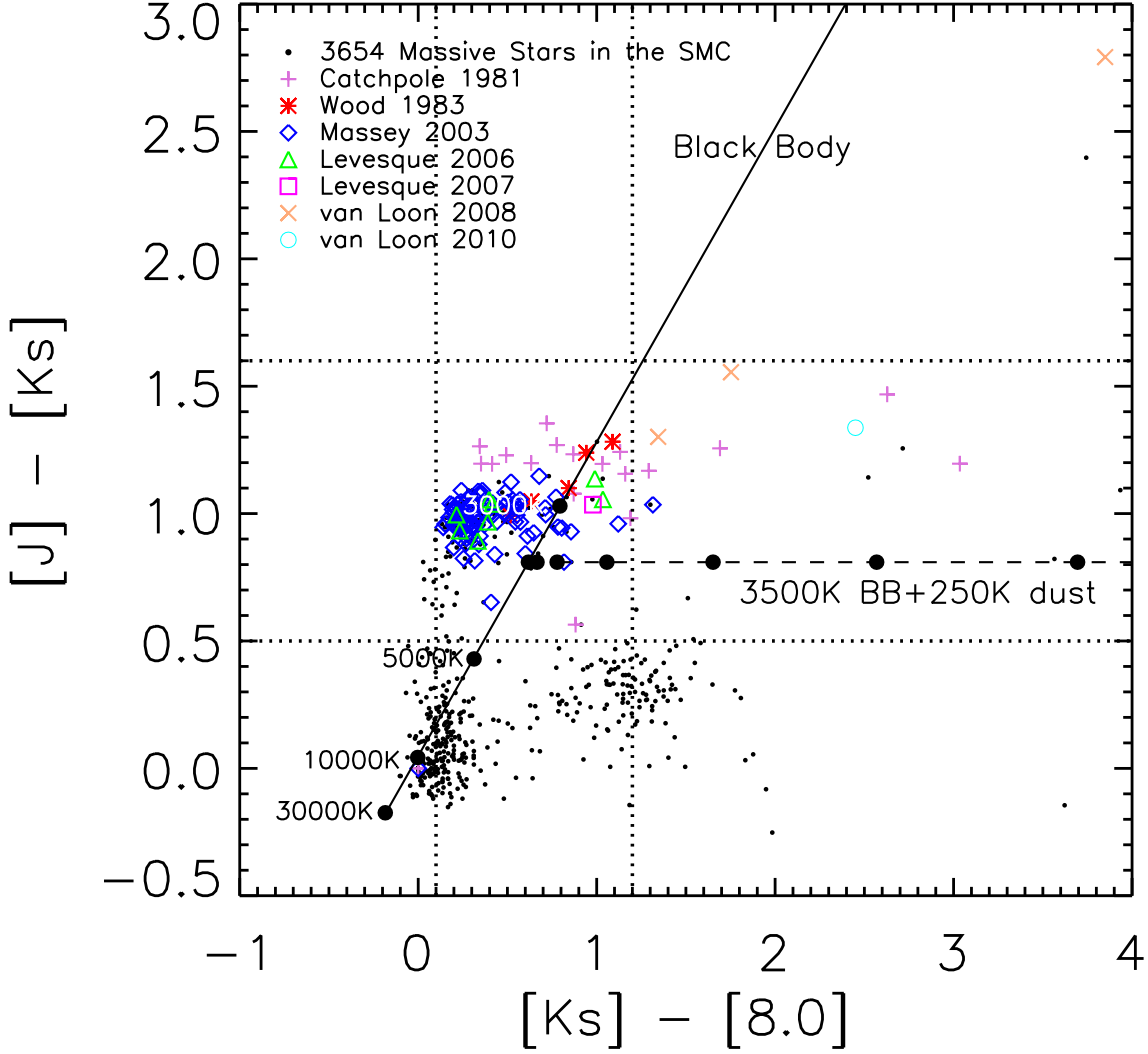}
\caption{
 The two-color diagram $\rm J - \rm K_{\rm S}$ versus K$_{\rm S}-[8.0]$. The black solid and dash-line represents the black body (BB) model at different temperatures and a model of a 3500 K BB plus 250 K dust from \citet{Bonanos10} respectively (see Section 4 of \citealt{Bonanos09} for detailed description). The limits of $0.5\sim1.6$ mag in $\rm J - \rm K_{\rm S}$ and $0.1\sim1.17$ mag in K$_{\rm S}-[8.0]$ are shown by dot-lines respectively. The outliers are the same as in Fig.~\ref{fig8k8} and not marked.
} \label{figjkk8}
\end{figure}

\clearpage


\begin{figure}
\centering
\includegraphics[width=\textwidth, bb=50 280 550 750]{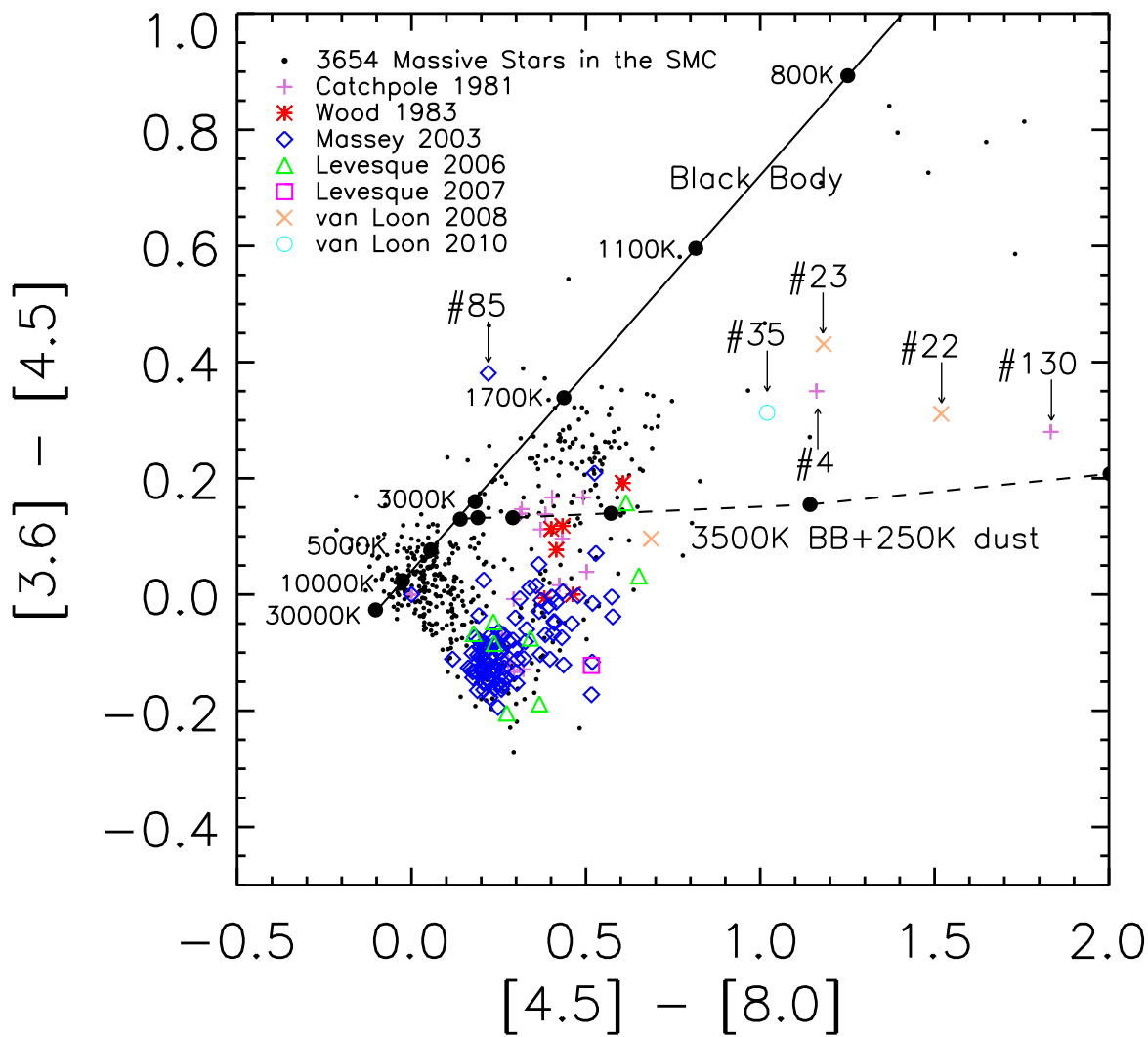}
\caption{
 The same as Fig.~\ref{figjkk8}, but for $[3.6]-[4.5]$ vs. $[4.5]-[8.0]$. Due to the impact of the CO line in the [4.5] band, most targets are far from the BB model. There is no certain limits for these color indexes, and only the obvious outliers are labeled by their IDs.
} \label{fig36454580}
\end{figure}

\clearpage

\begin{figure}
\centering
\includegraphics[width=\textwidth, bb=50 280 550 750]{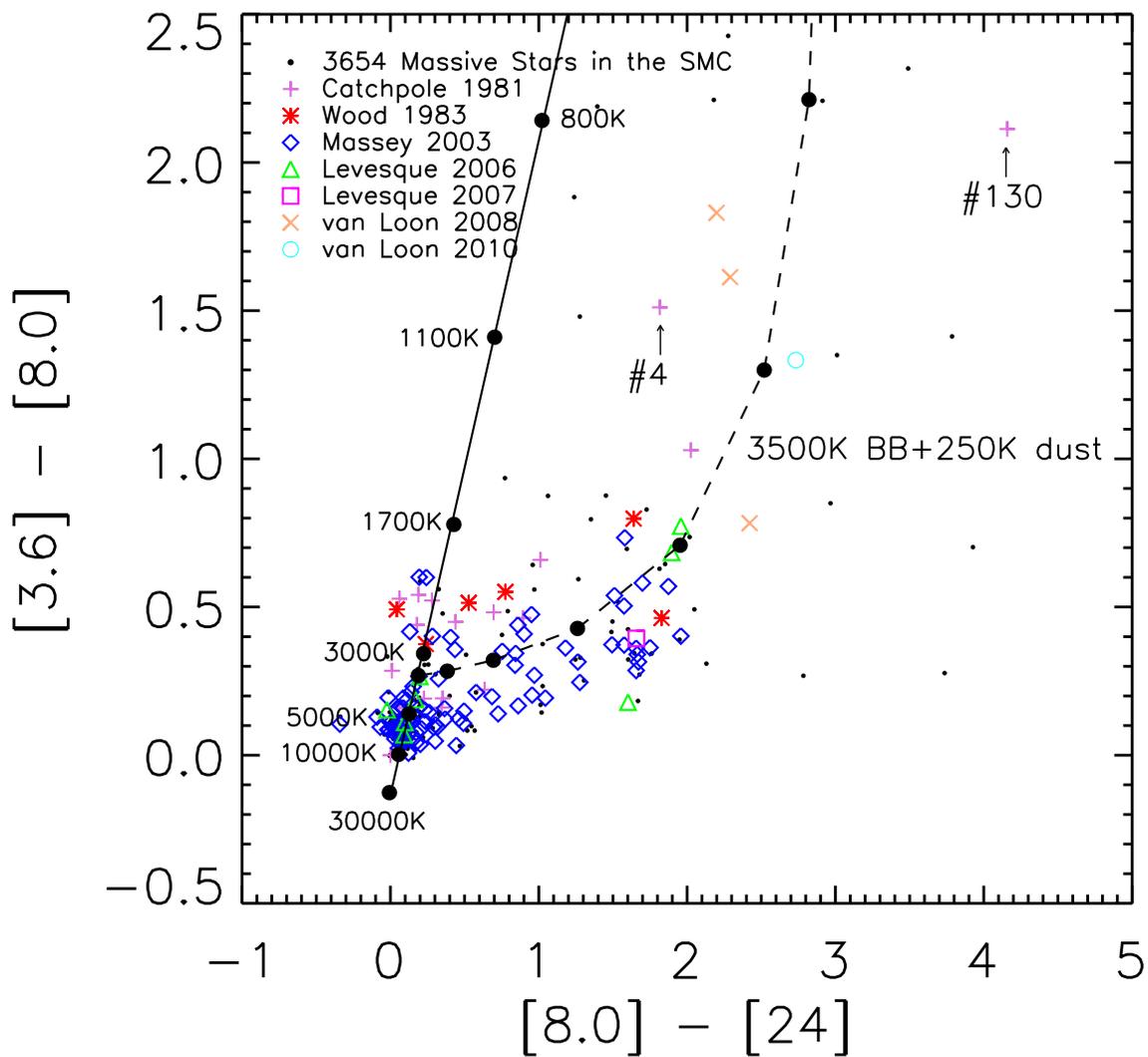}
\caption{
 The same as Fig.~\ref{figjkk8}, but for $[3.6]-[8.0]$ vs. $[8.0]-[24]$. As the circumstellar dust dims the [8.0]-band brightness, most targets are below the BB line but approximately follow the 3500 K BB plus 250 K dust model. The outliers in previous diagrams also roughly follow this model. Only Nos.4 and 130 are far from the model and marked.
} \label{fig36808024}
\end{figure}

\clearpage


\begin{figure}
\centering
\includegraphics[width=\textwidth, bb=50 325 550 750]{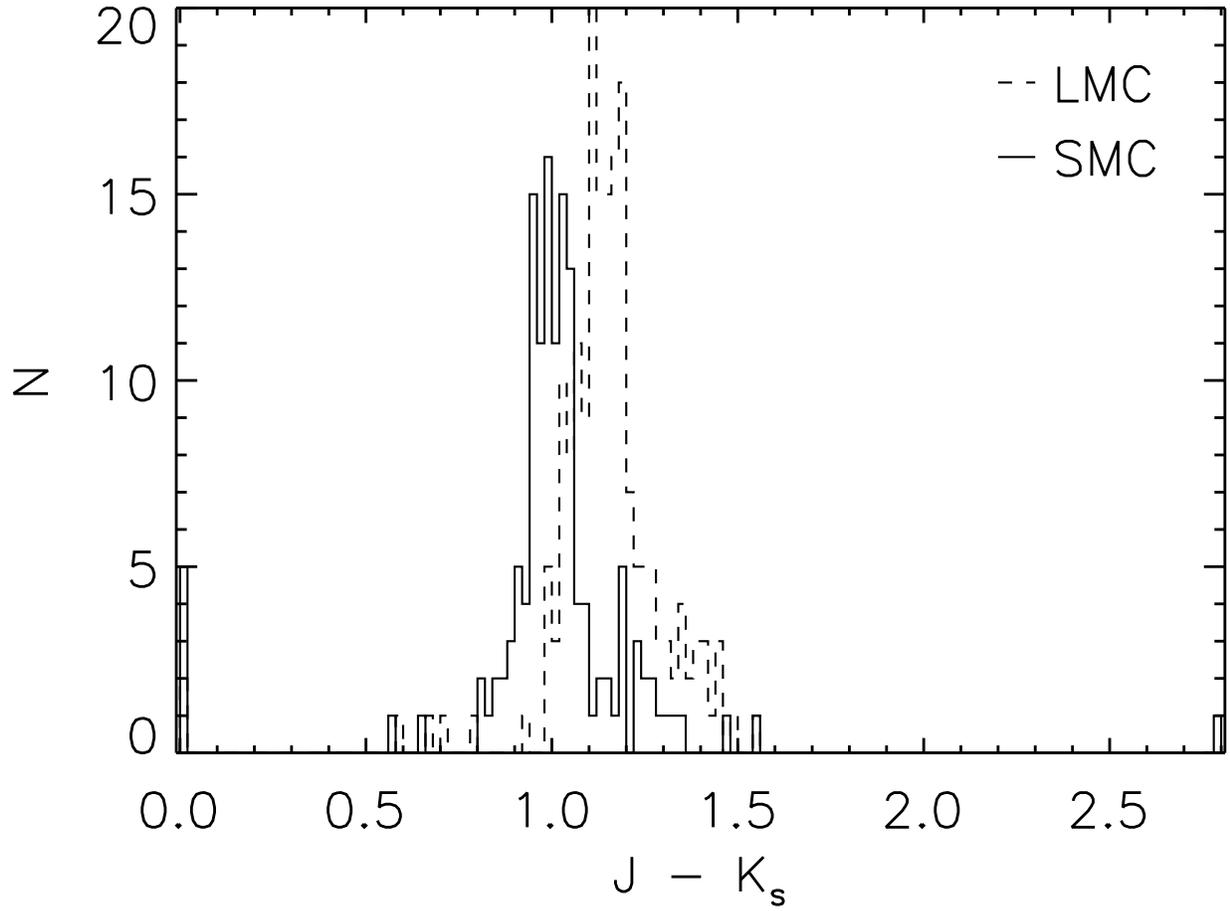}
\caption{
 Histogram of $\rm J - K_{S}$ of RSGs in SMC (solid line) and LMC (dash-line). The difference of $\rm J - K_{S}$ between them is about 0.1 mag.
} \label{figlmcsmcjk}
\end{figure}

\clearpage


\begin{figure}
\centering
\includegraphics[width=\textwidth, bb=35 290 515 740]{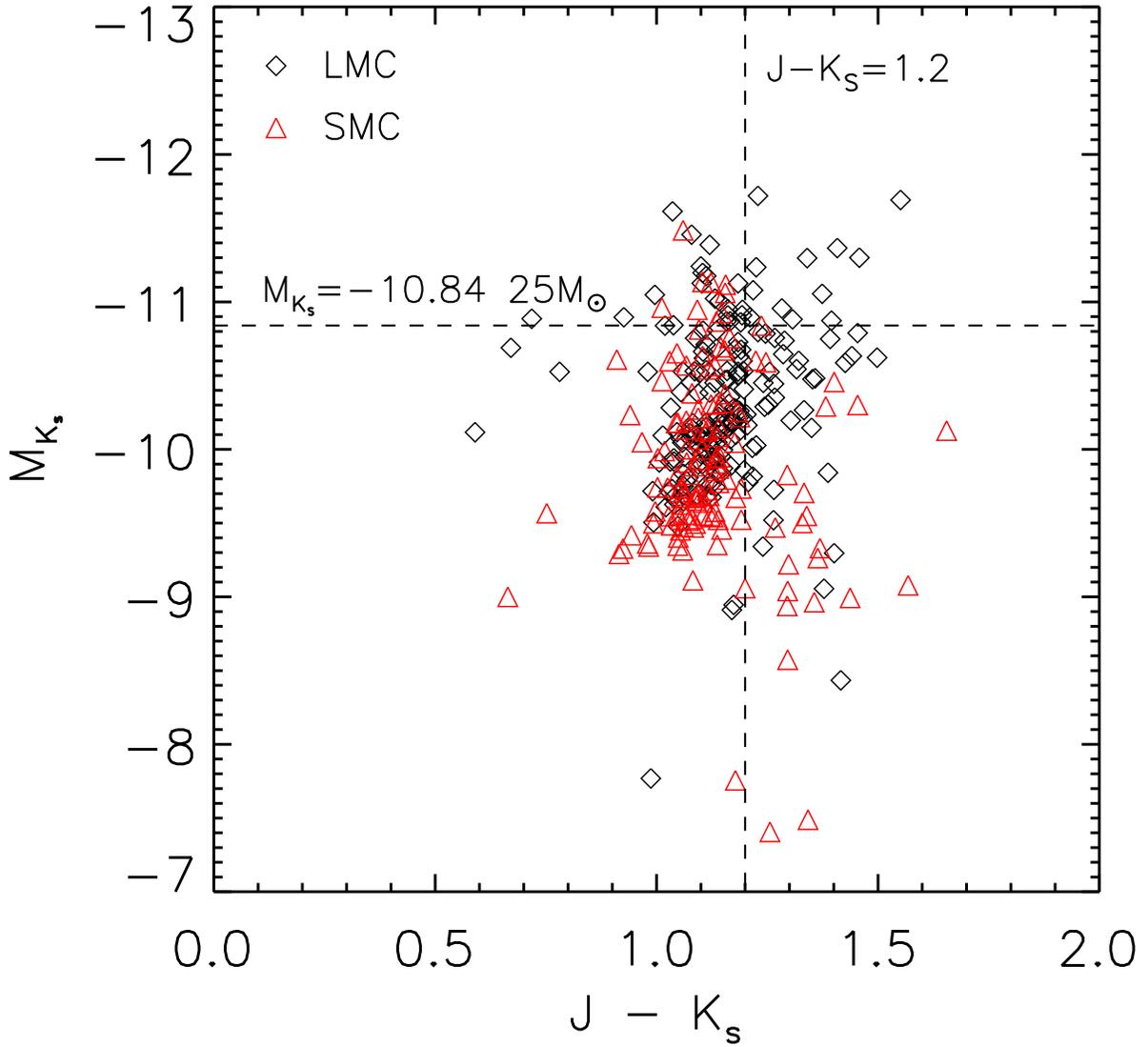}
\caption{
 The CMD for absolute magnitude in the $\rm K_{S}$-band versus $\rm J - K_{S}$ of RSGs in SMC (red triangle) and LMC (black diamond). The $\rm J - K_{S}$ in SMC is shifted redward by 0.1 mag to compensate for the metallicity effect. The SMC RSGs are slightly fainter and bluer than the LMC.
} \label{figlmcsmcmk}
\end{figure}

\clearpage


\begin{figure}
\centering
\includegraphics[width=\textwidth, bb=100 350 470 750]{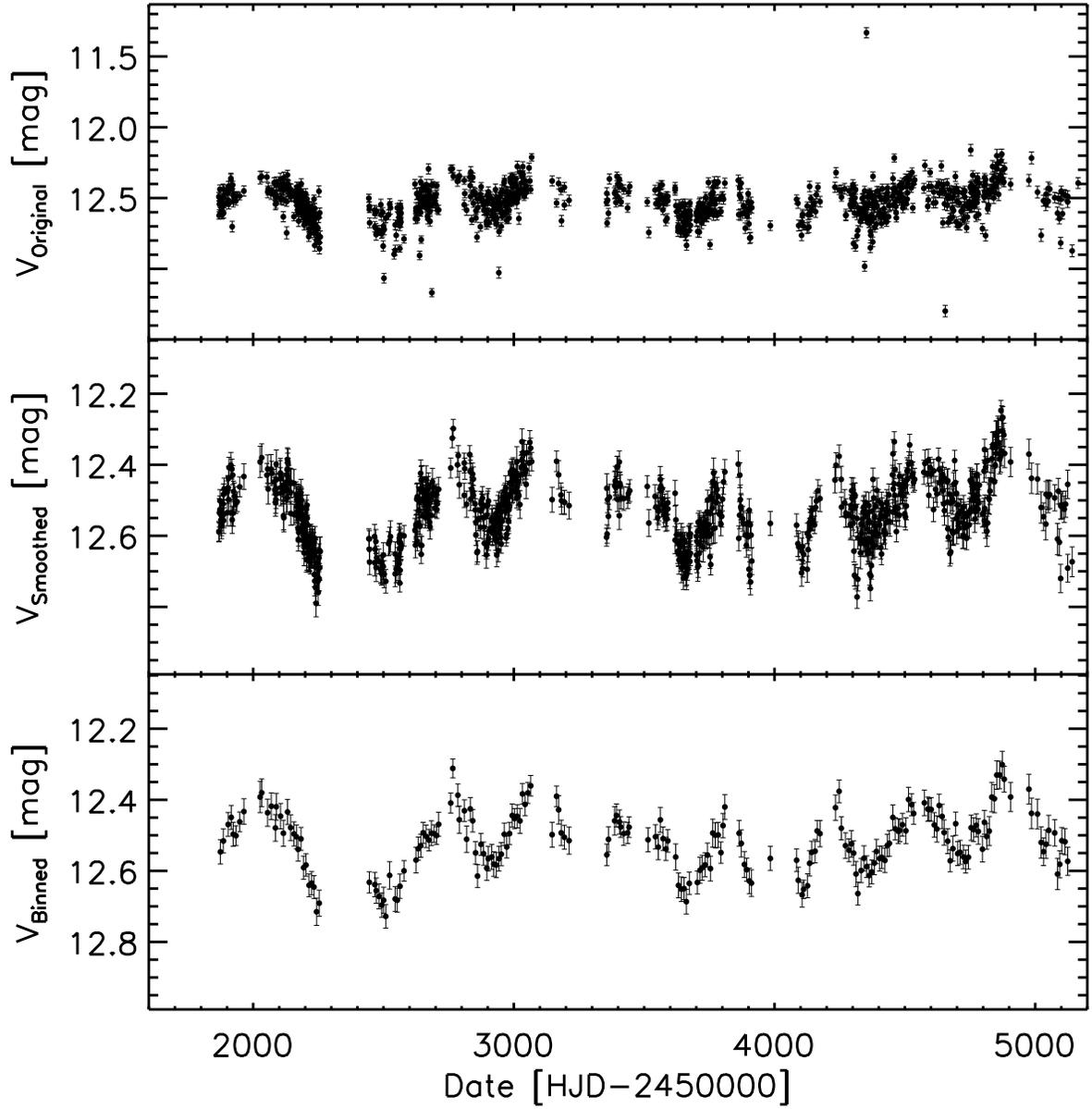}
\caption{
 An example of processing the ASAS light curve. Top to bottom: original light curve; least-square (Savitzky-Golay) polynomial smoothed light curve; 10 day-binned light curve.
} \label{asas_lightcurve_process}
\end{figure}

\clearpage


\begin{figure}
\centering
\includegraphics[width=\textwidth, bb=25 380 585 715]{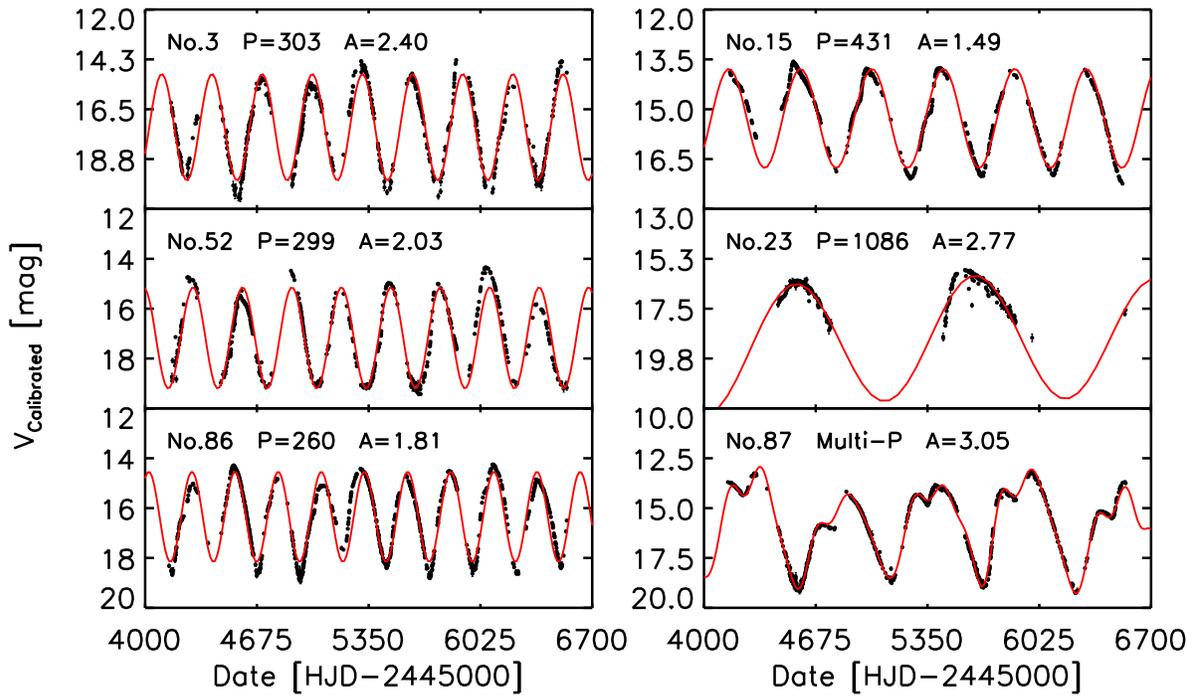}
\caption{
 The MACHO light curves. Three outliers that are too faint to be RSGs in the left column and three other similar targets in the right column are illustrated and shown with their ID number, period and amplitude of variation.
}
\label{macho_outliers}
\end{figure}

\clearpage


\begin{figure}
\centering
\includegraphics[width=0.48\textwidth, bb=100 340 465 730]{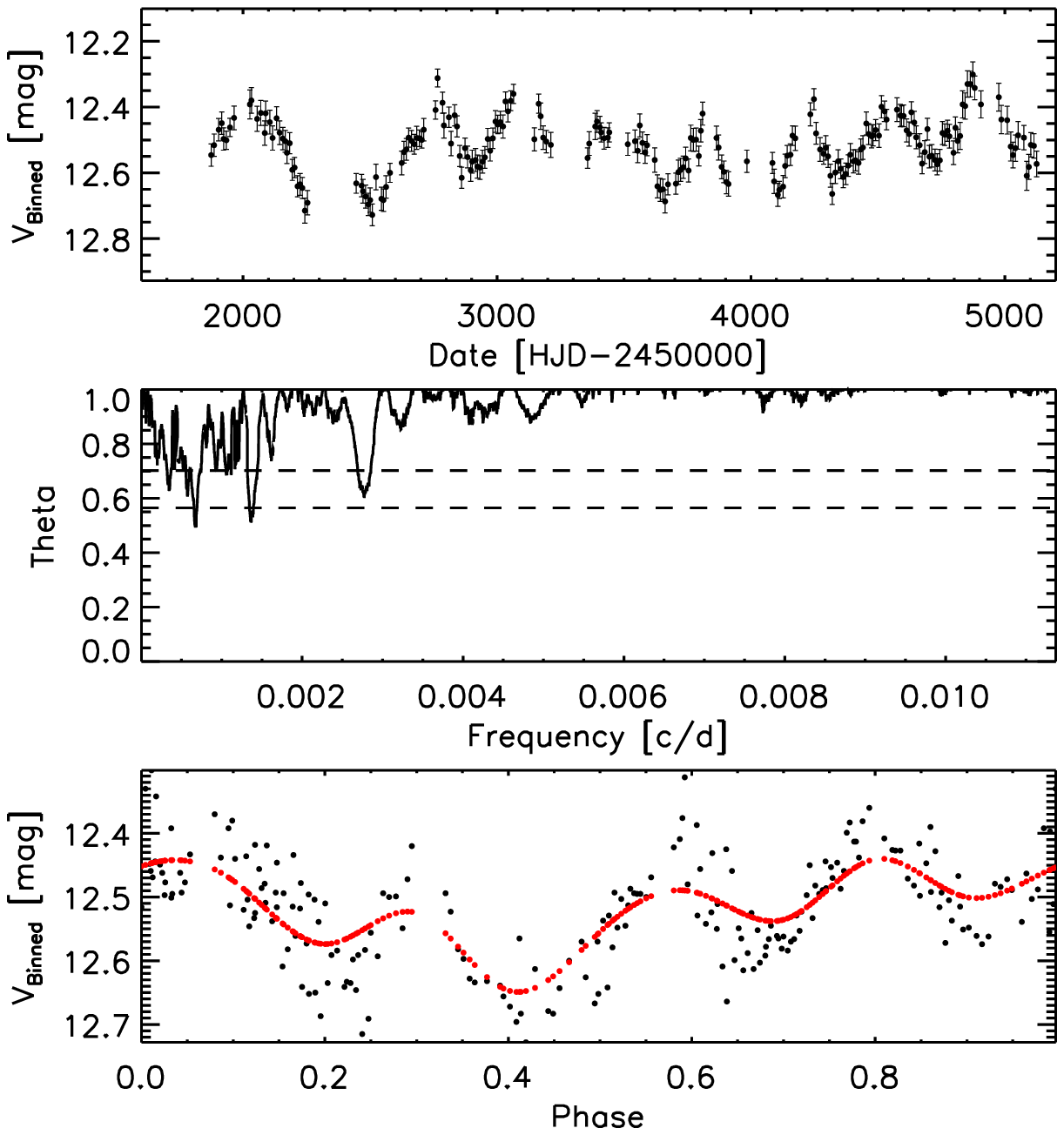}
\includegraphics[width=0.48\textwidth, bb=100 340 465 730]{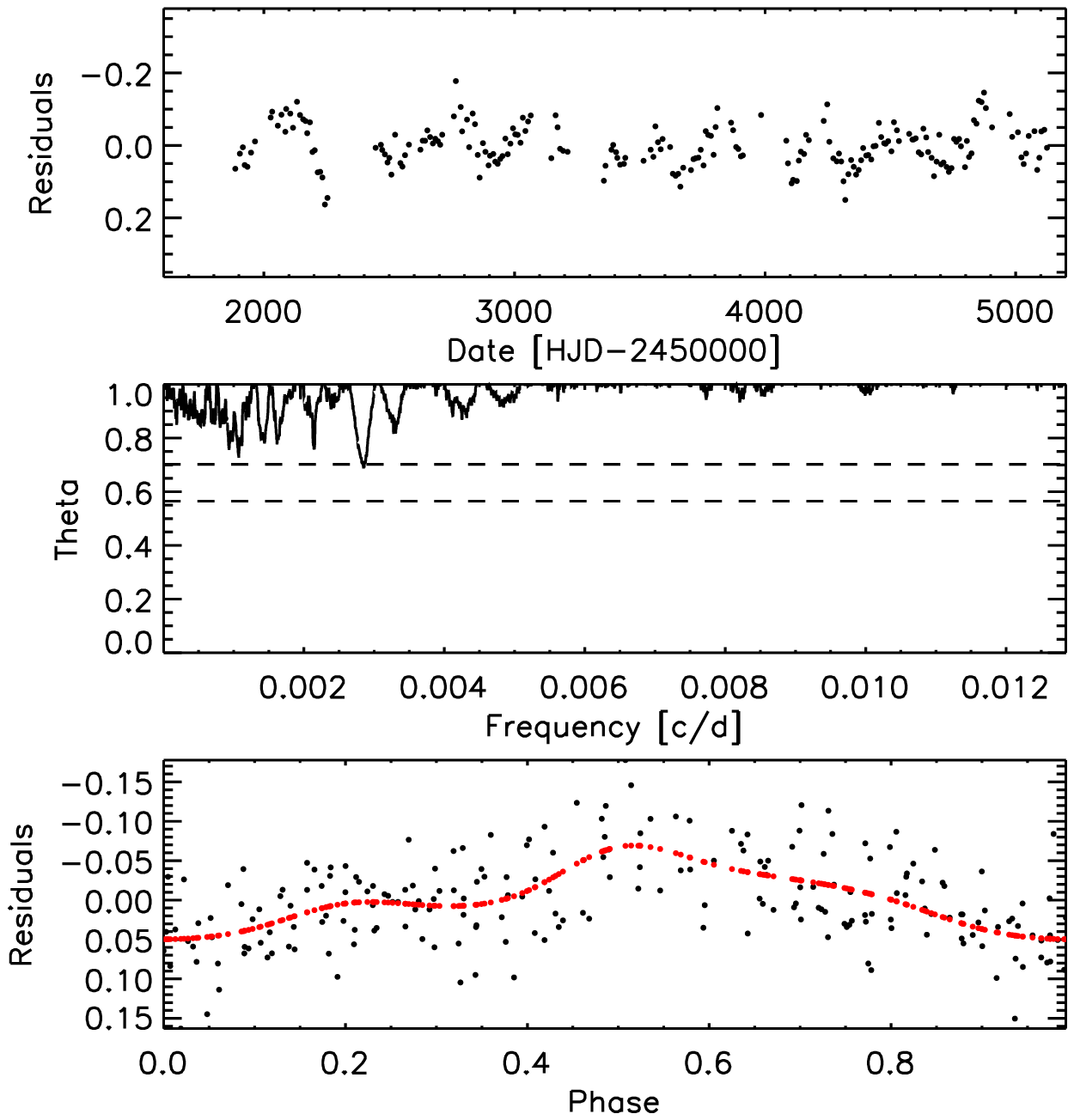}
\caption{
 The process to derive the period using the PDM2 method based on the ASAS data for the object No.12 which used also in following diagrams. Top to bottom: Left column : 10~day-binned light curve, theta diagram, phase diagram; right column: residual light curve, residual theta diagram, residual phase diagram. The dash-lines in the theta diagrams are significance levels of 0.05 and 0.01 respectively. The red lines in the bottom panel are the fitted curve.
}
\label{pdm_process}
\end{figure}

\clearpage


\begin{figure}
\centering
\includegraphics[width=0.8\textwidth, bb=160 360 450 730]{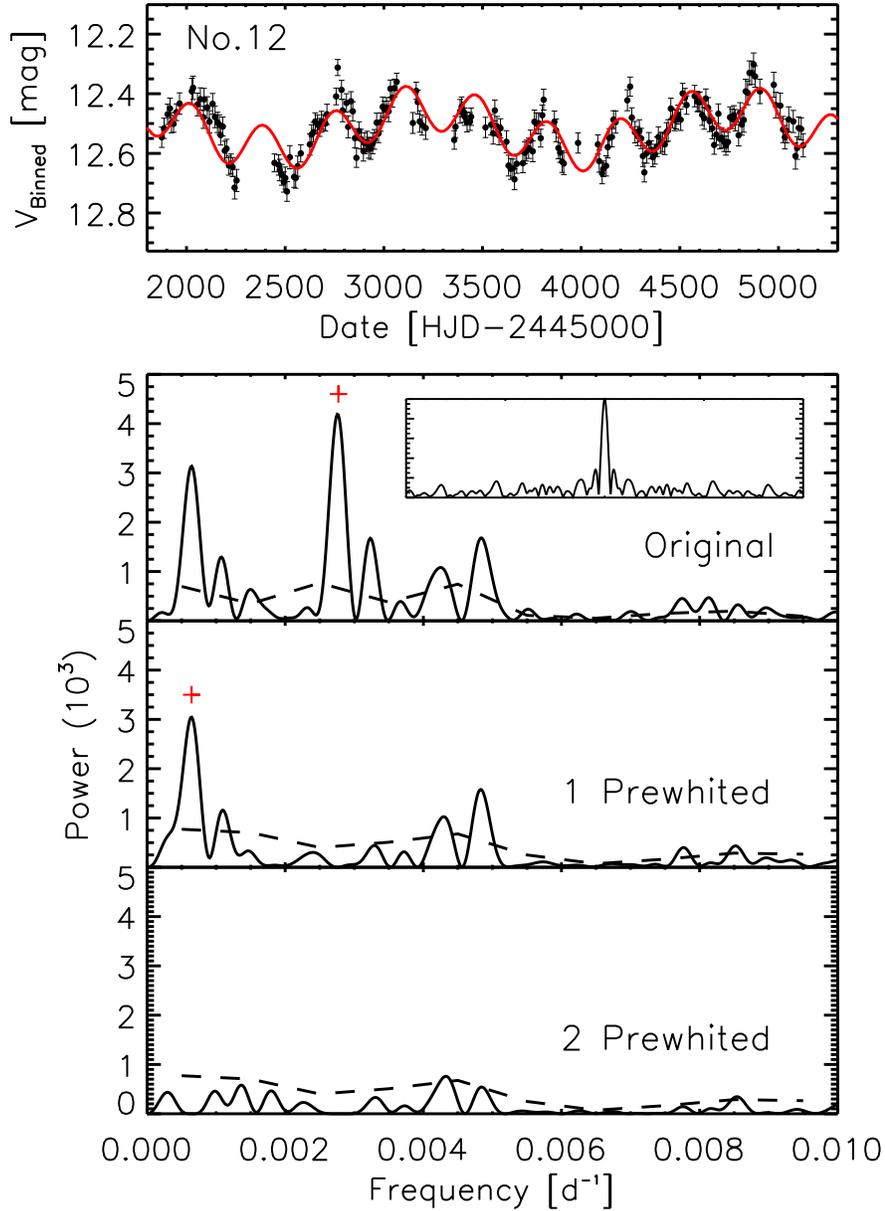}
\caption{
 The same as Fig.~\ref{pdm_process}, but using the Period04 method. Top panel is the 10~day-binned light curve with the fitted curve; bottom panels are power spectra. The spectral window in the first power spectrum is shown at the same scale as the power spectra. The dash-line shows the 4$\sigma$ level. The red cross marks the highest peak in the power spectra in each iteration.
} \label{p04_process}
\end{figure}

\clearpage


\begin{figure}
\centering
\includegraphics[width=0.8\textwidth, bb=165 360 485 715]{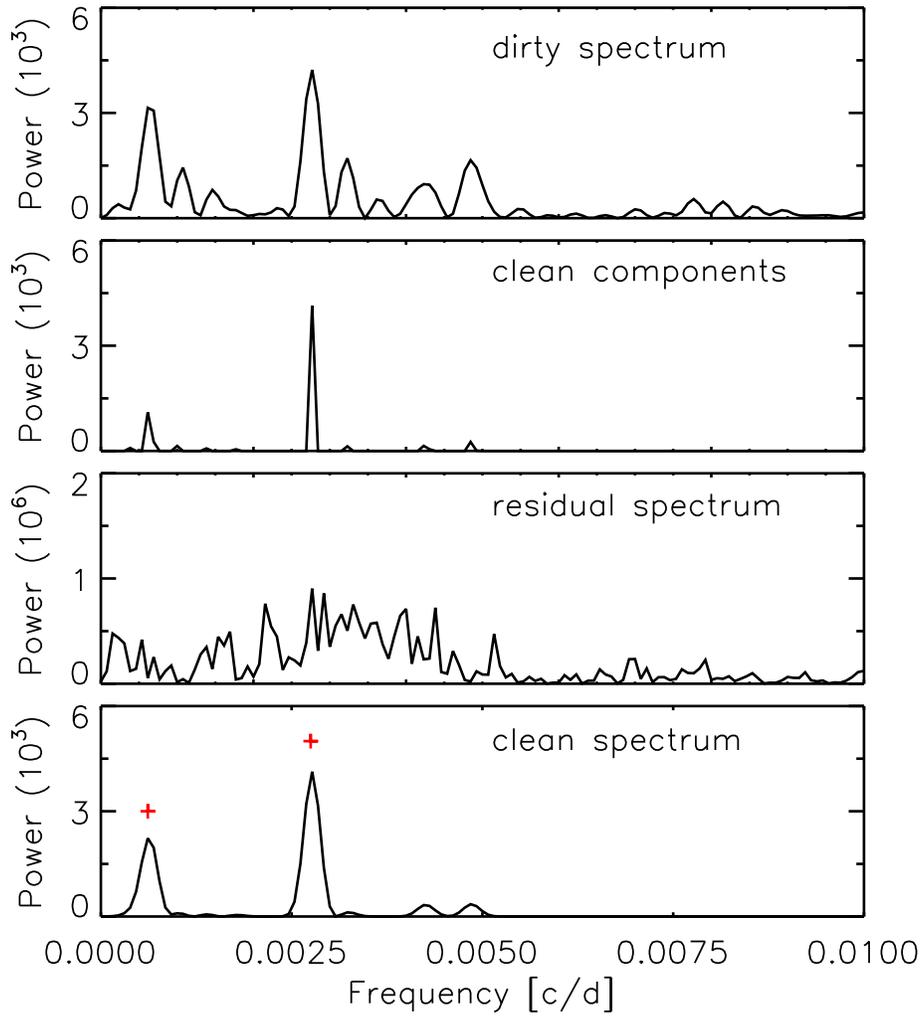}
\caption{
 An example of the CLEAN processing. From top to bottom: the ``dirty'' spectrum, clean components, residual spectrum, clean spectrum. The red cross marks the highest peak in the clean spectrum.
} \label{clean_process}
\end{figure}

\clearpage


\begin{figure}
\centering
\includegraphics[width=0.8\textwidth, bb=40 160 490 720]{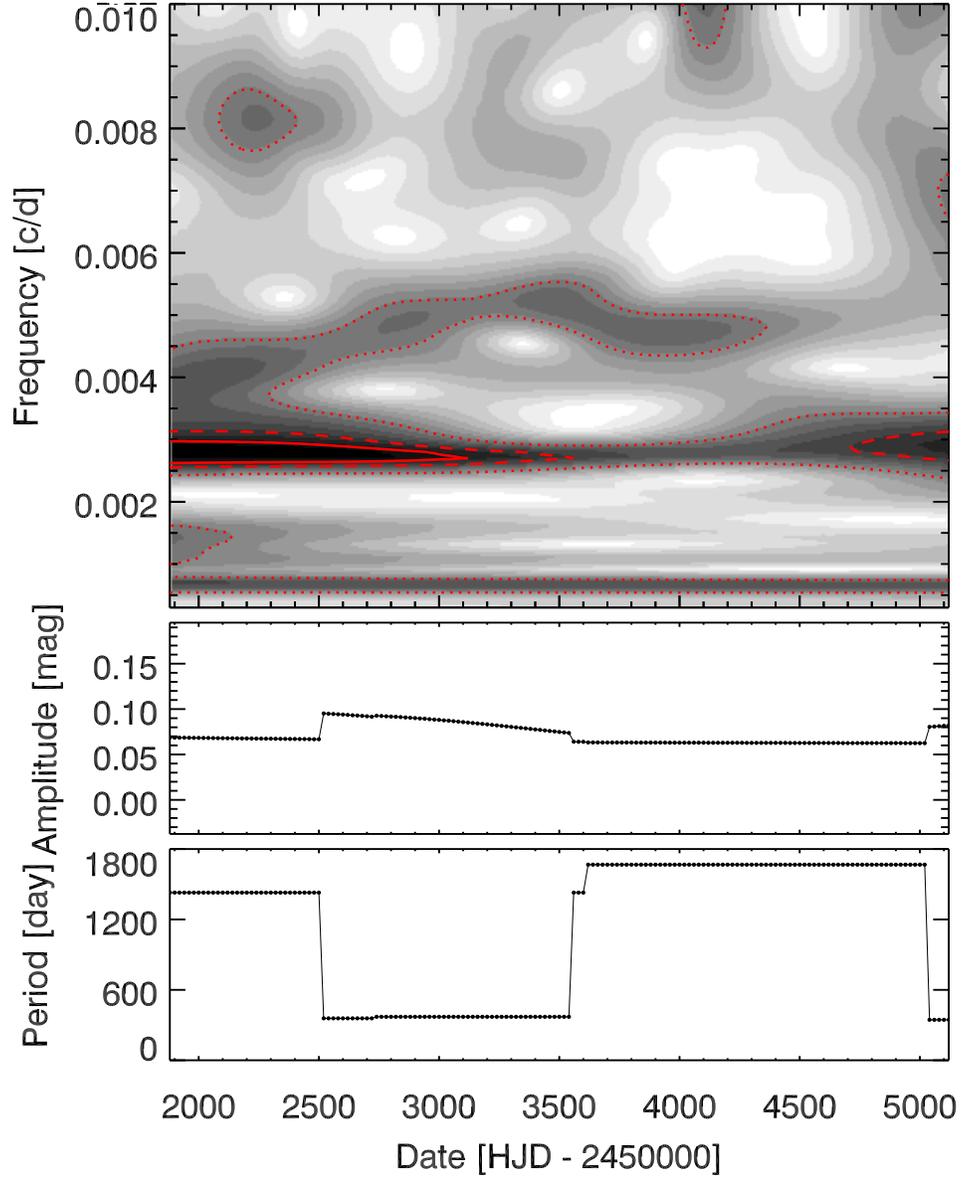}
\caption{
 An example of the WWZ processing. From top to bottom: the contour map of WWZ, the time-varying amplitude, the time-varying period. The dense regions in the contour map are marked by dot, dash and solid lines at the level of 32, 64 and 128 respectively.
} \label{wwz_process}
\end{figure}

\clearpage


\begin{figure}
\centering
\includegraphics[width=\textwidth, bb=30 400 555 666]{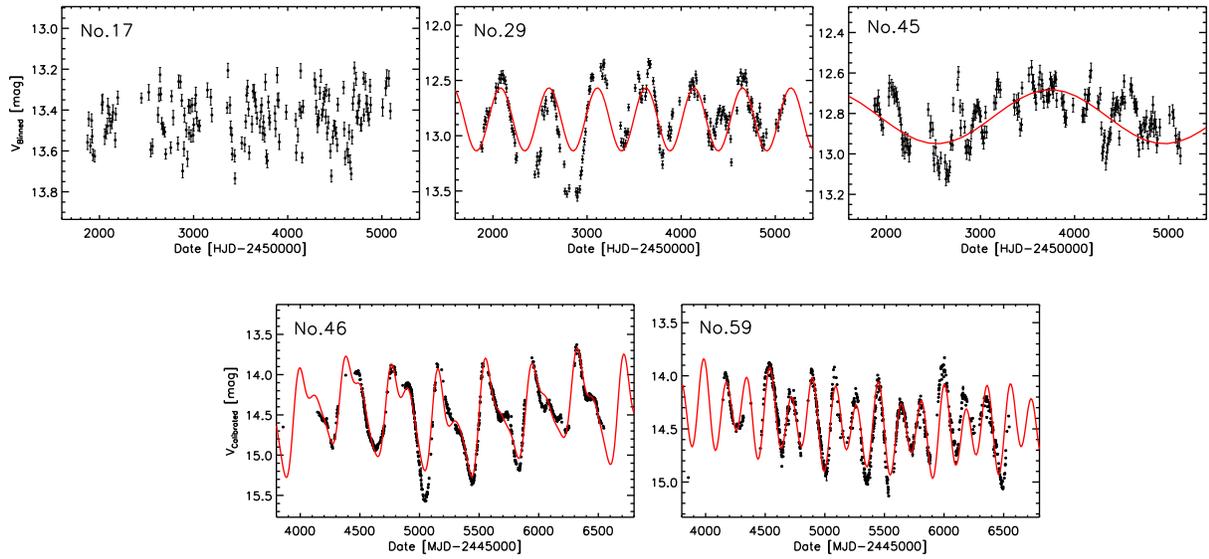}
\caption{
 The typical light curves of three kinds of variation in the ASAS data and two RSGs with the MACHO data. Top panel from left to right: irregular, semi-regular and long secondary period variables. Bottom panel: two MACHO targets with complex light variation. In each diagram, the target ID is present in the top left corner, and red line for the fitted curve.
} \label{lightcurve}
\end{figure}

\clearpage


\begin{figure}
\centering
\includegraphics[width=\textwidth, bb=50 290 510 730]{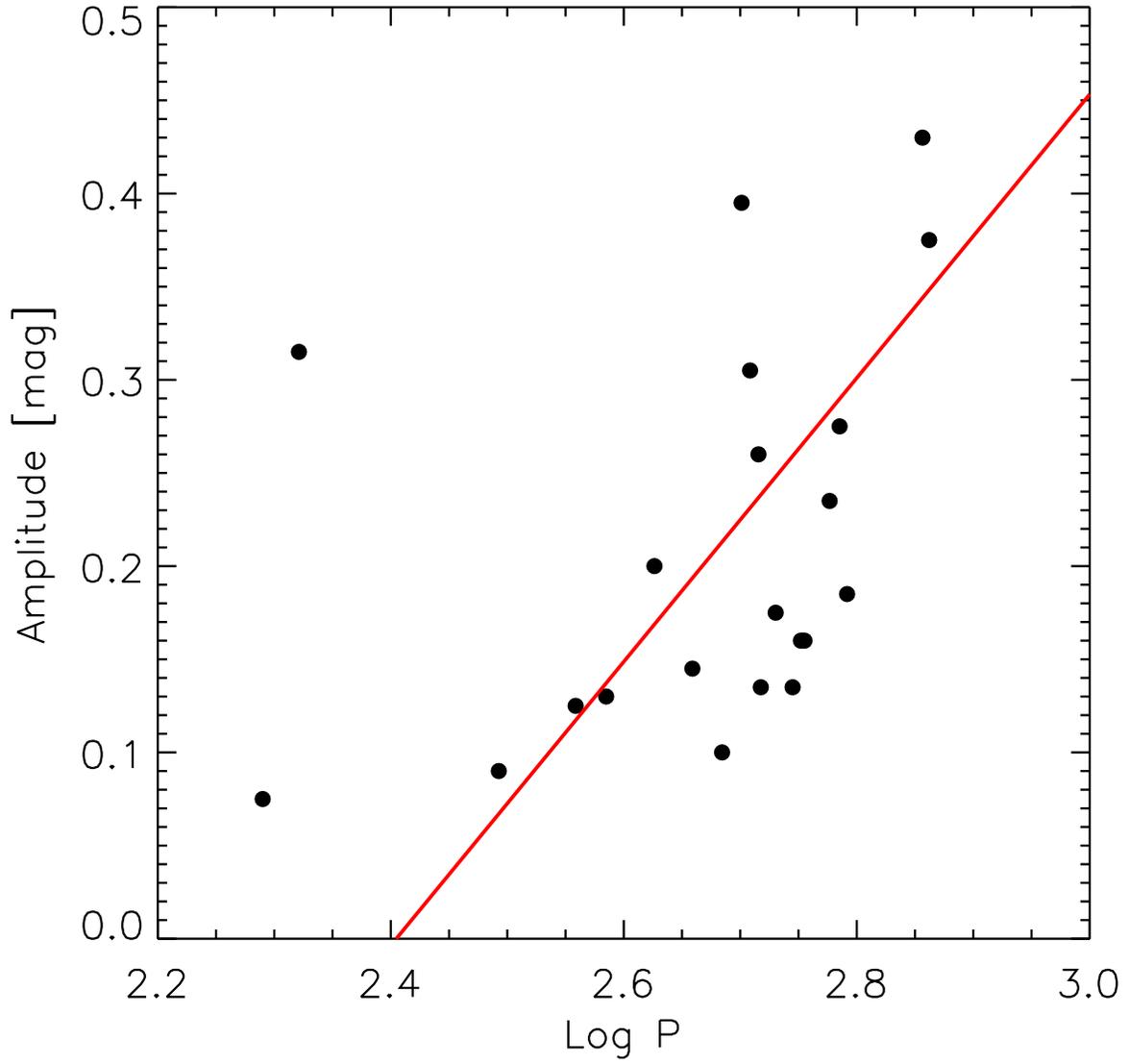}
\caption{
 The change of amplitude with period for the 6 semi-regular RSGs and 15 LSP RSGs with distinguishable short period, where the red solid line is the linearly fitted curve, $\rm \Delta V=(0.76\pm0.14)\times\log P-(1.83\pm0.24)$ with $\sigma=0.10$.
} \label{figpa}
\end{figure}

\clearpage


\begin{figure}
\centering
\includegraphics[width=\textwidth, bb=50 300 510 740]{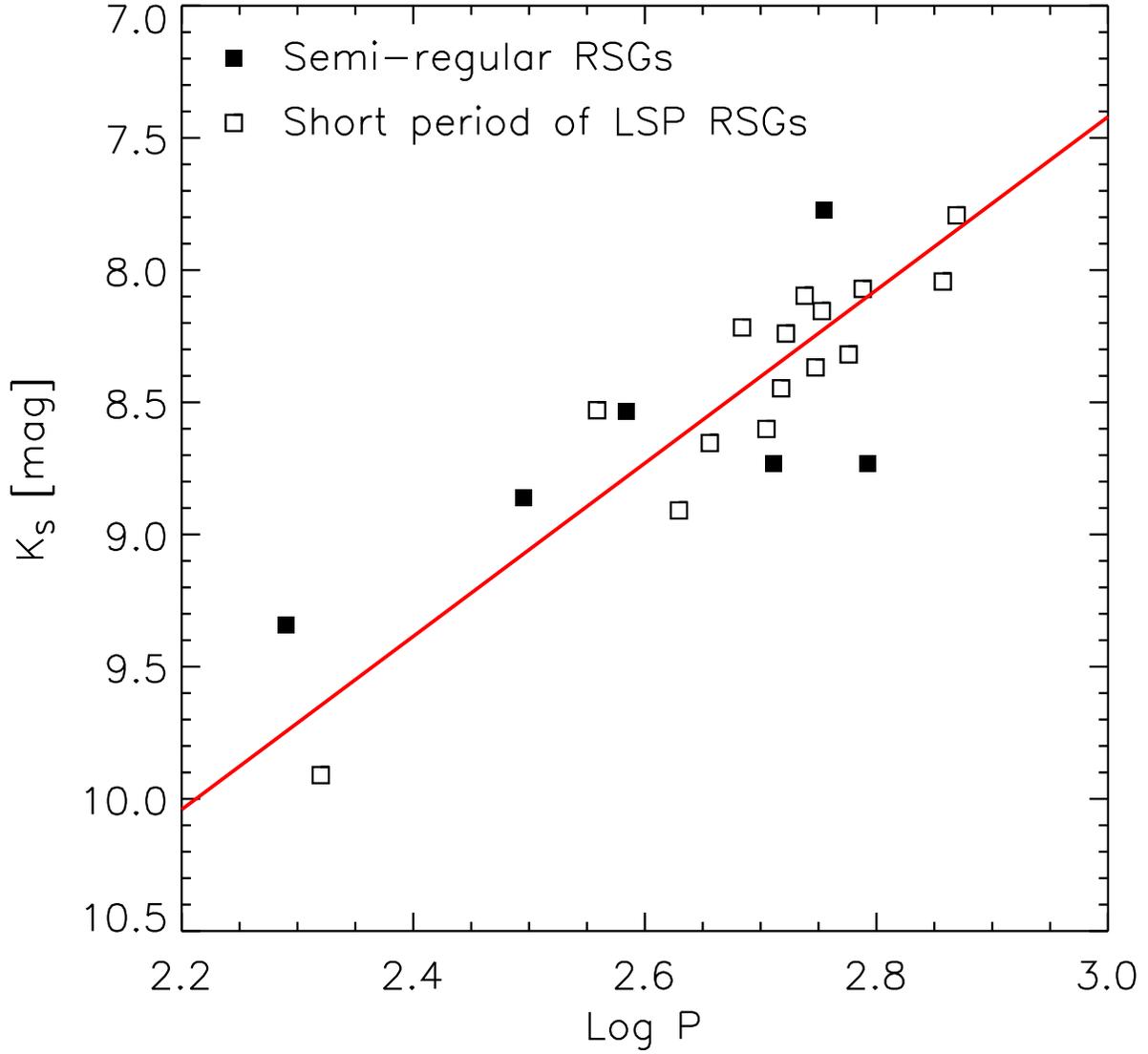}
\caption{
 The K$_{\rm S}$-band P-L relation of the RSGs in SMC, where the fitted line (red solid) writes K$_{\rm S}=(-3.28\pm0.39) \times \log \rm P+(17.25\pm0.64)$ with $\sigma=0.27$.
} \label{smcplk}
\end{figure}

\clearpage


\begin{figure}
\centering
\includegraphics[width=\textwidth, bb=5 315 555 715]{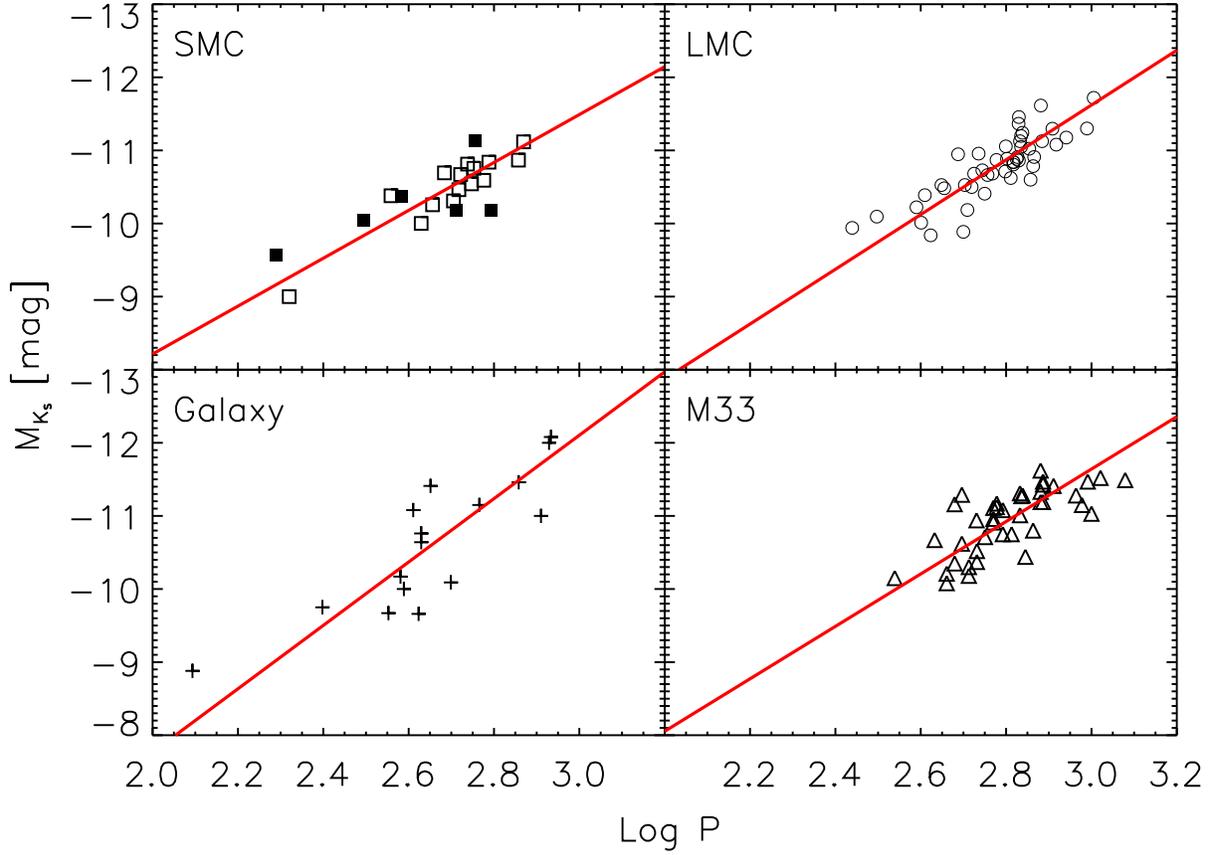}
\caption{
 The same as Fig.~\ref{smcplk} but for absolute magnitude in the K$_{\rm S}$-band and the RSGs in the Galaxy, LMC, SMC and M33 respectively. The convention of symbols and line styles are kept in following diagrams. The fitted parameters are listed in Table.~\ref{individualpl} as a form of $\rm K_{S}= a \times \log P +b$.
} \label{individualplk}
\end{figure}

\clearpage


\begin{figure}
\centering
\includegraphics[width=\textwidth, bb=50 300 510 740]{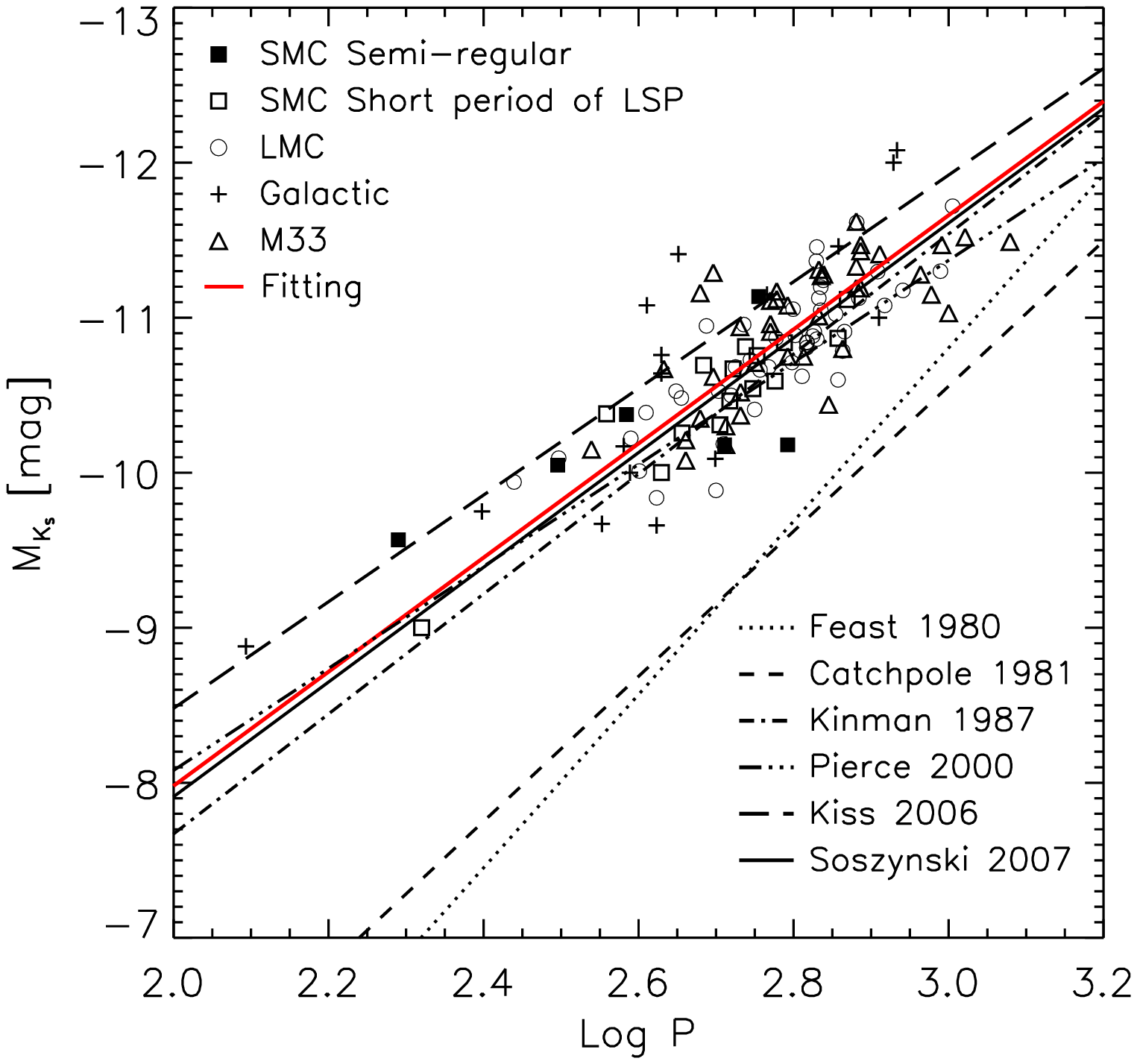}
\caption{
 The K$_{\rm S}$-band P-L relation of the RSGs in the Galaxy, LMC, SMC and M33, similar to Fig.~\ref{individualplk}, with the addition of the P-L relations of RSGs derived by other people \citep{Feast80, Catchpole81, Kinman87, Pierce00, Kiss06}. The AGB $a_{2}$ sequence from \citet{Soszynski07aca} is almost superposed on ours.
} \label{allplk}
\end{figure}

\clearpage


\begin{figure}
\centering
\includegraphics[width=\textwidth, bb=0 310 575 730]{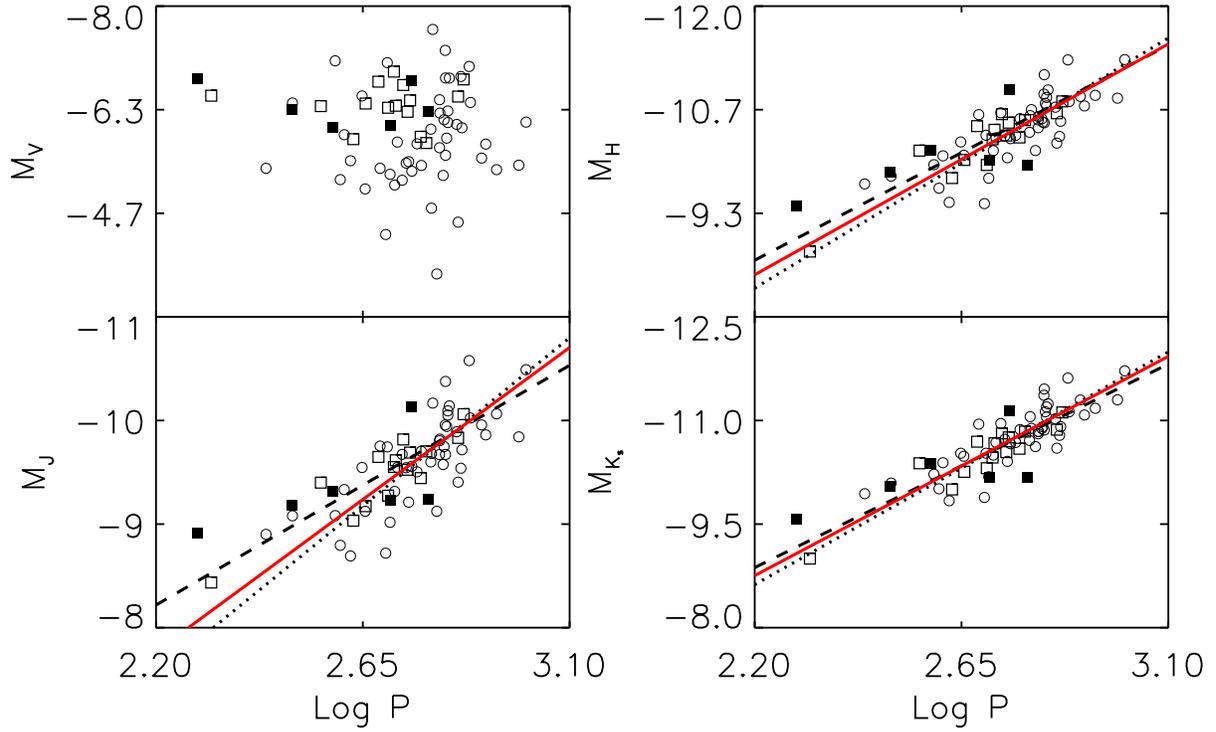}
\caption{
 Comparison of the P-L relation in the V, J, H, and K$_{\rm S}$ bands of RSGs with the LMC. The black dash- and dot-lines are the fitted line for SMC and LMC respectively, and the red solid line for a combination of SMC and LMC. The V band radiation suffers the interstellar and circumstellar extinction and exhibits no correlation with the period.
} \label{pl-semi-int1}
\end{figure}

\clearpage


\begin{figure}
\centering
\includegraphics[width=\textwidth, bb=0 310 575 730]{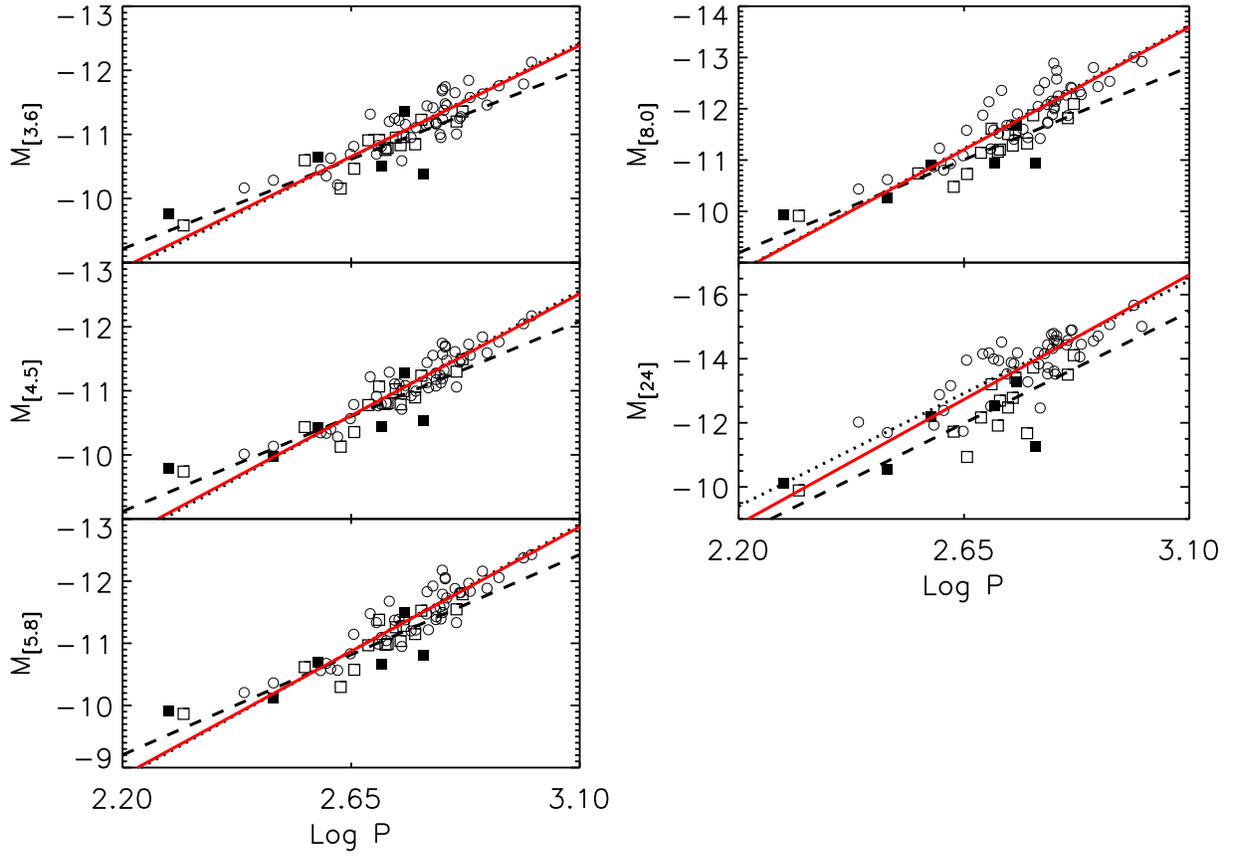}
\caption{
 The same as Fig.~\ref{pl-semi-int1} but for the [3.6], [4.5], [5.8], [8.0] and [24] bands.
} \label{pl-semi-int2}
\end{figure}

\clearpage


\begin{figure}
\centering
\includegraphics[width=\textwidth, bb=50 300 515 740]{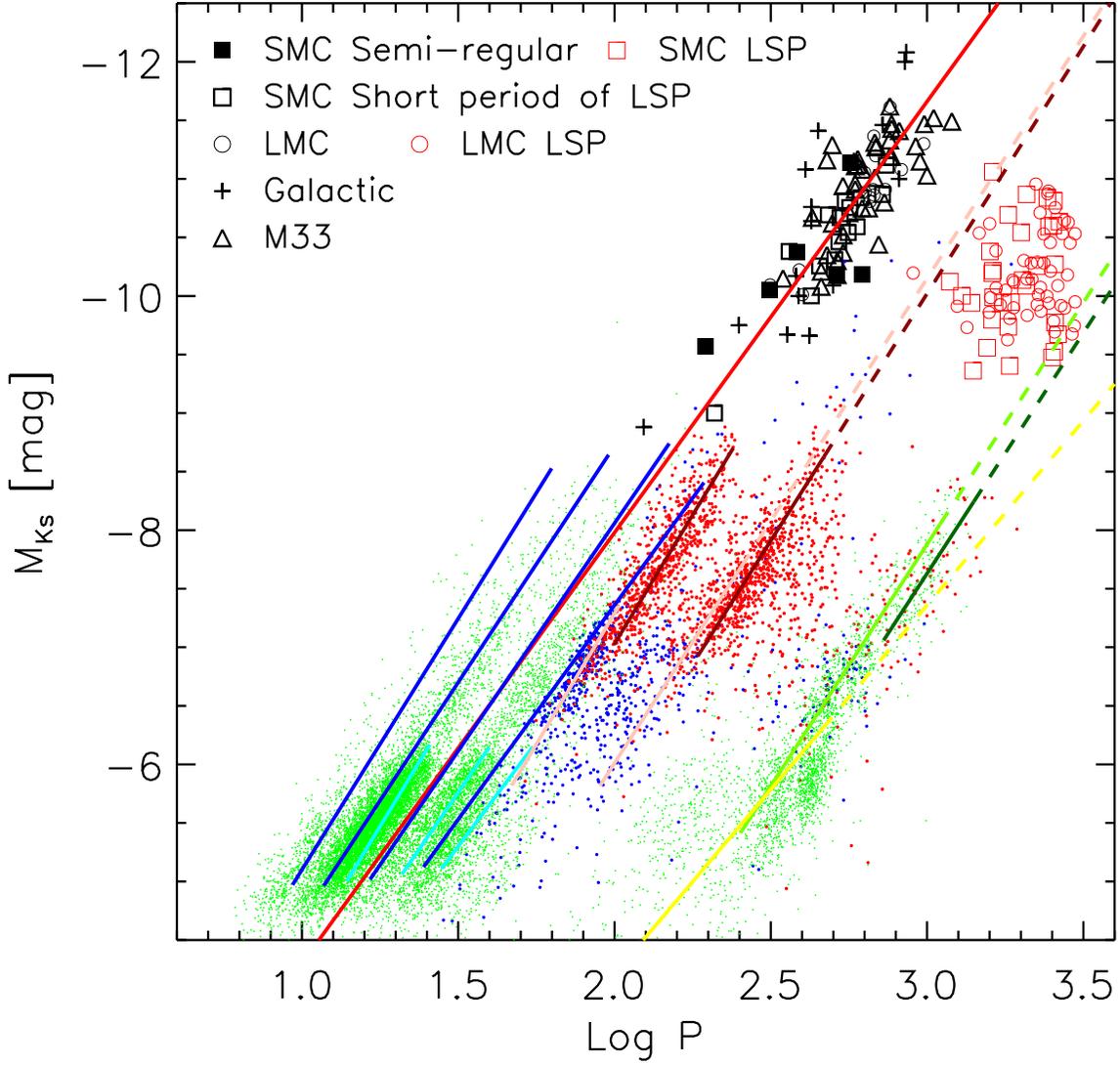}
\caption{
 The P-L relations for the LSP of LSP RSGs where the short periods are plotted for comparison. Superposed in this diagram are the P-L relations of LPVs (solid lines) from \citet{Soszynski07aca} and various LPVs targets (points) from \citet{Soszynski11aca} in the SMC, where the symbols are the same as the original papers. With the addition of more and fainter objects in \citet{Soszynski11aca}, there are small shifts of the P-L relations by \citet{Soszynski07aca} from the locations of the objects.
} \label{lspsmc}
\end{figure}

\clearpage


\begin{figure}
\centering
\includegraphics[width=\textwidth, bb=60 300 520 740]{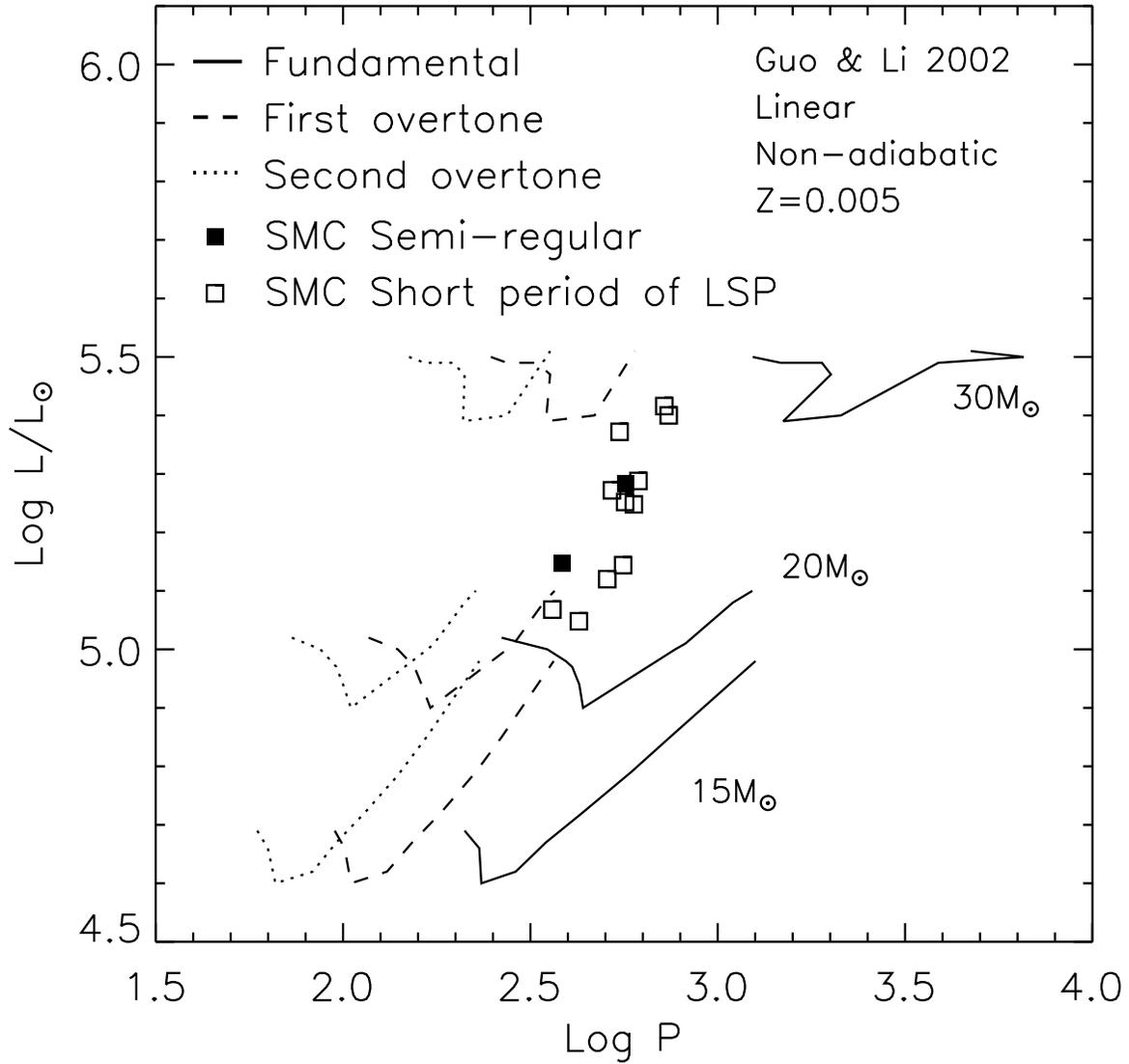}
\caption{
 The P-L relation for the first three radial modes (fundamental mode at the right and second overtone at the left) at 15, 20, and 30 M$_{\sun}$ stars with Z=0.005 from the model by \citet{Guo02}. The objects in our sample with reliable luminosity are shown mostly between 20 and 30 M$_{\sun}$ by squares.
} \label{guosmcpl}
\end{figure}

\clearpage


\begin{figure}
\centering
\includegraphics[width=\textwidth, bb=60 300 520 740]{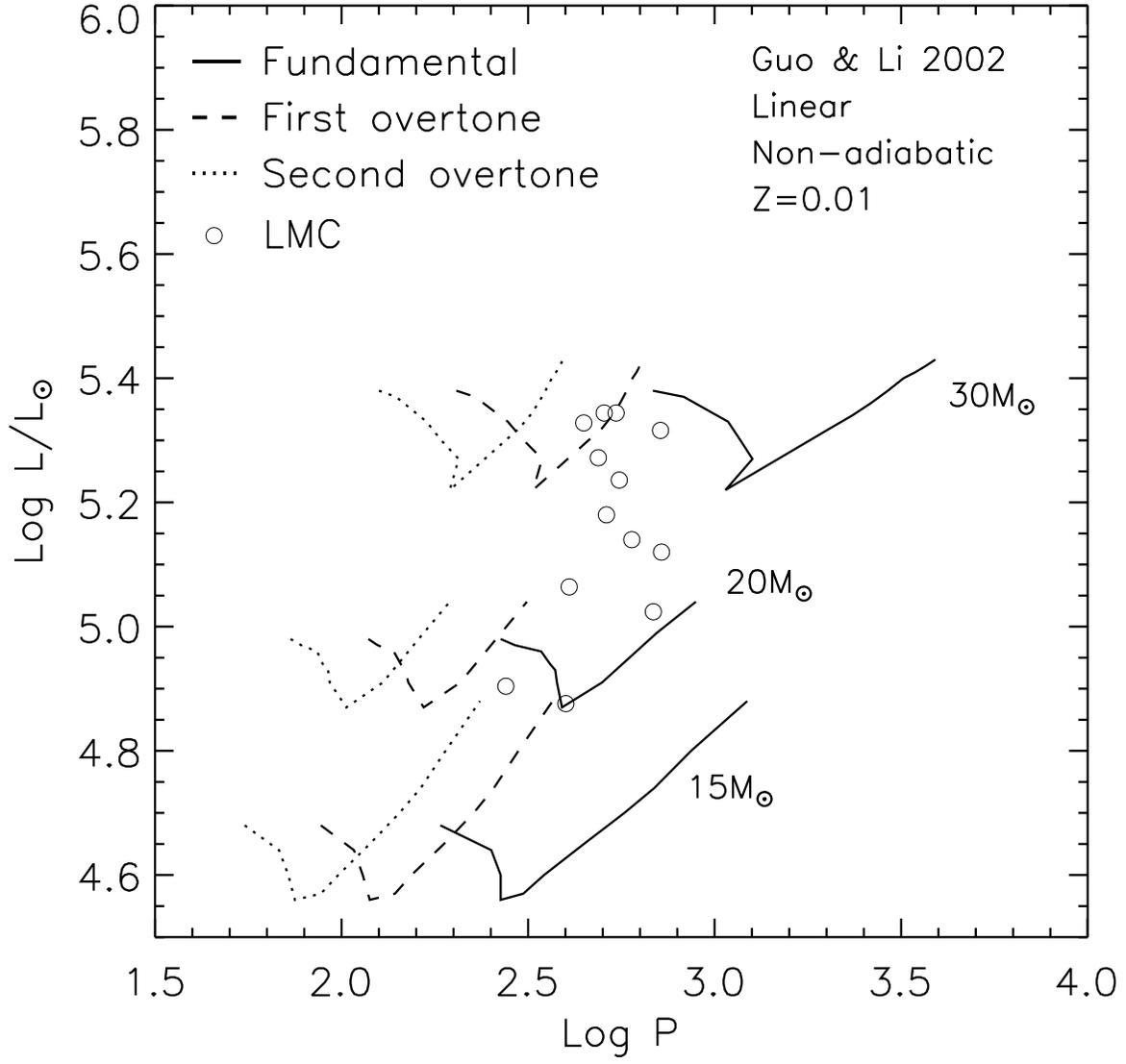}
\caption{
 The same as Fig.~\ref{guolmcpl} but for the models with Z=0.01 from \citet{Guo02} and RSGs in LMC.
} \label{guolmcpl}
\end{figure}

\clearpage


\begin{figure}
\centering
\includegraphics[width=\textwidth, bb=60 300 520 740]{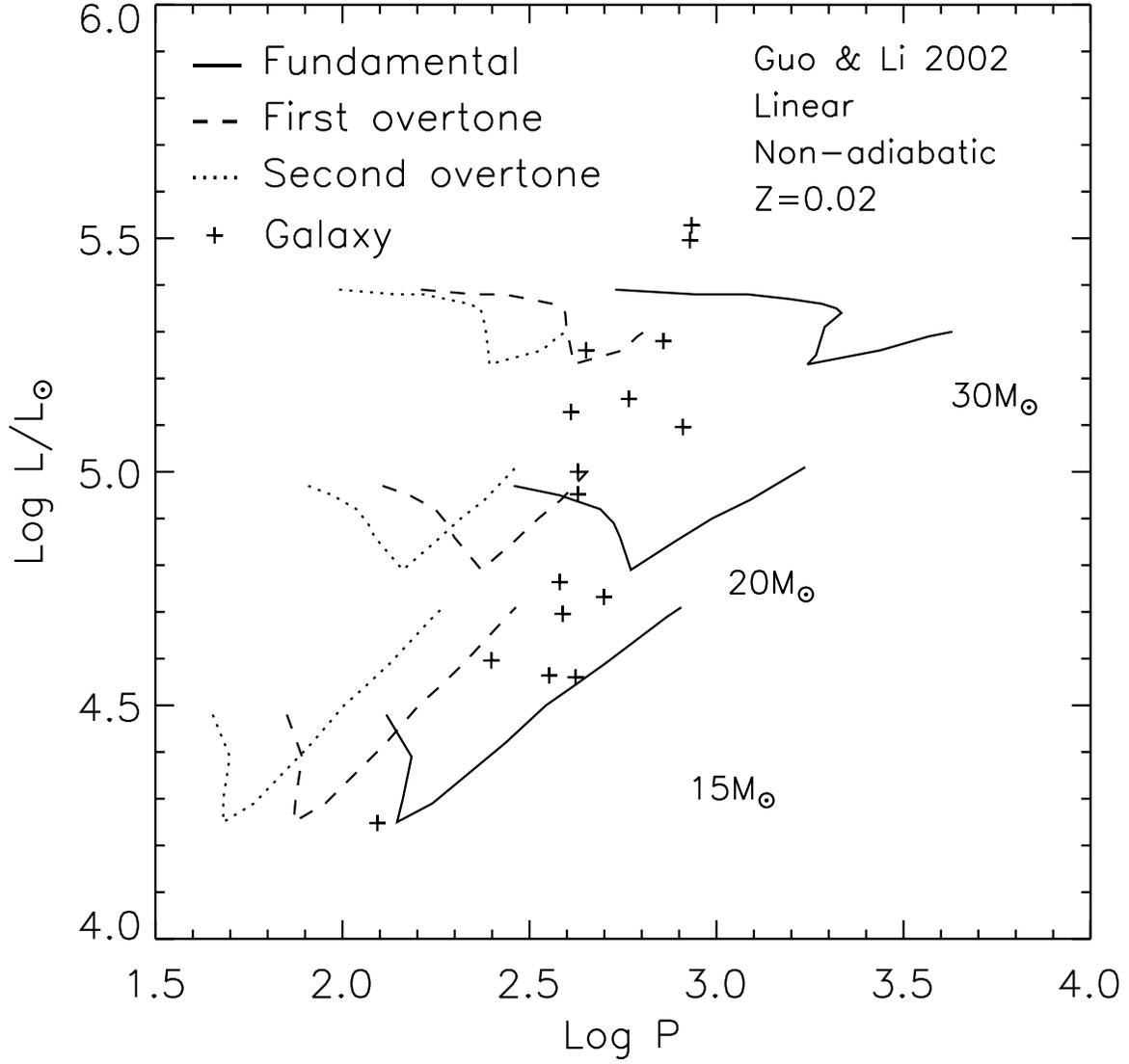}
\caption{
 The same as Fig.~\ref{guolmcpl} but for the models at Z=0.02 from \citet{Guo02} and RSGs in the Galaxy.
} \label{guogalaxypl}
\end{figure}

\clearpage


\begin{figure}
\centering
\includegraphics[width=\textwidth, bb=60 300 520 740]{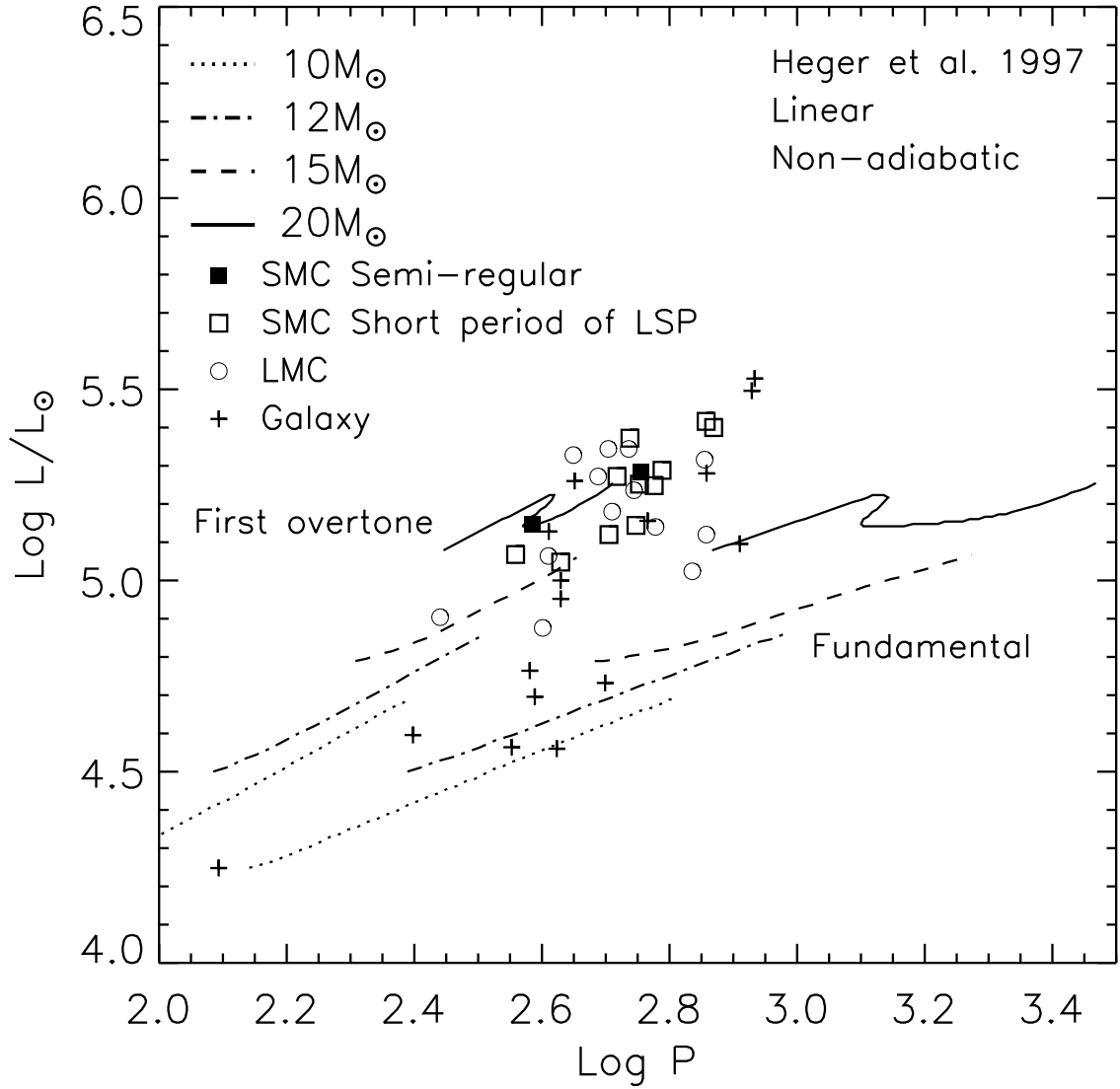}
\caption{
 The Same as Fig.~\ref{guolmcpl} but for the models by \citet{Heger97} which have no discrimination of the metallicity and for all the objects in SMC, LMC and the Galaxy.
} \label{hegerpl}
\end{figure}

\clearpage


\end{document}